\documentclass[aps,prd,showpacs,notitlepage,nofootinbib,preprintnumbers,amsmath,amssymb]{revtex4-1}

\usepackage{graphics,graphicx}% Include figure files
\usepackage{dcolumn}% Align table columns on decimal point
\usepackage{bm}% bold math
\usepackage{epsfig}
\usepackage[usenames]{color}
\usepackage[colorlinks=true,citecolor=blue,linkcolor=blue,urlcolor=blue]{hyperref}
\usepackage{mathbbol}
\usepackage{epstopdf}
\usepackage{simplewick}
\usepackage[utf8]{inputenc} 
\usepackage{slashed}
\usepackage{stackrel}
\usepackage{mathrsfs}
\usepackage{xcolor}
\usepackage{subcaption}
\usepackage[most]{tcolorbox}
\usepackage{physics}
\usepackage[normalem]{ulem}

\usepackage{ragged2e} % for the \justifying macro
\DeclareCaptionJustification{justified}{\justifying}

\def\eq#1{{Eq.~(\ref{#1})}}
\def\eqs#1{{Eqs.~(\ref{#1})}}
\def\fig#1{{Fig.~\ref{#1}}}
\newcommand{\ben}{\begin{eqnarray*}}
\newcommand{\een}{\end{eqnarray*}}
\newcommand{\un}[1]{\underline{#1}}

\newcommand{\thalf}{\tfrac{1}{2}}
\newcommand{\llangle}{\Big\langle \!\! \Big\langle}
\newcommand{\rrangle}{\Big\rangle \!\! \Big\rangle}
\newcommand{\cc}{\mbox{c.c.}}

\newcommand{\as}{\alpha_s}

\newcommand{\dhd}{{\textstyle d}
\lower.03ex\hbox{\kern-0.38em$^{\scriptstyle-}$}\kern-0.05em{}}
\newcommand{\dbar}{{\textstyle \delta}
\lower.03ex\hbox{\kern-0.38em$^{\scriptstyle-}$}\kern-0.05em{}}
\newcommand{\half}{{1\over 2}}

\newcommand{\ul}[1]{\underline{#1}}

\newcommand{\tord}{\textrm{T} \:}
\newcommand{\atord}{\overline{\textrm{T}} \:}

\makeatletter
\DeclareRobustCommand{\cev}[1]{%
  {\mathpalette\do@cev{#1}}%
}
\newcommand{\do@cev}[2]{%
  \vbox{\offinterlineskip
    \sbox\z@{$\m@th#1 x$}%
    \ialign{##\cr
      \hidewidth\reflectbox{$\m@th#1\vec{}\mkern4mu$}\hidewidth\cr
      \noalign{\kern-\ht\z@}
      $\m@th#1#2$\cr
    }%
  }%
}
\makeatother

\newcommand{\dprime}{{\prime \prime}}

\newcommand{\iUV}{
    \int \displaylimits^z_{\frac{1}{s x_{10}^2}} \frac{dz^\prime }{z^\prime} \int \displaylimits^{x_{10}^2}_{\frac{1}{z^\prime s}} \frac{dx_{21}^2}{x_{21}^2}
}
\newcommand{\iIR}{
\int \displaylimits^{z}_{\frac{\Lambda^2}{s}} \frac{dz^\prime}{z^\prime} \int \displaylimits^{\mathrm{min}[\frac{z}{z^\prime} x_{10}^2, \frac{1}{\Lambda^2}]}_{\mathrm{max}[x_{10}^2, \frac{1}{z^{\prime}s}]} \frac{d x_{21}^2}{x_{21}^2}
}

\newcommand{\inUV}{\int \displaylimits^{z^\prime}_{\frac{1}{sx_{10}^2}} \frac{dz^{\dprime}}{z^{\dprime}} \int \displaylimits^{\mathrm{min}[x_{10}^2, x_{21}^2 \frac{z^\prime}{z^{\dprime}}]}_{\frac{1}{z^{\dprime} s}} \frac{d x_{32}^2}{x_{32}^2}}

\newcommand{\inIR}{\int \displaylimits^{z^\prime \frac{x_{21}^2}{x_{10}^2}}_{\frac{\Lambda^2}{s}} \frac{dz^{\dprime}}{z^{\dprime}} 
    \int \displaylimits^{\mathrm{min}[\frac{z^\prime}{z^{\dprime}} x_{21}^2, \frac{1}{\Lambda^2}]}_{\mathrm{max}[x_{10}^2, \frac{1}{z^{\dprime}s}]} \frac{d x_{32}^2}{x_{32}^2} }

\newcommand{\Sec}[1]{Sec.~\ref{#1}}

\begin{document}

\title{Elastic Dijet Production in Electron Scattering on a Longitudinally Polarized Proton at Small $x$: A Portal to Orbital Angular Momentum Distributions}

\author{Yuri V. Kovchegov} 
         \email[Email: ]{kovchegov.1@osu.edu}
         \affiliation{Department of Physics, The Ohio State
           University, Columbus, OH 43210, USA}

\author{Brandon Manley}
  \email[Email: ]{manley.329@osu.edu}
	\affiliation{Department of Physics, The Ohio State
           University, Columbus, OH 43210, USA}
           
\date{\today}

\begin{abstract}
\noindent 
We calculate the elastic production of dijets from electron collisions with a longitudinally polarized proton target at small values of the Bjorken $x$ variable. Building on the pioneering proposals of \cite{Hatta:2016aoc,Bhattacharya:2022vvo, Bhattacharya:2023hbq, Bhattacharya:2024sck} for measuring the quark and gluon orbital angular momentum (OAM) distributions, our focus is on both the longitudinal double spin asymmetry (DSA) and longitudinal single spin asymmetry (SSA). We compute the numerators of these asymmetries 
in the small-$x$ formalism of the light cone operator treatment. Utilizing the small-$x$ expressions for the OAM distributions derived earlier in \cite{Kovchegov:2023yzd}, we demonstrate that the DSA provides a robust probe for both the quark and gluon OAM distributions within the proton. In contrast, we find that while the SSA is also sensitive to the OAM distributions, extraction of the latter from the SSA would require new developments in small-$x$ theory and phenomenology, and is probably not feasible at this point in time. These findings highlight the potential of DSA measurements in elastic dijet production at the future Electron-Ion Collider to provide the first-ever direct access to the quark and gluon OAM distributions at small $x$, paving the way for new insights into the proton spin puzzle.
\end{abstract}
\maketitle
\tableofcontents

%%%%%%%%%%%%%%%%%%%%%%%%%%%%%%%%%%%%%%%%%%%%%%%

\section{Introduction}

Orbital angular momentum (OAM) of the quarks and gluons in the proton has never been measured experimentally. At the same time, quark and gluon OAM is an integral part of the proton spin budget, which is quantified using the spin sum rules \cite{Jaffe:1989jz, Ji:1996ek}. The Jaffe--Manohar sum rule reads
\begin{equation}
S_{q + \bar q} +L_{q + \bar q}+S_G+L_G=\frac{1}{2}\, ,  
\label{eqn:JM}
\end{equation}
where the quark and gluon OAM contributions at momentum scale $Q$, $L_{q + \bar q} (Q^2)$ and $L_G (Q^2)$, can be written as the integrals of OAM distributions \cite{Bashinsky:1998if, Hagler:1998kg, Harindranath:1998ve, Hatta:2012cs, Ji:2012ba} over all momentum fractions $x$,
\begin{align}\label{eqn:LqLG}
    L_{q + \bar q} (Q^2) = \int\limits_0^1 dx \, L_{q + \bar q} (x,Q^2)\,, \qquad\qquad
    L_G(Q^2) = \int\limits_0^1 dx \, L_G (x,Q^2)\,.
\end{align}
Above, $S_{q + \bar q} (Q^2)$ and $S_G (Q^2)$ are the contributions of the quark and gluon helicity distributions to the proton spin, which have been studied much more than the OAM distributions $L_{q + \bar q} (x,Q^2)$ and $L_G (x,Q^2)$ in \eq{eqn:LqLG} (see \cite{EuropeanMuon:1987isl, Boer:2011fh, Aidala:2012mv, Accardi:2012qut, Leader:2013jra, Aschenauer:2013woa, Aschenauer:2015eha,  Proceedings:2020eah, Ji:2020ena, AbdulKhalek:2021gbh} and references therein). Clearly, a solution of the proton spin puzzle is impossible without a solid quantitative understanding of the OAM distributions. 

The evolution equations in $Q^2$ for the OAM distributions are known \cite{Hagler:1998kg,Hoodbhoy:1998yb}, but, to the best of the authors' knowledge, have never been used in phenomenology, owing to the lack of observables sensitive to the OAM distributions. The situation may change, however, due to the recent proposals for measuring the gluon and quark OAM distributions put forward in \cite{Hatta:2016aoc,Bhattacharya:2022vvo, Bhattacharya:2023hbq, Bhattacharya:2024sck}. The proposals involve elastic dijet and pion production cross sections in the longitudinally polarized lepton--proton scattering. The goal of the present work is to investigate the proposals \cite{Hatta:2016aoc,Bhattacharya:2022vvo, Bhattacharya:2024sck} in the small-$x$ regime, building on the earlier studies of OAM distributions in that kinematics conducted by the authors \cite{Kovchegov:2023yzd, Kovchegov:2019rrz, Manley:2024pcl}.   

In our previous paper \cite{Kovchegov:2023yzd} we derived the expressions for $L_{q + \bar q} (x,Q^2)$ and $L_G (x,Q^2)$ at small $x$, relating them to the so-called polarized dipole amplitudes and to the impact-parameter moments of those amplitudes, which will be defined below. The OAM distributions at small $x$ are subeikonal quantities: they are suppressed by a power of $x$ compared to the eikonal unpolarized parton distribution functions (PDFs) for quarks and gluons. Therefore, in order to calculate the OAM distributions in the small-$x$ $s$-channel/shock-wave formalism, one has to extend that standard approach of \cite{Mueller:1994rr,Mueller:1994jq,Mueller:1995gb,Balitsky:1995ub,Balitsky:1998ya,Kovchegov:1999yj,Kovchegov:1999ua,Jalilian-Marian:1997dw,Jalilian-Marian:1997gr,Weigert:2000gi,Iancu:2001ad,Iancu:2000hn,Ferreiro:2001qy} (see \cite{Gribov:1984tu,Iancu:2003xm,Weigert:2005us,JalilianMarian:2005jf,Gelis:2010nm,Albacete:2014fwa,Kovchegov:2012mbw,Morreale:2021pnn} for reviews) to include subeikonal contributions. Such corrections, and the subeikonal operator formalism that describes them, have been constructed over the past decade in \cite{Altinoluk:2014oxa,Balitsky:2015qba,Balitsky:2016dgz, Kovchegov:2017lsr, Kovchegov:2018znm, Chirilli:2018kkw, Jalilian-Marian:2018iui, Jalilian-Marian:2019kaf, Altinoluk:2020oyd, Boussarie:2020vzf, Boussarie:2020fpb, Kovchegov:2021iyc, Altinoluk:2021lvu, Kovchegov:2022kyy, Altinoluk:2022jkk, Altinoluk:2023qfr,Altinoluk:2023dww, Li:2023tlw}. Particularly relevant for this work is the case of helicity PDFs (hPDFs) in a longitudinally polarized proton, developed in \cite{Kovchegov:2015pbl, Hatta:2016aoc, Kovchegov:2016zex, Kovchegov:2016weo, Kovchegov:2017jxc, Kovchegov:2017lsr, Kovchegov:2018znm, Kovchegov:2019rrz, Boussarie:2019icw, Cougoulic:2019aja, Kovchegov:2020hgb, Cougoulic:2020tbc, Chirilli:2021lif, Kovchegov:2021lvz, Cougoulic:2022gbk, Borden:2023ugd, Adamiak:2023okq, Borden:2024bxa}. We will refer to the corresponding subeikonal operator formalism as the light cone Operator Treatment (LCOT).

Helicity PDFs are also subeikonal quantities and a novel formalism is also needed to study hPDFs at small $x$. They were first analyzed in the pioneering works of Bartels, Ermolaev and Ryskin (BER) \cite{Bartels:1995iu, Bartels:1996wc} employing the infrared evolution equations (IREE) approach \cite{Gorshkov:1966ht,Kirschner:1983di,Kirschner:1994rq,Kirschner:1994vc,Griffiths:1999dj, Blumlein:1996hb}. In the $s$-channel/shock-wave approach, hPDFs were shown to be related to the polarized dipole amplitudes in \cite{Kovchegov:2015pbl, Hatta:2016aoc, Kovchegov:2016zex, Cougoulic:2022gbk}. The small-$x$ evolution equations for these subeikonal polarized dipole amplitudes were derived in \cite{Kovchegov:2015pbl, Kovchegov:2016zex, Kovchegov:2018znm} by Kovchegov, Pitonyak, and Sievert  (KPS) and corrected in \cite{Cougoulic:2022gbk} by Cougoulic, Kovchegov, Tarasov, and Tawabutr (KPS-CTT). Very recently, another revision of the helicity evolution equations was derived by Borden, Kovchegov, and Li in \cite{Borden:2024bxa}, affecting {\sl only} the quark sector due to the quark to gluon and gluon to quark transition operators in the shock wave which were missing in the KPS-CTT evolution. Independently, small-$x$ helicity evolution was derived was derived by Chirilli in \cite{Chirilli:2021lif}, which appears to be similar to the KPS evolution in the gluon sector and contains the appropriate transition operators in the quark sector. Therefore, we will refer to the equations derived in \cite{Borden:2024bxa} in the large-$N_c \& N_f$ limit \cite{Veneziano:1976wm} as the KPS-CTT-BCL equations (with $N_c$ the number of quark colors and $N_f$ the number of quark flavors). Both these and BER equations resum powers of $\as \ln^2 (1/x)$ and $\as \, \ln (1/x) \, \ln (Q^2/\Lambda^2)$ with $\as$ the strong coupling constant and $\Lambda$ an infrared (IR) cutoff: we will refer to such resummation as the double-logarithmic approximation (DLA). Similar to the eikonal unpolarized evolution \cite{Mueller:1994rr,Mueller:1994jq,Mueller:1995gb,Balitsky:1995ub,Balitsky:1998ya,Kovchegov:1999yj,Kovchegov:1999ua,Jalilian-Marian:1997dw,Jalilian-Marian:1997gr,Weigert:2000gi,Iancu:2001ad,Iancu:2000hn,Ferreiro:2001qy}, the evolution equations for the polarized dipole amplitudes close in the large-$N_c$ limit \cite{tHooft:1973alw}. Such large-$N_c$ equations have been solved numerically in \cite{Cougoulic:2022gbk} and analytically in \cite{Borden:2023ugd}, with the resulting polarized gluon-gluon anomalous dimension being in agreement with the small-$x$ limit of the known perturbative results \cite{Altarelli:1977zs,Dokshitzer:1977sg,Zijlstra:1993sh,Mertig:1995ny,Moch:1999eb,vanNeerven:2000uj,Vermaseren:2005qc,Moch:2014sna,Blumlein:2021ryt,Blumlein:2021lmf,Davies:2022ofz,Blumlein:2022gpp} to the existing three loops. Helicity evolution equations \cite{Kovchegov:2015pbl, Kovchegov:2018znm, Cougoulic:2022gbk, Borden:2024bxa} also close in the large-$N_c \& N_f$ limit: this regime has only been explored numerically so far \cite{Kovchegov:2020hgb, Adamiak:2023okq}, though agreement with polarized anomalous dimensions up to and including three loops was established as well in \cite{Borden:2024bxa} using an iterative analytic solution. The solutions of the KPS-CTT and KPS-CTT-BCL helicity evolution equations derived in the LCOT framework appear to be in overall agreement with that in the BER framework, except for the minor (a few percent) difference in the resulting intercepts and in the terms in the expansion of the anomalous dimensions beyond three loops. The origin of the difference is not clear at this point. Phenomenology based on the helicity evolution equations \cite{Kovchegov:2015pbl, Kovchegov:2016zex, Kovchegov:2018znm, Cougoulic:2022gbk}, describing the world data on polarized deep inelastic scattering (DIS) and semi-inclusive DIS (SIDIS) at small $x$ has been developed in \cite{Adamiak:2021ppq, Adamiak:2023yhz}.      

The studies of OAM distributions at small $x$ started more recently: in one of the earlier works \cite{Hatta:2018itc} the small-$x$ behavior of the OAM distributions was studied in the the Dokshitzer-Gribov-Lipatov-Altarelli-Parisi (DGLAP) evolution equations \cite{Gribov:1972ri, Altarelli:1977zs, Dokshitzer:1977sg} framework. The small-$x$ limit of DGLAP evolution sums up powers of $\as \, \ln (1/x) \, \ln (Q^2/\Lambda^2)$. In the BER IREE framework the OAM distributions small-$x$ asymptotics was found in  
\cite{Boussarie:2019icw}. In the LCOT framework, the initial attempt to construct the small-$x$ evolution for OAM distributions \cite{Kovchegov:2019rrz} was corrected in our recent work~\cite{Kovchegov:2023yzd}. The evolution equations derived in \cite{Kovchegov:2023yzd} in the large-$N_c$ limit and in DLA give us the evolution of the moment dipole amplitudes, needed to calculate $L_{q + \bar q} (x,Q^2)$ and $L_G (x,Q^2)$. (Similar technique can be applied to the equations derived in \cite{Borden:2024bxa} to obtain DLA evolution for the moment dipole amplitudes at large $N_c \& N_f$.) The large-$N_c$ equations from \cite{Kovchegov:2023yzd} were solved numerically in that reference and analytically in \cite{Manley:2024pcl}: similar to hPDFs, the obtained small-$x$ asymptotics of OAM distributions appear to be similar to those resulting from the IREE approach \cite{Boussarie:2019icw}, albeit with a few percent difference in the intercepts. 

With the small-$x$ expressions for $L_{q + \bar q} (x,Q^2)$ and $L_G (x,Q^2)$ obtained in \cite{Kovchegov:2023yzd} and their DLA evolution derived in the same reference, it appears desirable to develop OAM phenomenology, attempting to devise a technique for constraining the OAM distributions from either the present or future data. To connect our formalism from \cite{Kovchegov:2023yzd} to the data, one needs to identify an observable sensitive to the OAM distributions. Here we employ the suggestion from \cite{Hatta:2016aoc,Bhattacharya:2022vvo,Bhattacharya:2024sck} and consider the forward elastic dijet production in the electron scattering on the polarized proton. We consider the cases when the electron is longitudinally polarized and unpolarized, with the corresponding observables being the longitudinal double- and single-spin asymmetries for elastic dijet production. Obtaining an explicit formula for the forward elastic dijet production cross section for polarized electron--proton scattering at small $x$, and expanding it for the small (transverse) momentum transfer $\Delta_\perp$, we observe that the terms linear in $\Delta_\perp$ in the double spin asymmetry (DSA) probe the moment dipole amplitudes and, hence, may be used to constrain the OAM distributions, as suggested in \cite{Bhattacharya:2022vvo, Bhattacharya:2024sck}. Analyzing the longitudinal single-spin asymmetry (SSA) in the same elastic dijet production we see that it also couples to the moment dipole amplitudes needed for the OAM distributions at small-$x$: however, the SSA also couples to other eikonal and subeikonal dipole amplitudes, whose phenomenology does not exist yet. This makes it very hard to separate the OAM contributions to the SSA from the other unknown terms. 

The paper is structured as follows. In \Sec{sec:sub-eik-tools} we review the essential ingredients needed for subeikonal calculations in the LCOT formalism. We set the stage for the elastic dijet production calculation in \Sec{sec:setup} by writing the numerators of the DSA and SSA in terms of six different Lorentz-invariant structures. The elastic dijet production cross section in ${\vec e} + {\vec p}$ and $e + {\vec p}$ collisions is constructed in \Sec{sec:production}, with the general expression for the cross section given in \eq{XS1}. Similar subeikonal calculations for dijet production have been carried out in \cite{Altinoluk:2022jkk, Agostini:2024xqs, Altinoluk:2024zom}: unlike those works, our emphasis here is on scattering on the longitudinally polarized proton and on elastic dijet production. (Gluon production at mid-rapidity in the longitudinally polarized proton--proton collision was recently calculated in \cite{Kovchegov:2024aus}: here we are interested in the forward elastic dijet production.) We follow the Trento convention \cite{Bacchetta:2004jz} and work in the frame where the virtual photon's transverse momentum is zero, $q_\perp =0$. The numerators of the DSA and SSA are extracted from our general expression for the elastic dijet cross section in \Sec{sec:DSA} and in \Sec{sec:ssa}, respectively. Particularly, in \Sec{sec:ssa} it becomes apparent that while the SSA does indeed couple to the polarized dipole amplitudes needed for the calculation of hPDFs and OAM distributions, extraction of those amplitudes from the SSA would require further theoretical and phenomenological developments in the field. The expansion of the DSA numerator to the first nontrivial order in $\Delta_\perp$ is carried out in \Sec{sec:DSA_exp}, where we find that two of the Lorentz-invariant quantities entering the expression for DSA do, indeed, couple to the polarized dipole amplitudes and moment amplitudes needed for the calculation of $L_{q + \bar q} (x,Q^2)$ and $L_G (x,Q^2)$ at small $x$. These two quantities are summarized in Eqs.~\eqref{integrated_res}, which is the main result of this work. We conclude that it may be possible to extract the small-$x$ OAM distributions from the elastic dijet DSA data by analyzing these two quantities as functions of the transverse momenta of the jets $\un p$, the (small) transverse momentum transfer ${\un \Delta}$, the electron's transverse momentum $\un k$ (which is the same before and after scattering in the $q_\perp=0$ frame we use), and the fractions $z$ and $1-z$ of the virtual photon's longitudinal momentum carried by the two jets. Of course, more detailed phenomenological studies are needed to validate this proposal for the future electron--proton colliders, such as the upcoming Electron-Ion Collider (EIC) \cite{Accardi:2012qut,Boer:2011fh,Proceedings:2020eah,AbdulKhalek:2021gbh}. We conclude in \Sec{sec:conclusions} by summarizing our findings.

%%%%%%%%%%%%%%%%%%%%%%%%%%%%%%%%%%%%%%%%%%%%%%%

\section{subeikonal quark and antiquark $S$-matrices in the background field}
\label{sec:sub-eik-tools}

Before we dive into the calculation of the dijet cross section, we will first review the tools used for subeikonal calculations developed in \cite{Altinoluk:2014oxa,Balitsky:2015qba,Balitsky:2016dgz, Kovchegov:2017lsr, Kovchegov:2018znm, Chirilli:2018kkw, Jalilian-Marian:2018iui, Jalilian-Marian:2019kaf, Boussarie:2020vzf, Boussarie:2020fpb, Altinoluk:2020oyd, Kovchegov:2021iyc, Altinoluk:2021lvu, Kovchegov:2022kyy, Altinoluk:2022jkk,Altinoluk:2023qfr,Altinoluk:2023dww, Li:2023tlw}. Specifically, we would like to revisit the $S$-matrix for a high energy quark or antiquark scattering on a longitudinally polarized proton up to subeikonal order. We will take our projectile to be moving along the $x^-$ light cone with a large minus momentum $p^-$. It scatters off a high-energy proton moving along the $x^+$ light cone with large plus momentum, $P^+$. Here we have used light cone coordinates 
$x^\pm =(x^0 \pm x^3)/{\sqrt{2}}$. Further, we denote transverse vectors by $\un x = (x^1, x^2)$ with the magnitude of the transverse vector denoted by $x_\perp = | \un x|$. Latin indices, e.g. $i, j=1,2$, indicate transverse coordinates. For the difference of two transverse vectors, we will write $\un x_{i} - \un x_{j} = \un x_{ij}$, where $i,j$ label the partons. We write four vectors as $x^\mu = (x^+, x^-, \un x)$ and the product of two four-vectors is then $a^\mu b_\mu = a^+ b^- + a^- b^+ - \un{a} \cdot \un {b}$.

Define the transverse position space $S$-matrix for a quark scattering off a proton as \cite{Kovchegov:2021iyc, Cougoulic:2022gbk}
\begin{align}
    V_{\un x, \un y; \sigma^\prime, \sigma} &\equiv \int \frac{d^2 p_{\mathrm{in}}}{(2\pi)^2} \frac{d^2 p_{\mathrm{out}}}{(2\pi)^2} 
    e^{i \un{p}_{\mathrm{out}} \cdot \un x - i \un{p}_{\mathrm{in}}\cdot \un y}
    \left[ \delta_{\sigma, \sigma^\prime} (2\pi)^2 \delta^2 \left(\un {p}_{\mathrm{out}} - \un{p}_{\mathrm{in}} \right)
    + i A_{\sigma^\prime, \sigma} \left(\un{p}_{\mathrm{in}}, \un{p}_{\mathrm{out}} \right)
    \right],
\end{align}
where $A = M/(2s)$, with $M$ the standard scattering amplitude \cite{Kovchegov:2012mbw} and $s$ the center of mass energy squared of the target-projectile system. Here ${\un p}_{in}$ and  ${\un p}_{out}$ denote the transverse momenta of the incoming and outgoing quark, respectively, with ${\un x}$ and $\un y$ their Fourier conjugate positions. The incident quark has helicity $\sigma$, while the outgoing quark has helicity $\sigma'$.

Isolating the subeikonal part of the scattering matrix, $V^\textrm{pol}$,  we write 
\begin{align} \label{Wilsonl}
    V_{\un x, \un y; \sigma^\prime, \sigma} = V_{\un x} \, \delta^2(\un x - \un y) \, \delta_{\sigma, \sigma^\prime} + V^{\mathrm{pol}}_{\un x, \un y; \sigma^\prime, \sigma}.
\end{align}
The leading (eikonal) term in \eq{Wilsonl} is given by the fundamental light cone Wilson line 
\begin{align} \label{Wilson_line}
   V_{\un x}[x_f^-, x_i^-] = \mathcal{P} \exp \left[ig \int \displaylimits^{x_f^-}_{x_i^-} dx^- A^+(0^+, x^-, \un x)  \right],
\end{align}
where $\mathcal{P}$ denotes path ordering and $A^\mu(x) = A^{a \mu}(x) t^a$ is the gluon field of the target, with $t^a$ the generators of $\mathrm{SU}(N_c)$ in the fundamental representation. We denote infinite Wilson lines by $V_{\un x} \equiv V_{\un x}[\infty, - \infty]$. 

\begin{figure}[ht!]
\centering
\begin{subfigure}[b]{0.33\textwidth}
    \includegraphics[scale=0.22]{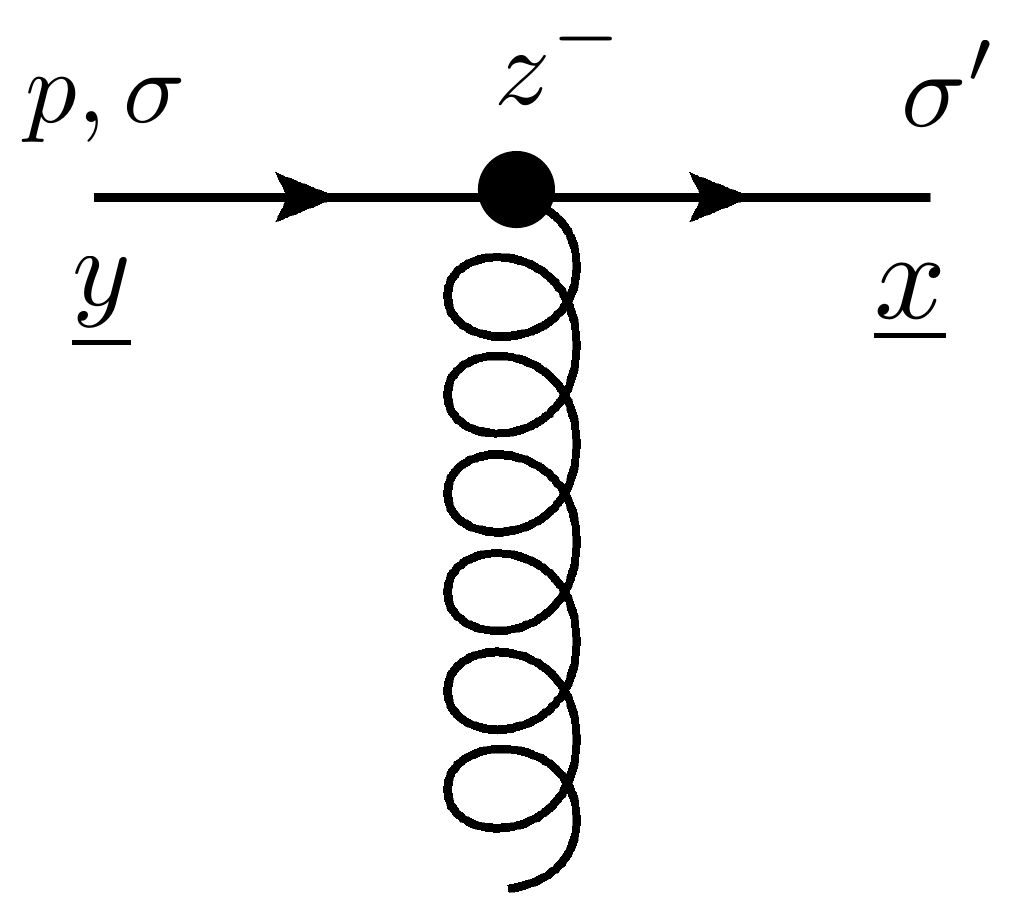}
    \caption{}
    \label{fig:quark-op}
\end{subfigure}
\begin{subfigure}[b]{0.33\textwidth}
    \includegraphics[scale=0.22]{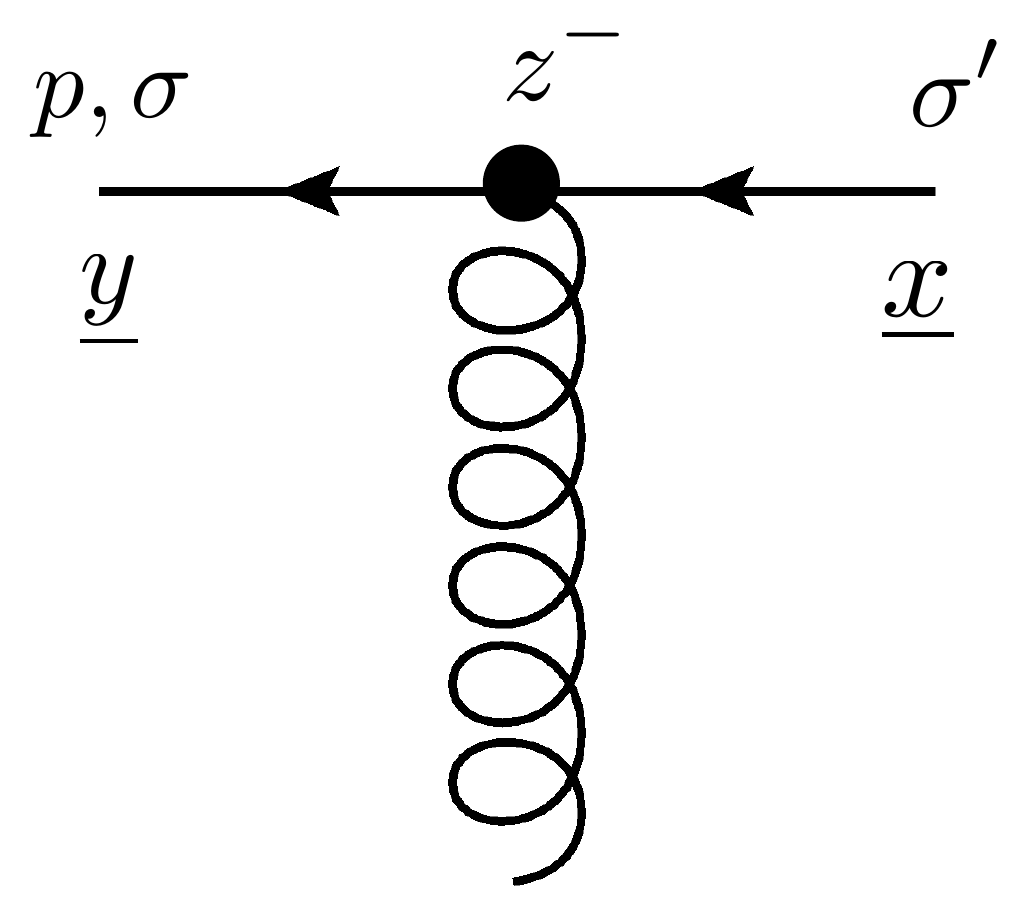}
    \caption{}
    \label{fig:aquark-op}
\end{subfigure}
    \caption{subeikonal gluon exchange between the target and a high energy (a) quark and (b) antiquark. The black circles denote the subeikonal vertices computed in \eqs{quark_glue} and (\ref{aquark_glue}).}
    \label{fig:ops}
\end{figure}

It has been shown \cite{Kovchegov:2017lsr, Kovchegov:2018znm, Kovchegov:2018zeq, Chirilli:2018kkw, Altinoluk:2020oyd, Kovchegov:2021iyc, Cougoulic:2022gbk} that subeikonal corrections come in with an insertion of one or two subeikonal operators anywhere along the quark's $x^-$ path, with such operator sandwiched between the eikonal light cone Wilson lines. Gluonic subeikonal operators come in via a single insertion, for which we have the generic form
\begin{align} \label{single_insert}
    V^{\mathrm{pol}}_{\un x, \un y; \sigma^\prime, \sigma} &= \int \displaylimits^\infty_{-\infty} dz^- d^2 z\, V_{\un x}[\infty, z^-] \, \delta^2(\un x - \un z) \, \mathcal{O}^{\mathrm{pol}}_{\sigma^\prime, \sigma}(z^-, \un z) \, V_{\un y}[z^-, -\infty] \, \delta^2(\un y - \un z),
\end{align}
while the quark subeikonal operators come in with two insertions, such that we have 
\begin{align} \label{double_insert}
    V^{\mathrm{pol}}_{\un x, \un y; \sigma^\prime, \sigma} &= \int \displaylimits^\infty_{-\infty} dz_1^- d^2 z_1
    \int \displaylimits^\infty_{z_1^-} dz_2^- d^2 z_2\, 
    V_{\un x}[\infty, z_2^-] \, \delta^2(\un x - \un z_2) \, \mathcal{O}^{\mathrm{pol}}_{\sigma^\prime, \sigma}(z_2^-, z_1^-, \un z_2, \un z_1) \, V_{\un y}[z_1^-, -\infty]  \, \delta^2(\un y - \un z_1).
\end{align}
For the longitudinally polarized proton we are considering here, the single insertion operator can be obtained by computing the subeikonal gluon vertex in \fig{fig:quark-op}. In any gauge where the gluon field vanishes at $x^- \to \pm \infty$, the resulting operator is, to subeikonal order \cite{Altinoluk:2014oxa,Balitsky:2015qba,Balitsky:2016dgz, Kovchegov:2017lsr, Kovchegov:2018znm, Chirilli:2018kkw, Jalilian-Marian:2018iui, Jalilian-Marian:2019kaf, Altinoluk:2020oyd, Kovchegov:2021iyc, Altinoluk:2021lvu, Kovchegov:2022kyy, Altinoluk:2022jkk,Altinoluk:2023qfr,Altinoluk:2023dww, Li:2023tlw},
\begin{align} \label{quark_glue}
    \mathcal{O}^{\mathrm{pol}\,G}_{\sigma^\prime, \sigma}(z^-, \un z)  = \frac{i}{2\, p^-} \left[  \sigma \delta_{\sigma, \sigma^\prime} \, g \, F^{12} 
     - \delta_{\sigma, \sigma^\prime} \cev{D}^i D^i 
    \right] + \mathcal{O}\left(\frac{1}{(p^-)^2} \right),
\end{align}
where $g$ is the strong coupling constant, $F^{12}$ is the gluon field strength tensor, and $p^-$ is the large momentum component of the scattering quark. Further, the right- and left-acting covariant derivatives are defined as $D^i = \partial^i - i g A^i$ and $\cev{D}^i = \cev{\partial}^i + i g A^i$, respectively.   
%%%%%%%%%%%%%%%%%%%%%%%%%%%%%%%%%%%%%%%%%%%%%
%%%%%%%%%%%%%%%%%%%%%%%%%%%%%%%%%%%%%%%%%%%%%

For antiquarks, the $S$-matrix can be written as
\begin{align} \label{antiS}
   {\overline V}_{\un x, \un y; \sigma^\prime, \sigma} = V_{\un x}^\dagger \, \delta^2(\un x - \un y) \, \delta_{\sigma, \sigma^\prime} + {\overline V}^{\mathrm{pol}}_{\un x, \un y; \sigma^\prime, \sigma}
\end{align}
with the single- and double-insertion subeikonal terms given by
\begin{subequations}\label{antiS2}
    \begin{align}
        & {\overline V}^{\mathrm{pol}}_{\un x, \un y; \sigma^\prime, \sigma} = \int \displaylimits_\infty^{-\infty} dz^- d^2 z\, V_{\un y}[-\infty, z^-] \, \delta^2(\un y - \un z) \, \overline{\mathcal{O}}^{\mathrm{pol}}_{\sigma^\prime, \sigma}(z^-, \un z) \, V_{\un x}[z^-, \infty]  \, \delta^2(\un x - \un z), \\
        & {\overline V}^{\mathrm{pol}}_{\un x, \un y; \sigma^\prime, \sigma} = \int \displaylimits_\infty^{-\infty} dz_2^- d^2 z_2
    \int \displaylimits^{-\infty}_{z_2^-} dz_1^- d^2 z_1 \, 
    V_{\un y}[- \infty, z_1^-] \, \delta^2(\un y - \un z_1) \, \overline{\mathcal{O}}^{\mathrm{pol}}_{\sigma^\prime, \sigma}(z_1^-, z_2^-, \un z_1, \un z_2) \, V_{\un x}[z_2^-, \infty]  \, \delta^2(\un x - \un z_2).
    \end{align}
\end{subequations}
Calculating the diagram in \fig{fig:aquark-op} one can show that the single-insertion gluonic operator is
\begin{align} \label{aquark_glue}
    \overline{\mathcal{O}}^{\mathrm{pol}\,G}_{\sigma^\prime, \sigma}(z^-, \un z) 
     =  \frac{i}{2\, p^-} \left[ \sigma \delta_{\sigma, \sigma^\prime} \, g F^{12} 
     + \delta_{\sigma, \sigma^\prime} \cev{D}^i D^i 
    \right] + \mathcal{O}\left(\frac{1}{(p^-)^2} \right).
\end{align}

\begin{figure}[ht!]
\centering
\begin{subfigure}[b]{0.48\textwidth}
    \includegraphics[scale=0.22]{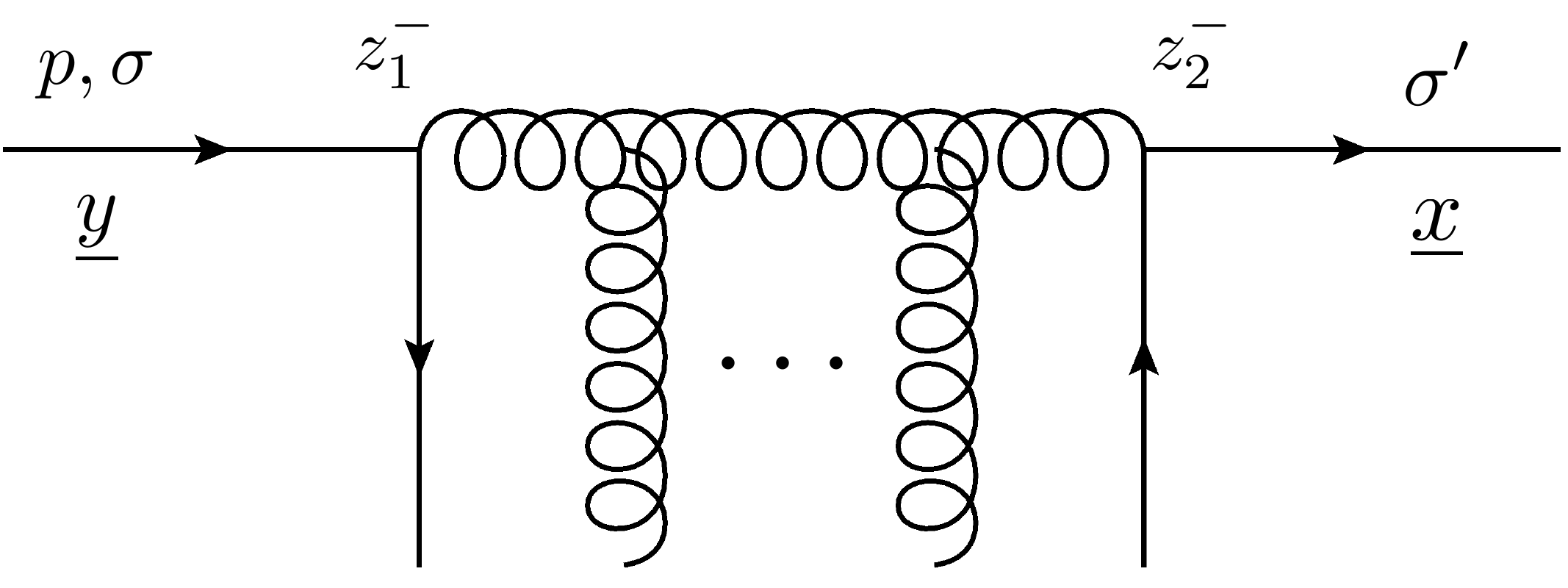}
    \caption{}
    \label{fig:quark-ex}
\end{subfigure}
\begin{subfigure}[b]{0.48\textwidth}
    \includegraphics[scale=0.22]{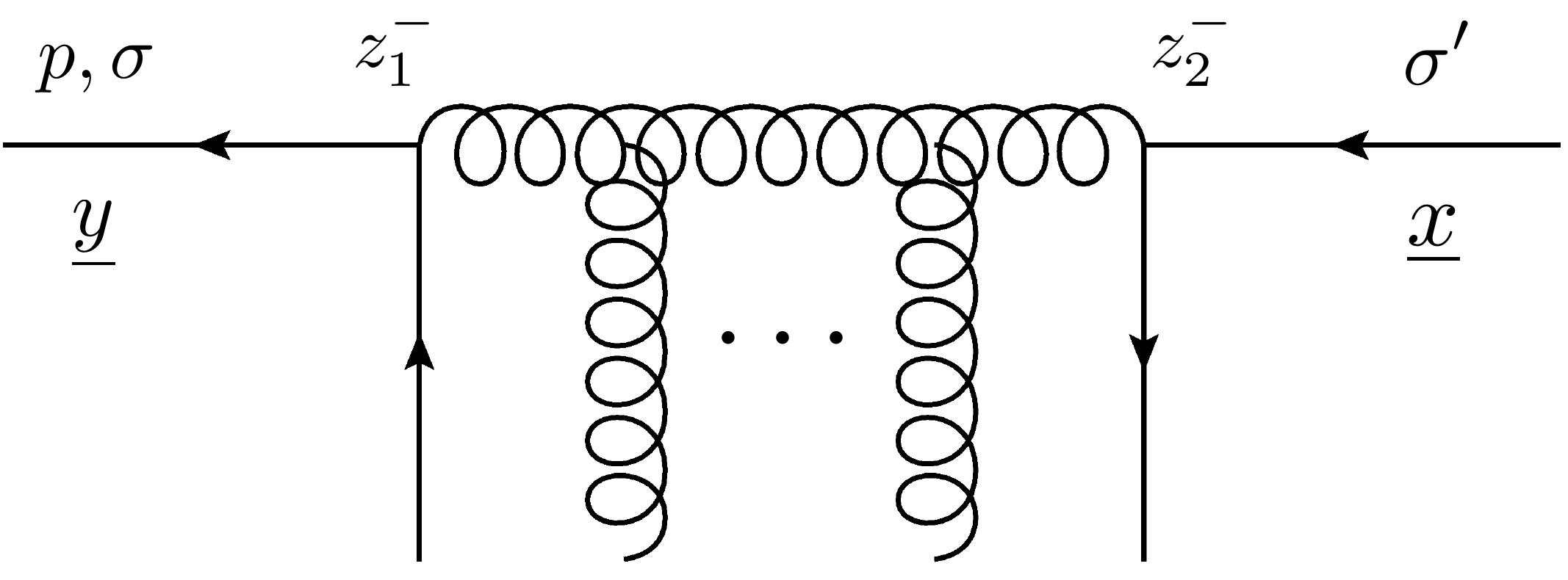}
    \caption{}
    \label{fig:aquark-ex}
\end{subfigure}
    \caption{subeikonal quark exchange between the target and a high energy (a) quark and (b) antiquark. The three dots in the centers of the diagrams indicate the additional (not shown) eikonal gluon exchanges summed up by the adjoint Wilson lines in \eqs{quark_exch} and (\ref{aquark_exch}).}
    \label{fig:ex}
\end{figure}

As mentioned above, in addition to the single-insertion operators, at the subeikonal level we also have double insertion operators. These operators describe the $t$-channel exchange of a quark and antiquark, as depicted in \fig{fig:ex}. For the quark $S$-matrix, the subeikonal operator is \cite{Altinoluk:2014oxa,Balitsky:2015qba,Balitsky:2016dgz, Kovchegov:2017lsr, Kovchegov:2018znm, Chirilli:2018kkw, Jalilian-Marian:2018iui, Jalilian-Marian:2019kaf, Altinoluk:2020oyd, Kovchegov:2021iyc, Altinoluk:2021lvu, Kovchegov:2022kyy, Altinoluk:2022jkk,Altinoluk:2023qfr,Altinoluk:2023dww, Li:2023tlw}
\begin{align} \label{quark_exch}
    \mathcal{O}^{\mathrm{pol}\,q\bar{q}}_{\sigma^\prime, \sigma}(z_2^-, z_1^-, \un z_2, \un z_1) = \frac{g^2}{4 p^-} t^b \psi_\beta (z_2^-, \un z_2) U_{\un z_2}^{ba} [z_2^-, z_1^-] \delta^2(\un z_2 - \un z_1) 
    \Big[\sigma \delta_{\sigma, \sigma^\prime} \gamma^+ \gamma^5  
     - \delta_{\sigma, \sigma^\prime} \gamma^+ 
    \Big]_{\alpha \beta} \bar{\psi}_\alpha (z_1^-, \un z_1) t^a 
    + \mathcal{O}\left(\frac{1}{(p^-)^2} \right), 
\end{align}
where we have defined the adjoint light cone Wilson line, 
\begin{align}
    U_{\un x}[x_f^-, x_i^-] = \mathcal{P} \exp \left[i g \int \displaylimits^{x_f^-}_{x_i^-} dx^- \mathcal{A}^+(0^+, x^-, \un x) \right],
\end{align}
with $\mathcal{A}^\mu = \mathcal{A}^{a \mu} T^a$ the target gluon field in the adjoint representation. 

%%%%%%%%%%%%%%%%%%%%%%%%%%%%%%%%%%%%%%%%%
%%%%%%%%%%%%%%%%%%%%%%%%%%%%%%%%%%%%%%%%%

Similarly, the operator for the subeikonal antiquark $S$-matrix in \fig{fig:aquark-ex} is 
\begin{align} \label{aquark_exch}
    \overline{\mathcal{O}}^{\mathrm{pol}\,q\bar{q}}_{\sigma^\prime, \sigma}(z_1^-, z_2^-, \un z_1, \un z_2) = \frac{g^2}{4 p^-} t^a \psi_\alpha (z_1^-, \un z_1) \, U_{\un z_1}^{ab} [z_1^-, z_2^-] \delta^2(\un z_2 - \un z_1) 
    \Big[\sigma \delta_{\sigma, \sigma^\prime} \gamma^+ \gamma^5  
     + \delta_{\sigma, \sigma^\prime} \gamma^+ 
    \Big]_{\beta \alpha} {\bar \psi}_\beta (z_2^-, \un z_2) t^b 
    + \mathcal{O}\left(\frac{1}{(p^-)^2} \right). 
\end{align}
Using \eqs{quark_glue} and (\ref{quark_exch}) in \eqs{single_insert} and (\ref{double_insert}) respectively, we arrive at the quark $S$-matrix, including eikonal and subeikonal terms, 
\begin{align} \label{q-Smatrix}
    V_{\un x, \un y; \sigma^\prime, \sigma} = \delta_{\sigma, \sigma^\prime} V_{\un x} \delta^2(\un x - \un y) + \sigma \delta_{\sigma, \sigma^\prime} \Big[V^{\mathrm{G}[1]}_{\un x} + V^{\mathrm{q}[1]}_{\un x}  \Big] \delta^2(\un x - \un y)
    + \delta_{\sigma, \sigma^\prime} \Big[V^{\mathrm{G}[2]}_{\un x, \un y} + V^{\mathrm{q}[2]}_{\un x} \delta^2(\un x - \un y)  \Big],
\end{align}
where we have defined the polarized Wilson lines \cite{Cougoulic:2022gbk}
\begin{subequations}\label{VqG}
\begin{align}
& V_{\un x}^{\textrm{G} [1]}  = \frac{i \, g \, P^+}{s} \int\limits_{-\infty}^{\infty} d{x}^- V_{\un{x}} [ \infty, x^-] \, F^{12} (x^-, {\un x}) \, \, V_{\un{x}} [ x^-, -\infty]  , \label{VG1} \\
& V_{\un x}^{\textrm{q} [1]}  = \frac{g^2 P^+}{2 \, s} \int\limits_{-\infty}^{\infty} \!\! d{x}_1^- \! \int\limits_{x_1^-}^\infty d x_2^- V_{\un{x}} [ \infty, x_2^-] \, t^b \, \psi_{\beta} (x_2^-,\un{x}) \, U_{\un{x}}^{ba} [x_2^-, x_1^-] \, \left[ \gamma^+ \gamma^5 \right]_{\alpha \beta} \, \bar{\psi}_\alpha (x_1^-,\un{x}) \, t^a \, V_{\un{x}} [ x_1^-, -\infty] , \label{Vq1} \\
& V_{{\ul x}, {\un y}}^{\textrm{G} [2]}  = - \frac{i \, P^+}{s} \int\limits_{-\infty}^{\infty} d{z}^- d^2 z \ V_{\un{x}} [ \infty, z^-] \, \delta^2 (\un{x} - \un{z}) \, \cev{D}^i (z^-, {\un z}) \, D^i  (z^-, {\un z}) \, V_{\un{y}} [ z^-, -\infty] \, \delta^2 (\un{y} - \un{z}) , \label{VxyG2} \\
& V_{{\ul x}}^{\textrm{q} [2]} = - \frac{g^2 P^+}{2 \, s} \int\limits_{-\infty}^{\infty} \!\! d{x}_1^- \! \int\limits_{x_1^-}^\infty d x_2^- V_{\un{x}} [ \infty, x_2^-] \, t^b \, \psi_{\beta} (x_2^-,\un{x}) \, U_{\un{x}}^{ba} [x_2^-, x_1^-] \, \left[ \gamma^+ \right]_{\alpha \beta} \, \bar{\psi}_\alpha (x_1^-,\un{x}) \, t^a \, V_{\un{x}} [ x_1^-, -\infty] \label{Vq2}.
\end{align}
\end{subequations}
In \eqs{VqG}, we have used $2 P^+ p^- = s$ with $P^+$ the large momentum component of the target proton. The polarized Wilson lines in \eqs{VG1} and (\ref{Vq1}) are referred to as type-1 polarized Wilson lines, and, therefore, have a superscript of ``[1]" \cite{Cougoulic:2022gbk}. Similarly, \eqs{VxyG2} and (\ref{Vq2}) are called the type-2 polarized Wilson lines and come with the ``[2]" superscript \cite{Cougoulic:2022gbk}.

Likewise, we can construct the antiquark $S$-matrix by using \eqs{aquark_glue} and (\ref{aquark_exch}) in \eqs{antiS2}, which gives
\begin{align} \label{aquark-Smatrix}
       \overline{V}_{\un x, \un y; \sigma^\prime, \sigma} = \delta_{\sigma, \sigma^\prime} V^\dagger_{\un x} \delta^2(\un x - \un y) + \sigma \delta_{\sigma, \sigma^\prime} \Big[V^{\mathrm{G}[1] \dagger}_{\un x} + V^{\mathrm{q}[1]\dagger}_{\un x} \Big] \delta^2(\un x - \un y)
    - \delta_{\sigma, \sigma^\prime} \Big[V^{\mathrm{G}[2]\dagger}_{\un x, \un y} + V^{\mathrm{q}[2]\dagger}_{\un x} \delta^2(\un x - \un y)  \Big].
\end{align}
The $S$-matrix for a high energy gluon can be constructed analogously, and is given in \cite{Cougoulic:2022gbk}. 

The polarized dipole amplitudes that contribute to the flavor-singlet helicity PDFs and OAM distributions are \cite{Kovchegov:2015pbl, Kovchegov:2016weo, Kovchegov:2018zeq, Cougoulic:2022gbk, Kovchegov:2019rrz, Kovchegov:2023yzd}
\begin{subequations} \label{pdas}
    \begin{align}
        Q_{10}(s) &= \frac{1}{2N_c} \mathrm{Re}\,  \llangle \tord 
        \tr \left[ V_{\un 0} V_{\un 1}^{\mathrm{pol}[1] \dagger} 
        \right] 
        +
        \tord 
        \tr \left[ V_{\un 1}^{\mathrm{pol}[1]}  V_{\un 0}^\dagger 
        \right] 
        \rrangle(s),
        \\ 
        G^i_{10}(s) &= \frac{1}{2N_c} \mathrm{Re}\,  \llangle \tord 
        \tr \left[ V_{\un 0} V_{\un 1}^{i\mathrm{G}[2] \dagger} 
        \right] 
        +
        \tord 
        \tr \left[ V_{\un 1}^{i\mathrm{G}[2]}  V_{\un 0}^\dagger 
        \right] 
        \rrangle(s),
    \end{align}
\end{subequations}
where $V^{\mathrm{pol}[1]}_{\un x} = V^{\mathrm{G}[1]}_{\un x} + V^{\mathrm{q}[1]}_{\un x}$, $\tord$ is the time-ordering operator, and 
\begin{align} \label{d-d_pw}
    V_{\un z}^{i\mathrm{G}[2]} = \frac{P^+}{2s} \int \displaylimits^\infty_{-\infty} dz^- V_{\un z}[\infty, z^-] \left[ D^i(z^-, \un z) - \cev{D}^i(z^-, \un z) \right] V_{\un z}[z^-, -\infty].
\end{align}
The polarized Wilson line in \eq{d-d_pw} is related to $V^{\mathrm{G}[2]}_{\un x, \un y}$ in \eq{VxyG2} above (see, e.g., \cite{Cougoulic:2022gbk}). The double angle brackets in \eqs{pdas} are related to the standard angle brackets that denote averaging in the unpolarized target in the eikonal small-$x$ physics. The double angle brackets denote averaging in the (polarized) proton state via \cite{Kovchegov:2015pbl, Kovchegov:2017lsr}
\begin{align} \label{double-bracket}
    \llangle \ldots \rrangle(s) = s\,  \Big\langle \ldots \Big\rangle(s) = s \, \frac{1}{2} \sum_{S_L} S_L \frac{1}{2 P^+ V^-} \left \langle P, S_L | \ldots | P, S_L \right \rangle.
\end{align}
Here $S_L$ and $P^+$ are the proton's helicity and momentum, $V^- = \int d x^- d^2 x_\perp$ is the infinite volume, and $s$ is the center-of-mass energy squared between the subeikonal line and the target.

Since the helicity PDFs depend on the impact-parameter integrated polarized dipoles, we integrate \eqs{pdas} over all impact parameters while keeping the dipole size $x_{10}$ fixed \cite{Kovchegov:2015pbl, Kovchegov:2017lsr, Kovchegov:2018znm, Cougoulic:2022gbk}, defining\footnote{To make our definitions more uniform and in agreement with our previous paper \cite{Kovchegov:2023yzd} on the subject, we will employ integration over $x_1$ as the impact parameter integration, instead of the more conventional integration over the center of the dipole in the transverse plane. The final result of the calculation is independent of the convention adopted for the definition of moments: see Sec.~IIC of \cite{Kovchegov:2023yzd}, along with the footnote in that Section.} 
\begin{subequations} \label{integrated}
    \begin{align}
        \int d^2 x_1 
        \, Q_{10}(s) &= Q(x_{10}^2, s), 
        \\ 
        \int d^2 x_1 
        \, G^i_{10}(s) &= x_{10}^i G_1(x_{10}^2, s) + \epsilon^{ij} x_{10}^j G_2(x_{10}^2, s),
    \end{align}
\end{subequations}
where $\epsilon^{ij}$ is the two-dimensional Levi-Civita tensor. Only the functions $Q(x_{10}^2, s)$ and $G_2(x_{10}^2, s)$ contribute to the helicity PDFs \cite{Cougoulic:2022gbk}.

It was shown in \cite{Kovchegov:2023yzd} (see also \cite{Kovchegov:2019rrz}) that to construct the OAM distributions at small $x$, in addition to the polarized dipole amplitudes integrated over all impact parameters, one also needs the first impact-parameter moments of the polarized dipole amplitudes, dubbed the ``moment amplitudes" in \cite{Kovchegov:2023yzd}, defined as
\begin{subequations} \label{moments}
    \begin{align}
        \int d^2 x_1 \, x_1^m \, Q_{10}(s) &= x_{10}^m I_3(x_{10}^2, s) + \epsilon^{mj} x_{10}^j J_3(x_{10}^2, s),
        \\ \label{Gi_moms}
        \int d^2 x_1 \, x_1^m \, G^i_{10}(s) &= \epsilon^{mi} x_{10}^2 I_4(x_{10}^2, s) + 
        \epsilon^{mk} x_{10}^k x_{10}^i I_5(x_{10}^2, s) + 
        \delta^{im} x_{10}^2 J_4(x_{10}^2, s) + x_{10}^i x_{10}^m J_5(x_{10}^2, s).
    \end{align}
\end{subequations}
Only the ``$I$"-type moments ($I_3, I_4, I_5$) contribute to the OAM distributions \cite{Kovchegov:2023yzd}. Note that in \cite{Kovchegov:2023yzd}, the decomposition of the impact-parameter moment of $G^i_{10}$ contained a term proportional to $\epsilon^{ik}x_{10}^k x_{10}^m$ in addition to the ones shown on the right of \eq{Gi_moms}, with the corresponding moment denoted by $I_6(x_{10}^2, s)$ (see Eq.~(28b) there). As was already suspected in that reference, this structure is redundant since\footnote{We thank Ming Li for pointing this out to us.}
\begin{align} 
\epsilon^{mk} x_{10}^k \, x_{10}^i - \epsilon^{ik} x_{10}^k \, x_{10}^m = \epsilon^{mi} x_{10}^2. 
\end{align}
We therefore omit the $\epsilon^{ik}x_{10}^k x_{10}^m$ structure and redefine the impact-parameter moment of $G^i_{10}$ using \eq{Gi_moms} above. One may translate between the definitions in \eqs{moments} and the ones from \cite{Kovchegov:2023yzd}, defined by Kovchegov and Manley (KM), via 
\begin{subequations} \label{translation}
    \begin{align}
        I_4^{\mathrm{here}}(x_{10}^2, s) &= I_4^{\mathrm{KM}}(x_{10}^2, s) - I_6^{\mathrm{KM}}(x_{10}^2, s), 
        \\ 
         I_5^{\mathrm{here}}(x_{10}^2, s) &= I_5^{\mathrm{KM}}(x_{10}^2, s) + I_6^{\mathrm{KM}}(x_{10}^2, s). 
    \end{align}
\end{subequations}
We see that the large-$N_c$ DLA evolution equations for the moment amplitudes derived in \cite{Kovchegov:2023yzd} contain an additional dependent degree of freedom, $I_6^{\mathrm{KM}}$: the equations are correct, albeit slightly more complicated than they should be. For completeness, in Appendix \ref{app:updated_evolution} we include the more-compact large-$N_c$ DLA evolution equations for the moment amplitudes $I_3, I_4, I_5$ as defined above in \eqs{moments}.

In addition to the dipole amplitudes in \eqs{integrated} and \eqref{moments}, we will employ the following object and its impact-parameter integrated form, which is related to the quark to gluon and gluon to quark transition operators studied in \cite{Borden:2024bxa}, 
\begin{subequations} \label{objects}
\begin{align} \label{obj}
    \widetilde{Q}_{10}(s) &\equiv \llangle \frac{g^2}{16 \sqrt{k^- p^-}} \int \displaylimits^\infty_{-\infty} dy^-  \int \displaylimits^\infty_{-\infty} dz^- \Bigg[ 
    \bar{\psi}(y^-, \un x_0) \left(\frac{1}{2} \gamma^+ \gamma^5 \right) V_{\un 0}[y^-, \infty] V_{\un 1}[\infty, z^-] \psi(z^-, \un x_1) 
    \\ \notag 
    & \hspace{3cm}
    +  \bar{\psi}(y^-, \un x_0) \left(\frac{1}{2} \gamma^+ \gamma^5 \right) V_{\un 0}[y^-, -\infty] V_{\un 1}[-\infty, z^-] \psi(z^-, \un x_1) 
    \Bigg] \rrangle(s),
    \\
     \widetilde{Q}(x_{10}^2, s) &= \int d^2 x_1 \, \widetilde{Q}_{10}(s). \label{Q_tilde_b}
\end{align}
\end{subequations}
Unlike the polarized dipole amplitudes in \eqs{integrated} and \eqref{moments}, the object $\widetilde{Q}_{10}(s)$ contains quark and antiquark fields located at two different transverse positions, with the appropriate Wilson lines attached to them. Therefore, the double angle brackets in \eq{obj} are \cite{Borden:2024bxa}
\begin{align}\label{double_def}
\llangle \ldots  \rrangle \equiv  2 \, P^+ \, \sqrt{k^- p^-} \, \Big\langle \ldots \Big\rangle ,
\end{align} 
with $k^-$ and $p^-$ the longitudinal momenta carried by the two quark lines represented by the Wilson lines in \eq{obj}. This is a generalization of \eq{double-bracket}.

The flavor-singlet helicity PDFs at DLA can be expressed in terms of the amplitudes in \eqs{integrated} and (\ref{objects}) using 
\cite{Kovchegov:2015pbl, Kovchegov:2016zex, Kovchegov:2017lsr, Kovchegov:2018znm, Cougoulic:2022gbk, Borden:2024bxa}
\begin{subequations} \label{pdfs}
    \begin{align}
    \label{quark_hel}
    \Delta \Sigma(x, Q^2) &= \frac{N_f}{\as \pi^2} \, \widetilde{Q}\left(x_{10}^2 = \frac{1}{Q^2}, s = \frac{Q^2}{x} \right) ,
    \\ 
    \label{gluon_hel}
    \Delta G(x,Q^2) &= \frac{2 N_c}{\as \pi^2}\, G_2 \left(x_{10}^2 = \frac{1}{Q^2}, s = \frac{Q^2}{x} \right) .
    \end{align}
\end{subequations}
The formula \eqref{quark_hel} employing $\tilde Q$ from \eq{Q_tilde_b} is a slight modification on the earlier expression \cite{Kovchegov:2015pbl, Kovchegov:2016zex, Kovchegov:2017lsr, Kovchegov:2018znm, Cougoulic:2022gbk} obtained recently in \cite{Borden:2024bxa}. The difference between \eq{quark_hel} and the earlier expressions for $\Delta \Sigma$ from \cite{Kovchegov:2015pbl, Kovchegov:2016zex, Kovchegov:2017lsr, Kovchegov:2018znm, Cougoulic:2022gbk} can be attributed to a scheme change, with \eq{quark_hel} appearing to be in agreement with the $\overline{\text{MS}}$ scheme.

Analogous to \eqs{moments}, we define the first impact-parameter moment of \eq{obj} as 
\begin{align} \label{object_moment}
    \int d^2 x_1 \, x_1^m \, \widetilde{Q}_{10}(s) &= x_{10}^m \, \widetilde{I}(x_{10}^2, s) + \epsilon^{mi} x_{10}^i \, \widetilde{J}(x_{10}^2, s). 
\end{align}
Using \eqs{object_moment}, we can derive a relation similar to \eq{quark_hel} for the quark OAM distribution. First, we recall the DLA evolution equation for $\widetilde{Q}_{10}(s)$. Starting from Eq.~(58) of \cite{Borden:2024bxa}, we write
\begin{align}\label{Qtilde_evol}
    \widetilde{Q}_{10}(zs) &= \widetilde{Q}_{10}^{(0)}(zs) 
    - \frac{\as N_c}{2 \pi^2} \int \displaylimits^z_{\frac{\Lambda^2}{s}} \frac{dz'}{z'} \int d^2 x_2 \Bigg\{ 
    \frac{\un x_{20}}{x_{20}^2} \cdot \frac{\un x_{21}}{x_{21}^2} 
    \, Q_{21}(z's) 
    \\ \notag 
    & \hspace{6cm}
    + 
    \left[
        \epsilon^{ik} \frac{x_{20}^k + x_{21}^k}{x_{21}^2 x_{20}^2} - 
        2 \frac{\un x_{21} \times \un x_{20}}{x_{21}^2 x_{20}^2} \left(\frac{x_{21}^i}{x_{21}^2} - \frac{x_{20}^i}{x_{20}^2} \right)
    \right] G^i_{21}(z's) 
    \Bigg\}.
\end{align}
Following the procedure to construct moment evolution equations established in \cite{Kovchegov:2023yzd}, we multiply both sides by $x_1^m$ and integrate over $\un x_1$, keeping $\un x_{10}$ fixed. By identifying the coefficients of the resulting tensor structures and keeping only the DLA terms, we obtain an evolution equation for $\tilde{I}(x_{10}^2, s)$, 
\begin{align}\label{Itilde_evol}
    \widetilde{I}(x_{10}^2, zs) &= \widetilde{I}^{(0)}(x_{10}^2, zs) 
    - \frac{\as N_c}{4 \pi} \int \displaylimits^z_{\frac{\Lambda^2}{s}} \frac{dz'}{z'} 
    \int \displaylimits^{\mathrm{min}\left[ \frac{z}{z'} x_{10}^2, \frac{1}{\Lambda^2} \right]}_{\mathrm{max}\left[ x_{10}^2, \frac{1}{z's} \right]} \frac{dx_{21}^2}{x_{21}^2}
    \Big[ 
            Q + 3\, G_2 - I_3 + 2\, I_4 - I_5
    \Big](x_{21}^2, z's).
\end{align}
Here $\widetilde{I}^{(0)}(x_{10}^2, zs)$ is defined using \eq{object_moment} with the initial condition/inhomogeneous term $\widetilde{Q}_{10}^{(0)}$ for the $\tilde Q$ evolution \eqref{Qtilde_evol} on the left-hand side of \eq{object_moment}. Variables $z$ and $z'$ are the minus momentum fractions of the softer parton in the parent and daughter dipole \cite{Kovchegov:2015pbl, Kovchegov:2016zex, Kovchegov:2018znm}, while $s$ is the energy of the target--projectile system.

Comparing our \eq{Itilde_evol} with Eq.~(30) in \cite{Kovchegov:2023yzd} (and noting the difference in definition of the moments via \eqs{translation}), we observe the following relation 
\begin{align} \label{Lq_relation}
    L_{q+\bar{q}}(x, Q^2) = - \frac{2\, N_f}{\as \pi^2} \, \widetilde{I} \left(x_{10}^2 = \frac{1}{Q^2}, s = \frac{Q^2}{x} \right),
\end{align} 
which mirrors \eqs{pdfs}. Note that, similar to \eq{quark_hel}, \eq{Lq_relation} includes the initial condition $\widetilde{I}^{(0)}$ into the definition of $L_{q+\bar{q}}$.

Combining \eq{Lq_relation} with \eqs{pdfs}, and with the expression for $L_G(x,Q^2)$ in terms of the moments defined in \eq{Gi_moms} obtained in Eq.~(36) of \cite{Kovchegov:2023yzd}, the flavor-singlet helicity PDFs and OAM distributions in DLA at small $x$ can be summarized as
\begin{subequations}\label{PDF+OAM_summ}
    \begin{align}
        & \Delta \Sigma(x, Q^2) = \frac{N_f}{\as \pi^2} \, \widetilde{Q}\left(x_{10}^2 = \frac{1}{Q^2}, s = \frac{Q^2}{x} \right) ,
        \\ 
        & \Delta G(x,Q^2) = \frac{2 N_c}{\as \pi^2} \, G_2 \left(x_{10}^2 = \frac{1}{Q^2}, s = \frac{Q^2}{x} \right),
        \\ 
        & L_{q+\bar{q}}(x, Q^2) = - \frac{2\, N_f}{\as \pi^2} \, \widetilde{I} \left(x_{10}^2 = \frac{1}{Q^2}, s = \frac{Q^2}{x} \right),
        \\ 
        & L_{G}(x, Q^2) = - \frac{2\, N_c}{\as \pi^2} \,\Big[2 \, I_4 + 3\, I_5  \Big]\left(x_{10}^2 = \frac{1}{Q^2}, s = \frac{Q^2}{x} \right).
    \end{align}
\end{subequations}
We see that while the quark hPDF is given by the impact-parameter integrated quantity $\widetilde Q$, the quark OAM distribution it given by its first impact-parameter moment $\widetilde I$. Similarly, while the gluon hPDF is given by $G_2$, which results from integrating the dipole amplitude $G^i$ over all impact parameters, the gluon OAM distribution is given by the impact-parameter moments of $G^i$. The origin of this rather simple correspondence between hPDFs and OAM distributions is not clear to us at the moment.

With the tools for subeikonal calculations in hand, let us now turn to the main focus of this paper: dijet production in polarized electron-proton collisions.

%%%%%%%%%%%%%%%%%%%%%%%%%%%%%%%%%%%%%%%%%%%

\section{Dijet production in polarized electron-proton collisions: general discussion}
\label{sec:setup}

We begin our study of dijet production with a general discussion of the cross section. Our strategy is as follows. We will first compute the general dijet cross section in terms of the leptonic and hadronic tensors in the basis of the virtual photon's polarizations. To probe the longitudinal proton spin, we will then consider both the double- and single-spin asymmetries. By evaluating the leptonic tensor explicitly, we will identify explicit structures of the hadronic tensor that couple to the DSA and SSA. In the next Section, we will evaluate these structures using the tools from Section~\ref{sec:sub-eik-tools}. 

\begin{figure}[ht!]
    \centering
    \includegraphics[scale=0.2]{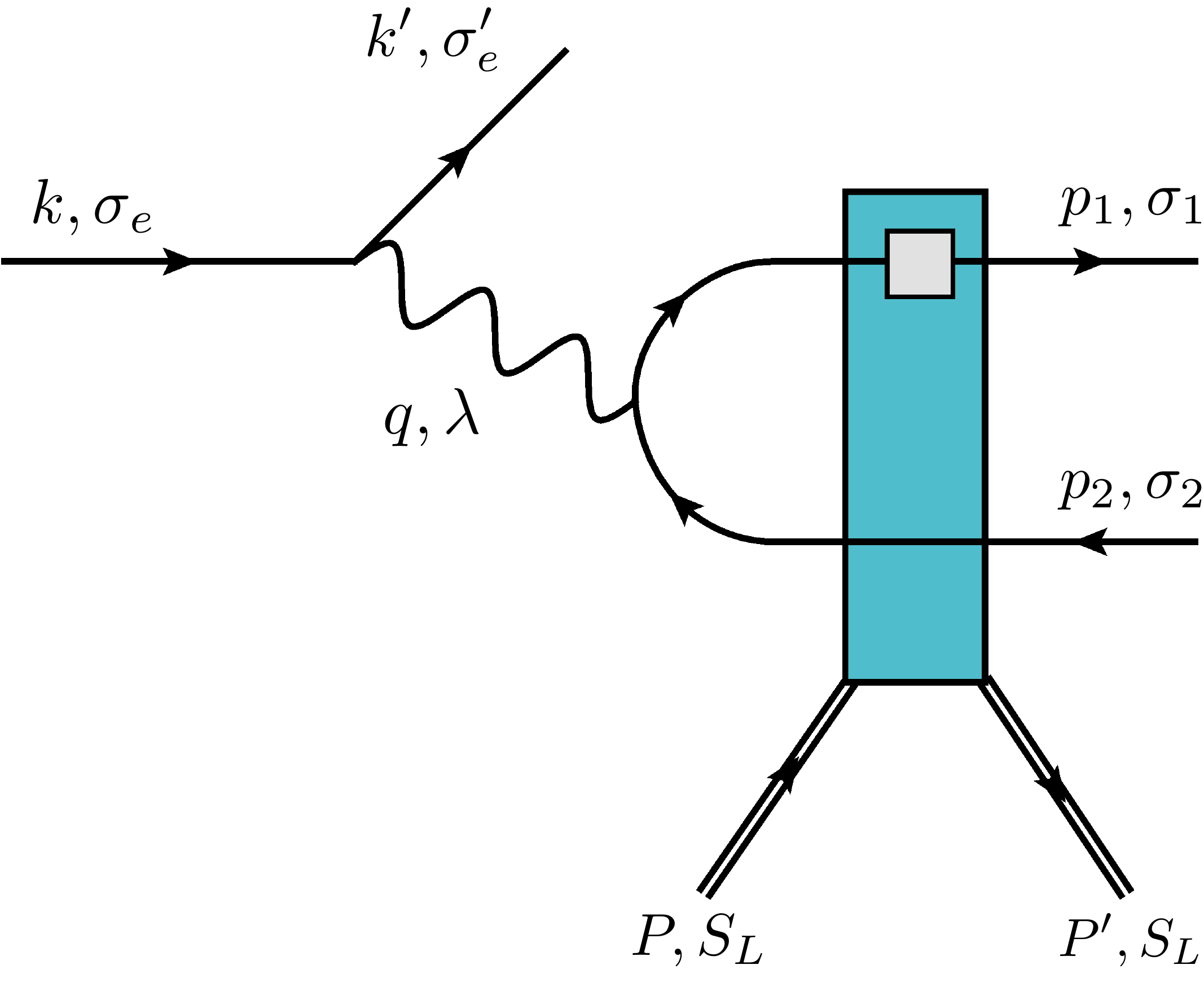}
    \caption{A diagram depicting elastic dijet electroproduction on a proton target at the leading order in $\alpha_{EM}$. An example diagram for how the dipole interacts with the proton shockwave (blue rectangle) is shown. The gray square box indicates the subeikonal quark $S$-matrix from \eq{q-Smatrix}. The other diagrams for the interaction with the proton are shown below in Fig. \ref{fig:dijet_diagrams}. The outgoing proton has the same polarization $S_L$ as the incoming proton, since there is no longitudinal spin reversal in our subeikonal calculation (per \eqs{q-Smatrix} and \eqref{aquark-Smatrix}).}
    \label{fig:dijet_xsec}
\end{figure}

Consider forward elastic dijet production in an electron scattering off of a longitudinally polarized proton. To leading order in the electromagnetic coupling, $\alpha_{EM}$, and at small $x$, the diagram for this process is shown in Fig. \ref{fig:dijet_xsec}.

%%%%%%%%%%%%%%%%%%%%%%%%%%%%%%%%%%%%%%%%%%%%%
%%%%%%%%%%%%%%%%%%%%%%%%%%%%%%%%%%%%%%%%%%%%%

Following \cite{Bacchetta:2004jz, Braun:2005rg, Mantysaari:2020lhf, Bhattacharya:2022vvo,Bhattacharya:2024sck,Hatta:2016aoc}, we will work in the ``dipole" frame, where the virtual photon's transverse momentum is zero, $q_\perp =0$, such that 
\begin{subequations} \label{kinematics}
\begin{align}
P^\mu &\approx (P^+, 0^-, {\un 0}),  \\
 q^\mu &= \left( - \frac{Q^2}{2 q^-}, q^-, {\un 0} \right), \\
 k^\mu &= \left( \frac{k_\perp^2}{2 k^-} , k^-, {\un k} \right), \\
 k'^\mu &= \left( \frac{k_\perp^2}{2 (k^- - q^-)} , k^- - q^-, {\un k} \right).
\end{align}
\end{subequations}
We will further parametrize the transverse momentum of the incoming electron by writing $\un k = k_\perp (\cos \phi, \sin \phi)$. The momenta in \eqs{kinematics} are depicted in \fig{fig:dijet_xsec}. Additionally, we define the following standard invariants
\begin{subequations}
    \begin{align}
        Q^2 &= -q^2, \\ 
        y &= \frac{P \cdot q}{P \cdot k} = \frac{q^-}{k^-}, 
        \\ 
        x &= \frac{Q^2}{2 \, P\cdot q}, 
        \\ 
        s &= (P + k)^2 \approx 2\,  P^+k^-.
    \end{align}
\end{subequations}
Above and all throughout this paper, we neglect the electron and the proton mass (the latter will appear in some prefactors though, when required by the definition of the quantities involved). Conservation of the ``+" momentum component in the electron--photon system implies that
\begin{align}\label{relation1}
Q^2 = \frac{k_\perp^2 y^2}{1-y}.
\end{align}

The cross section for electron--proton scattering,
in any frame, can be written as 
\begin{align}
\label{DIS1}
E'  \, \frac{d \sigma}{d^3 k'} = \frac{2 M_p \alpha_{EM}^2}{s \, Q^4} \, L_{\mu\nu}(k, k^\prime) \, W^{\mu\nu}(P, q),
\end{align}
where we have defined the leptonic and hadronic tensors as 
\begin{subequations}
    \begin{align}
        L^{\mu\nu}(k, k^\prime) &= 2 \, \left[ k^\mu k'^\nu + k^\nu k'^\mu - g^{\mu\nu} k \cdot k' + i \, \sigma_e \, \epsilon^{\mu\nu\rho\sigma} \, k_\rho \, k'_\sigma \right],
        \\ 
        2 M_p \, W^{\mu \nu} (P, q) &= \frac{1}{2\pi} \, \sum_X \, \bra{P, S_L} j^\mu (0) \ket{X} \, \bra{X} j^\nu (0) \ket{P, S_L} \, (2 \pi)^4 \, \delta^4 (P + q - p_X). \label{hadronic_tensor}
    \end{align}
\end{subequations}
Here $M_p$ is the proton mass. For production cross sections, we do not sum over {\sl all} the final states denoted by $X$. In our particular calculation of elastic dijet production, $\ket{X}$ consists of the recoiling proton and two (quark and antiquark) jets, with a rapidity gap between the proton and the jets: in such a case, the sum over $X$ in \eq{hadronic_tensor} goes over the momenta and quantum numbers of the specified final-state particles.  

Using the transverse (T) and longitudinal (L) virtual photon polarizations 
\begin{align}
\epsilon_T^\lambda = (0,0, {\un \epsilon}_\lambda ) , \ \ \ \epsilon_L = \left( \frac{Q}{2 q^-}, \frac{q^-}{Q}, {\un 0} \right)
\end{align}
with ${\un \epsilon}_\lambda = - (1/\sqrt{2}) (-\lambda, i)$, we can write 
\begin{align}\label{metric}
g_{\mu\nu} = - \sum_{\lambda = \pm 1} \epsilon_{T \, \mu}^{\lambda *} \, \epsilon_{T \nu}^\lambda + \epsilon_{L \, \mu}^{*} \, \epsilon_{L \nu} + \frac{q_\mu \, q_\nu}{q^2} .
\end{align}
The last term in \eq{metric} vanishes when multiplying the leptonic tensor, $L^{\mu\nu}$, via the Ward identity. We are left with the transverse and longitudinal polarizations. In the polarization basis, we define 
\begin{align}
L_{\lambda\lambda'} \equiv L^{\mu\nu} \, \epsilon_\mu^\lambda \, \epsilon_\nu^{\lambda' *}, \ \ \ W^{\lambda\lambda'} \equiv W^{\alpha\beta} \, \epsilon_\alpha^{\lambda *} \, \epsilon_\beta^{\lambda'}.
\end{align}
Then, \eq{DIS1} can be recast as \cite{Mantysaari:2020lhf}
\begin{align}\label{DIS2}
E'  \, \frac{d \sigma}{d^3 k'} =  \frac{2 M_p \alpha_{EM}^2}{s \, Q^4} \, \sum_{\lambda, \lambda' = 0, \pm 1} \, (-1)^{\lambda + \lambda'} \, L^{\mu\nu} \, \epsilon_\mu^\lambda \, \epsilon_\nu^{\lambda' *} \, W^{\alpha\beta} \, \epsilon_\alpha^{\lambda *} \, \epsilon_\beta^{\lambda'}  = \frac{2 M_p \alpha_{EM}^2}{s \, Q^4} \, \sum_{\lambda, \lambda' = 0, \pm 1} \, (-1)^{\lambda + \lambda'} \, L_{\lambda\lambda'} \, W^{\lambda\lambda'},
\end{align}
where we have denoted the longitudinal polarization by $\lambda =0$.

We may rewrite \eq{DIS2} in terms of the virtual photon--proton cross section by noting that the latter is related to the hadronic tensor via \cite{Kovchegov:2012mbw}\footnote{Note that, strictly speaking the quantity $\sigma_{\lambda \lambda'}^{\gamma^* p}$ becomes a cross section only for $\lambda = \lambda'$.}
\begin{align}
\sigma_{\lambda \lambda'}^{\gamma^* p} 
= \frac{4 \pi^2 \alpha_{EM} \, x}{Q^2} \, 2 M_p \, W^{\lambda \lambda'} .
\end{align}
Thus,
\begin{align}
2 M_p \, W^{\lambda \lambda'} = \frac{Q^2}{4 \pi^2 \alpha_{EM} \, x} \, \sigma_{\lambda \lambda'}^{\gamma^* p}, 
\end{align}
and \eq{DIS2} becomes
\begin{align}\label{DIS3}
E'  \, \frac{d \sigma}{d^3 k'} = \frac{\alpha_{EM} \, y}{4\pi^2 \, Q^4} \, \sum_{\lambda, \lambda' = 0, \pm 1} \, (-1)^{\lambda + \lambda'} \, L^{\lambda\lambda'} \, \sigma^{\gamma^* p}_{\lambda\lambda'}.
\end{align}

Now, explicit computation of the leptonic tensor gives (cf. \cite{Mantysaari:2020lhf}) 
\begin{subequations}\label{Lij}
\begin{align}
& L_{\lambda = T, \lambda' = T} = \frac{Q^2}{y^2}  \, \delta_{\lambda \lambda'} \,  \left\{   \left[ 1 + (1-y)^2 \right] - \sigma_e \, \lambda \,  \left[ (1-y)^2 -1 \right] \right\}  -  Q^2 \, \delta_{\lambda, - \lambda'} \, \frac{2 (1-y)}{y^2} \, e^{-2 i \lambda \, \phi}, \\
& L_{\lambda = T, \lambda' = L} = - e^{-i \lambda \, \phi} \, \sqrt{2 (1-y)} \, \frac{Q^2}{y^2} \, \left[ \lambda \,  (2-y) + \sigma_e \, y \right] = \left( L_{\lambda = L, \lambda' = T} \right)^* , \\
& L_{\lambda = L, \lambda' = L} = Q^2 \, \frac{4 (1-y)}{y^2},
\end{align}
\end{subequations}
where we have labeled the transverse virtual photon polarizations by $\lambda = T$, and the longitudinal polarization by $\lambda = L$. In arriving at \eq{Lij}, we have used \eq{relation1}. 

Using Eqs.~\eqref{Lij} in \eq{DIS3}, and noticing that
\begin{align}
    \left( \sigma^{\gamma^* p}_{\lambda\lambda'} \right)^* = \sigma^{\gamma^* p}_{\lambda' \lambda}
\end{align}
we obtain
\begin{align}\label{DIS20}
& E' \, \frac{d \sigma}{d^3 k'} =\frac{\alpha_{EM}}{4 \pi^2  \, Q^2 \, y} \, \Bigg\{ \left[ 1 + (1-y)^2 \right]  \, \sum_{\lambda = \pm 1} \sigma^{\gamma^* p}_{\lambda \lambda}  + \sigma_e \,  \left[ 1 - (1-y)^2 \right] \, \sum_{\lambda = \pm 1} \lambda \, \sigma^{\gamma^* p}_{\lambda \lambda} - 2 (1-y) \, \sum_{\lambda = \pm 1}  e^{-2 i \lambda \, \phi} \, \sigma^{\gamma^* p}_{\lambda, - \lambda} \\
& + 4 (1-y) \, \sigma^{\gamma^* p}_{00} + (2-y) \, \sqrt{2 (1-y)} \,  \sum_{\lambda = \pm 1} \lambda \, \left[ e^{i \lambda \, \phi} \, \sigma^{\gamma^* p}_{0\lambda} + \cc \right] + \sigma_e \, y \, \sqrt{2 (1-y)} \, \sum_{\lambda = \pm 1} \left[ e^{i \lambda \, \phi} \, \sigma^{\gamma^* p}_{0\lambda} + \cc \right] \Bigg\} . \notag
\end{align}

For a longitudinally polarized proton, we distinguish the double- and single-spin asymmetries in the dipole frame, in the notation of \cite{Diehl:2005pc}, 
\begin{subequations}
\begin{align}
 A_{LL}^{\gamma^*} &= \frac{ d\sigma (++) - d\sigma(+-) }{d\sigma (++) + d\sigma(+-)} \equiv \frac{ d\sigma (++) - d\sigma(+-) }{2 \, d\sigma_{unpol}} \equiv   \frac{d \sigma^{DSA} }{d\sigma_{unpol}} , \\
 A_{UL}^{\gamma^*} &= \frac{ d\sigma (+) - d\sigma(-) }{d\sigma (+) + d\sigma(-)} = \frac{ d\sigma (+) - d\sigma(-) }{2 \, d\sigma_{unpol}} \equiv   \frac{d \sigma^{SSA} }{d\sigma_{unpol}} ,
\end{align}
\end{subequations}
such that\footnote{Strictly speaking, our lepton helicity $\sigma_e$ is measured with respect to the lepton beam direction, while the proton's longitudinal polarization $S_L$ is calculated in the dipole ($q_\perp=0$) frame. However, per \cite{Diehl:2005pc}, the difference between the longitudinal proton polarizations in the $q_\perp =0$ and $k_\perp =0$ frames is a higher-order in $x$ effect which we can neglect.}
\begin{subequations}
\begin{align} \label{DSA_def}
 d \sigma^{DSA} &= \frac{1}{4} \, \sum_{\sigma_e, S_L} \, \sigma_e \, S_L \, d \sigma (\sigma_e, S_L), \\
 \label{SSA_def}
 d \sigma^{SSA} &= \frac{1}{4} \, \sum_{\sigma_e, S_L} \, S_L \, d \sigma (\sigma_e, S_L), \\
 d\sigma_{unpol} &= \frac{1}{4} \, \sum_{\sigma_e, S_L} \, d \sigma (\sigma_e, S_L) .
\end{align}
\end{subequations}
Substituting \eqs{DIS3} and (\ref{Lij}) in \eqs{DSA_def} and (\ref{SSA_def}), we get
\begin{subequations} \label{observables}
    \begin{align} \label{DSA}
        E_{k^\prime} \frac{d \sigma^{DSA}}{d^3 k^\prime} &= 
        \frac{\alpha_{EM}}{4\pi^2 \, Q^2}
         \half \, \sum_{S_L} S_L 
        \Bigg\{ 
            (2-y) \, \sum_{\lambda = \pm 1} \lambda \, \sigma^{\gamma^*p}_{\lambda \lambda} +  \sqrt{2(1-y)} \sum_{\lambda = \pm 1} 
            \Big[
            e^{i \lambda \phi} \sigma^{\gamma^*p}_{0 \lambda} + \mathrm{c.c.} 
            \Big] 
        \Bigg\},
        \\  \label{SSA}
    E_{k^\prime} \frac{d \sigma^{SSA}}{d^3 k^\prime} &= 
    \frac{\alpha_{EM}}{4 \pi^2 \, Q^2 \, y} \, \half \, 
    \sum_{S_L} S_L 
        \Bigg\{ 
           \left[1+ (1-y)^2 \right] \sum_{\lambda = \pm 1} \sigma^{\gamma^*p}_{\lambda \lambda} - 2 (1-y) \Big[ e^{-2i\phi} \sigma^{\gamma^*p}_{1,-1} +\mathrm{c.c.}
        \Big]
        \\ \notag 
        & \hspace{3cm}
          + 4 (1-y) \, \sigma^{\gamma^*p}_{00} +  (2-y) \sqrt{2(1-y)} \sum_{\lambda = \pm 1} \Big[
            \lambda \, e^{i \lambda \phi} \sigma^{\gamma^*p}_{0 \lambda} + \mathrm{c.c.}
          \Big]  
        \Bigg\},
    \end{align}
\end{subequations}
for the numerators of the DSA and SSA (see \cite{Diehl:2005pc, Diehl:1996st, Arens:1996xw} for similar decompositions). \eqs{observables} contain the main quantities we would like to calculate\footnote{Note that, despite the explicit appearance of the angle $\phi$ describing the transverse motion of the electron, \eqs{observables} can be written in a Lorentz-invariant form along the lines of Eq.~(17) of \cite{Bacchetta:2004jz}.}. Upon a quick inspection, there appear to be six independent structures in \eqs{observables}:
\begin{align} \label{structures}
    \sum_{\lambda = \pm 1} \lambda \, \sigma^{\gamma^*p}_{\lambda \lambda}, \hspace{0.25cm}
    \sum_{\lambda = \pm 1} 
            \Big[
            e^{i \lambda \phi} \sigma^{\gamma^*p}_{0 \lambda} + \mathrm{c.c.} 
            \Big] ,
    \hspace{0.25cm}     
    \sum_{\lambda = \pm 1} \sigma^{\gamma^*p}_{\lambda \lambda},
     \hspace{0.25cm}
    e^{-2i\phi} \sigma^{\gamma^*p}_{1,-1} +\mathrm{c.c.} ,
     \hspace{0.25cm}
    \sigma^{\gamma^*p}_{00},
     \hspace{0.25cm}
    \sum_{\lambda = \pm 1} \Big[
            \lambda \, e^{i \lambda \phi} \sigma^{\gamma^*p}_{0 \lambda} + \mathrm{c.c.}
          \Big]  .
\end{align}
However, as we will see below, given the current state of small-$x$ physics, only the two structures in the expression for the DSA in \eq{DSA} are a viable probe for the moment amplitudes defined above. 

Now that we have general expressions for the DSA and SSA, we turn to calculating the cross section for the forward elastic dijet production in virtual photon--proton scattering: substituting it for $\sigma^{\gamma^*p}_{\lambda \lambda'}$ in \eqs{observables} would give us the numerators of the double and single-spin asymmetries for these elastic dijet production processes.

%%%%%%%%%%%%%%%%%%%%%%%%%%%%%%%%%%%%%%%%%%%%%%%%%%%%%%%%%%%%%%%%%%%%%%%%%%%%%%%
\section{General elastic dijet production calculation}
\label{sec:production}

Since all of the structures in \eqs{observables} stem from the virtual photon-proton cross section in the polarization basis, $\sigma^{\gamma^* p}_{\lambda \lambda^\prime}$, we start with a general calculation, without restricting ourselves to any one structure in \eqs{observables}. Furthermore, although our discussion in the previous Section applied to elastic as well as inelastic dijet production, let us restrict ourselves to elastic dijet production now. To do so, introduce the four-momentum of the outgoing proton,
\begin{align}
P'^\mu = \left( P^{\prime \, +}, \frac{P^{\prime \, 2}_\perp}{2 P^{\prime \, +}}, {\un P}' \right) ,
\end{align}
where ${\un P}'  = - {\un p}_1 - {\un p}_2$ with $p_1$ and $p_2$ the momenta of the two produced jets, as shown in \fig{fig:dijet_xsec}. Further, we assume that $P^{\prime \, +}$ is still very large, such that $P' \cdot q \approx P^{\prime \, +} \, q^- \approx P^+ \, q^- \approx P \cdot q$. The corrections to this approximation are subeikonal, but they do not couple to the proton longitudinal spin. 

\begin{figure}[ht!]
\centering
\includegraphics[scale=0.45]{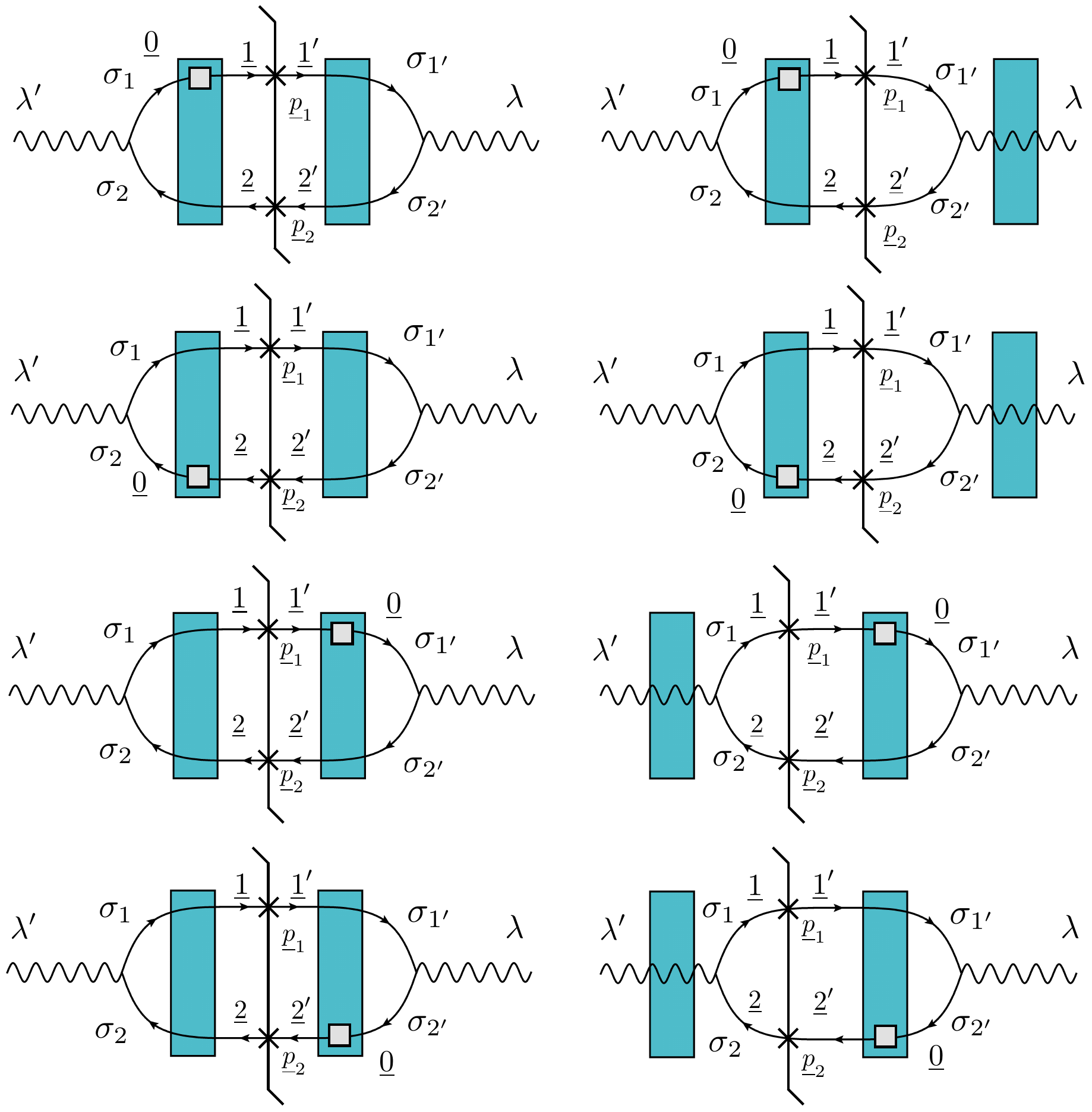}  
\caption{Dijet production diagrams at the subeikonal level. The vertical shaded rectangles represent the proton shockwave and the gray square boxes indicate the subeikonal parts of the quark and antiquark $S$-matrices from \eqs{q-Smatrix} and (\ref{aquark-Smatrix}). The crosses indicate the measured jets.}
\label{fig:dijet_diagrams}
\end{figure}

The elastic dijet production cross section in the $\gamma^* + p$ scattering is, in our infinite momentum frame,
\begin{align}
d\sigma^{\gamma^* p \to q {\bar q} p'} & = \frac{1}{4 q^0 E_P} \, |M|^2 \, \frac{d^2 p_1 \, d p_1^-}{(2 \pi)^3 \, 2 p_1^-} \, \frac{d^2 p_2 \, d p_2^-}{(2 \pi)^3 \, 2 p_2^-} \, \, \frac{d^2 P' \, d P^{\prime \, +}}{(2 \pi)^3 \, 2 P^{\prime \, +}} \, (2 \pi)^4 \, \delta^4 (q+P - p_1 - p_2 - P') \\ &  = \frac{1}{4 \pi} |A|^2 \, \frac{d^2 p_1 \, d^2 p_2}{(2 \pi)^4} \, \frac{d z}{z (1-z)} \notag
\end{align}
with $A = M/(2s)$, $E_p$ the energy of the incoming proton,  and $z = p_1^- /q^-$, the fraction of the photon's minus momentum carried by the produced quark. Utilizing the quark and antiquark $S$-matrices from \eqs{q-Smatrix} and (\ref{aquark-Smatrix}), while noticing that ${\overline V}^{\mathrm{pol}}_{\un x, \un y; \sigma^\prime, \sigma} = - \left( V^{\mathrm{pol}}_{\un x, \un y; -\sigma^\prime, - \sigma} \right)^\dagger$, we write, to subeikonal order,\footnote{Note that the same angle brackets are used in \eq{XS1} to indicate both the spin-dependent and spin-independent averaging in the proton state. Where the polarized Wilson line, $V^{\mathrm{pol}}$, appears in the correlation function, the spin-dependent averaging should be used, as defined in the single-bracket part of \eq{double-bracket}. By contrast, when $V^{\mathrm{pol}}$ does not appear in a correlator, it is understood as being unpolarized, with the spin-independent definition of the angle brackets used, that is, with the single-bracket part of \eq{double-bracket} but without the factor of $S_L$ under the sum.} 
\begin{align}\label{XS1}
 & z (1-z) \, \frac{1}{2} \! \sum_{S_L = \pm 1} \!\! S_L \, \frac{d\sigma_{\lambda \lambda'}^{\gamma^* p \to q {\bar q} p'}}{d^2 p_1 \, d^2 p_2 \, d z} = \frac{1}{2 (2 \pi)^5} \,  \frac{1}{2} \sum_{S_L= \pm 1} S_L\, A_{\lambda'} \, A^*_{\lambda} 
 \\ 
& = \frac{1}{2 (2 \pi)^5} \, \int d^2 x_1 \, d^2 x_{1'} \, d^2 x_2 \, d^2 x_{2'} \, d^2 x_0 \, e^{- i {\un p}_1 \cdot {\un x}_{11'} - i {\un p}_2 \cdot {\un x}_{22'}} \, \sum_{\sigma_1, \sigma_2, \sigma'_1, \sigma'_2, i , j}
\notag
\\ & 
\times \Bigg\{ \Psi_{\lambda', \sigma_1, \sigma_2; i, i}^{\gamma^* \to q {\bar q}} ({\un x}_{02}, z) \, \left[ \Psi_{\lambda, \sigma'_1, \sigma'_2; j, j}^{\gamma^* \to q {\bar q}} ({\un x}_{1'2'}, z) \right]^* \, \frac{1}{N_c^2} \left\langle \tord \tr \left[ V^\textrm{pol}_{{\un 1}, {\un 0}; \sigma'_1, \sigma_1} \, V_{{\un x}_2}^\dagger \right] \right\rangle \, \left\langle \atord \tr \left[ V_{{\un 2'}} \, V_{{\un 1'}}^\dagger - 1 \right]  \right\rangle \, \delta_{\sigma_2 , \sigma'_2} 
\notag 
\\ & 
 -  \Psi_{\lambda', \sigma_1, \sigma_2; i, i}^{\gamma^* \to q {\bar q}} ({\un x}_{10}, z) \, \left[ \Psi_{\lambda, \sigma'_1, \sigma'_2; j, j}^{\gamma^* \to q {\bar q}} ({\un x}_{1'2'}, z) \right]^* \frac{1}{N_c^2} \left\langle \tord \tr \left[ V_{\un 1} \,  V^{\textrm{pol} \, \dagger}_{{\un 2}, {\un 0}; - \sigma'_2, - \sigma_2} \right] \right\rangle \, \left\langle \atord \tr \left[ V_{{\un 2'}} \, V_{{\un 1'}}^\dagger - 1 \right]  \right\rangle \, \delta_{\sigma_1 , \sigma'_1} 
 \notag 
 \\& 
 +  \Psi_{\lambda', \sigma_1, \sigma_2; i, i}^{\gamma^* \to q {\bar q}} ({\un x}_{12}, z) \, \left[ \Psi_{\lambda, \sigma'_1, \sigma'_2; j, j}^{\gamma^* \to q {\bar q}} ({\un x}_{02'}, z) \right]^* \frac{1}{N_c^2} \left\langle \tord \tr \left[   V_{{\un 1}} \, V_{{\un 2}}^\dagger - 1 \right]  \right\rangle \, \left\langle \atord \tr \left[ V_{{\un 2'}} \,  V^{\textrm{pol} \, \dagger}_{{\un 1'}, {\un 0}; \sigma_1, \sigma'_1}  \right]  \right\rangle \, \delta_{\sigma_2 , \sigma'_2} 
 \notag 
 \\& 
 -  \Psi_{\lambda', \sigma_1, \sigma_2; i, i}^{\gamma^* \to q {\bar q}} ({\un x}_{12}, z) \, \left[ \Psi_{\lambda, \sigma'_1, \sigma'_2; j, j}^{\gamma^* \to q {\bar q}} ({\un x}_{1'0}, z) \right]^* \frac{1}{N_c^2} \left\langle \tord \tr \left[ V_{{\un 1}} \, V_{{\un 2}}^\dagger - 1 \right] \right\rangle \, \left\langle \atord \tr \left[  V^{\textrm{pol} }_{{\un 2'}, {\un 0}; -\sigma_2, - \sigma'_2} \, V_{{\un 1'}}^\dagger \right]  \right\rangle \, \delta_{\sigma_1 , \sigma'_1} \Bigg\} .\notag
\end{align}
The diagrams corresponding to \eq{XS1} are shown in \fig{fig:dijet_diagrams}, leading to production of two quark jets in the current fragmentation region. In general, we could also produce gluon jets. This would require the quark-to-gluon transition operators from \cite{Borden:2024bxa, Chirilli:2021lif}, containing quark or antiquark exchanges with the shock wave. However, since we are considering an elastic scattering process here, the exchange between the projectile and the target has to be color-singlet. This means that, both in the amplitude and in the complex conjugate amplitude, we cannot have a single quark exchange between the projectile dipole and the target (in addition to any number of gluon exchanges), and the interaction should start with a two-quark exchange, one on a quark line and another on the antiquark line. The cross section would then contain two quark exchanges each in the amplitude and in the complex conjugate amplitude: the interaction would be subsubeikonal,  that is, suppressed by an additional power of $x$ compared to the diagrams shown in \fig{fig:dijet_diagrams}. We do not consider such small corrections to our calculation. In addition, the contributions with the $\gamma^* \to q {\bar q}$ splitting happening inside the shock wave have also been neglected in \eq{XS1} and in \fig{fig:dijet_diagrams} (see \cite{Altinoluk:2022jkk} for an analysis of such subeikonal contributions for the inclusive dijet production). One can show that the in-shock wave splittings come in with a delta-function forcing the dipole size to be zero: since any of the traces in \eq{XS1} is zero when either $\un x_{1} = \un x_2$ or $\un x_{1'} = \un x_{2'}$, such contributions are zero for the elastic process at hand. 

%%%%%%%%%%%%%%%%%%%%%%%%%%%%%%%%%%%%%%%%%%%%%%%%%%%%%%%%%%%%%%%%%%%%%%%%%%%%%%%%%%%%
%%%%%%%%%%%%%%%%%%%%%%%%%%%%%%%%%%%%%%%%%%%%%%%%%%%%%%%%%%%%%%%%%%%%%%%%%%%%%%%%%%%%

In \eq{XS1}, $\Psi^{\gamma^* \to q\bar{q}}_{\lambda, \sigma_1, \sigma_2; i, j}(\un x, z)$ is the light cone wave function for the virtual photon to fluctuate into a quark-antiquark pair. For a transversely polarized photon, this wave function is given by (see e.g. \cite{Kovchegov:2012mbw} and references therein)
\begin{align} \label{t-psi}
    \Psi^{\gamma^* \to q\bar{q}}_{\lambda=\pm 1, \sigma_1, \sigma_2; i, j}({\un x}_{12}, z) &= \frac{e Z_f}{2\pi} \delta_{ij} \delta_{\sigma_1, - \sigma_2} z (1-z) (1- 2z + \sigma_1 \lambda) \, i Q \, \frac{\un{\epsilon}_{\lambda} \cdot \un x_{12}}{x_{12}} \, K_1 \left( x_{12} Q \sqrt{z(1-z)}\right),
\end{align}
where $e$ is the electron charge and $Z_f$ is the fraction of the magnitude of the electron charge carried by a given quark flavor $f$. We have neglected the mass of the quarks here. The indices $i,j$ label the colors of the quark and antiquark. For a longitudinally polarized photon, the wave function is
\begin{align} \label{l-psi}
    \Psi^{\gamma^* \to q\bar{q}}_{\lambda=0 , \sigma_1, \sigma_2; i, j}({\un x}_{12}, z) &=   
    -\frac{e Z_f}{2\pi} \delta_{ij} \delta_{\sigma_1, - \sigma_2} \big[ z (1-z) \big]^{3/2} \, 2 Q \, K_0 \left( x_{12} Q \sqrt{z(1-z)}\right).
\end{align}
Both in \fig{fig:dijet_diagrams} as well as in \eq{XS1} and below, we abbreviate $\un x_{1} = \un 1$, etc., for the transverse coordinates in the subscripts of the (standard or polarized) Wilson lines. As usual, $\atord$ denotes anti-time ordering in the complex conjugate amplitudes.

To evaluate the expression further, we employ the specific polarized Wilson lines given above in \eqs{VqG}. We will separately consider the type-1 and type-2 polarized Wilson lines.

%%%%%%%%%%%%%%%%%%%%%%%

\subsection{Type-1 operators}

First, consider the terms in \eqs{XS1} that arise from $V^{\textrm{pol} [1]}_{\un x} = V_{\un x}^{\textrm{G} [1]} + V_{\un x}^{\textrm{q} [1]}$. Integrating over ${\un x}_0$, summing over $\sigma'_1, \sigma'_2$ and over $\sigma_2$ while employing the fact that the light cone wave functions $\Psi \sim \delta_{\sigma_1, - \sigma_2}$ in the massless quark case at hand, we obtain
\begin{align}\label{XS3}
 z (1-z) \, \frac{1}{2} \! \sum_{S_L = \pm 1} \!\! S_L \ & \frac{d\sigma_{\lambda \lambda'}^{[1] \, \gamma^* p \to q {\bar q} p'}}{d^2 p_1 \, d^2 p_2 \, d z} = \frac{1}{2 (2 \pi)^5} \, \int d^2 x_1 \, d^2 x_{1'} \, d^2 x_2 \, d^2 x_{2'} \,  e^{- i {\un p}_1 \cdot {\un x}_{11'} - i {\un p}_2 \cdot {\un x}_{22'}} 
 \\ \notag 
 &
 \times 
 \sum_{\sigma_1, i , j} \, \sigma_1 \,  \Psi_{\lambda', \sigma_1, - \sigma_1; i, i}^{\gamma^* \to q {\bar q}} ({\un x}_{12}, z) \, \left[ \Psi_{\lambda, \sigma_1, - \sigma_1; j, j}^{\gamma^* \to q {\bar q}} ({\un x}_{1'2'}, z) \right]^* 
 \\ \notag 
 & 
 \times 
 \frac{1}{N_c^2}  \Bigg\{ 
 \left[ \left\langle \tord \tr \left[ V^{\textrm{pol}[1]}_{{\un 1}} \, V_{{\un 2}}^\dagger \right] \right\rangle
 \big(zs \big)
 -
 \left\langle \tord \tr \left[ V_{{\un 1}} \,  V^{\textrm{pol} [1] \, \dagger}_{{\un 2}} \right] \right\rangle \big((1-z)s \big)
 \right] \, \left\langle \atord \tr \left[ V_{{\un 2'}} \, V_{{\un 1'}}^\dagger - 1 \right]  \right\rangle(s)
 \\ \notag 
 & \hspace{1cm} +  
 \left[ \left\langle \atord \tr \left[ V_{\un 2^\prime} V^{\textrm{pol}[1]\dagger}_{{\un 1^\prime}} \right] \right\rangle \big(zs \big)
 -
 \left\langle \atord \tr \left[ V^{\textrm{pol} [1]}_{{\un 2^\prime}} V_{{\un 1^\prime}}^\dagger    \right] \right\rangle \big((1-z)s \big)
 \right] \, \left\langle \tord \tr \left[ V_{{\un 1}} \, V_{{\un 2}}^\dagger - 1 \right]  \right\rangle(s)
 \Bigg\} .\notag
\end{align}

Defining the unpolarized dipole $S$-matrix \cite{Mueller:1994rr,Mueller:1994jq,Mueller:1995gb,Balitsky:1995ub,Balitsky:1998ya,Kovchegov:1999yj,Kovchegov:1999ua,Jalilian-Marian:1997dw,Jalilian-Marian:1997gr,Weigert:2000gi,Iancu:2001ad,Iancu:2000hn,Ferreiro:2001qy}
\begin{align} \label{unpol-dip-amp}
S_{10} (s) =  \frac{1}{N_c} \, \left\langle \mbox{T} \, \tr \left[ V_{\un 1} \,  V_{\un 0}^{\dagger} \right] \right\rangle (s),
\end{align}
we rewrite \eq{XS3} as 
\begin{align}\label{XS4}
 z (1-z) \, \frac{1}{2} \! \sum_{S_L = \pm 1} \!\! S_L \, \frac{d\sigma_{\lambda \lambda'}^{[1] \, \gamma^* p \to q {\bar q} p'}}{d^2 p_1 \, d^2 p_2 \, d z} &= -\frac{1}{2 (2 \pi)^5} \, \int d^2 x_1 \, d^2 x_{1'} \, d^2 x_2 \, d^2 x_{2'} \,  e^{- i {\un p}_1 \cdot {\un x}_{11'} - i {\un p}_2 \cdot {\un x}_{22'}} 
 \\ \notag 
 &
 \times 
 \sum_{\sigma_1, i , j} \, \sigma_1 \,  \Psi_{\lambda', \sigma_1, - \sigma_1; i, i}^{\gamma^* \to q {\bar q}} ({\un x}_{12}, z) \, \left[ \Psi_{\lambda, \sigma_1, - \sigma_1; j, j}^{\gamma^* \to q {\bar q}} ({\un x}_{1'2'}, z) \right]^* 
 \\ \notag 
 & 
 \times 
 \frac{1}{N_c}  \Bigg\{ 
 \left[ \left\langle \tord \tr \left[ V^{\textrm{pol}[1]}_{{\un 1}} \, V_{{\un 2}}^\dagger \right] \right\rangle
 \big(zs \big)
 -
 \left\langle \tord \tr \left[ V_{{\un 1}} \,  V^{\textrm{pol} [1] \, \dagger}_{{\un 2}} \right] \right\rangle \big((1-z)s \big)
 \right] \, 
 \Big[
  1 - S^*_{ 1^\prime 2^\prime}(s)
 \Big]
 \\ \notag 
 & \hspace{1cm} +  
 \left[ \left\langle \atord \tr \left[ V_{\un 2^\prime} V^{\textrm{pol}[1]\dagger}_{{\un 1^\prime}} \right] \right\rangle \big(zs \big)
 -
 \left\langle \atord \tr \left[ V^{\textrm{pol} [1]}_{{\un 2^\prime}} V_{{\un 1^\prime}}^\dagger  \right] \right\rangle
 \big( (1- z) s \big)
 \right] \, 
 \Big[
  1 - S_{1  2}(s)
 \Big]
 \Bigg\} .\notag
\end{align}
Employing the double angle bracket notation from \eq{double-bracket}, we write
\begin{align}\label{XS5}
 z (1-z) \, \frac{1}{2} \! \sum_{S_L = \pm 1} \!\! S_L \ & \frac{d\sigma_{\lambda \lambda'}^{[1] \, \gamma^* p \to q {\bar q} p'}}{d^2 p_1 \, d^2 p_2 \, d z} = -\frac{1}{2 (2 \pi)^5 s} \, \int d^2 x_1 \, d^2 x_{1'} \, d^2 x_2 \, d^2 x_{2'} \,  e^{- i {\un p}_1 \cdot {\un x}_{11'} - i {\un p}_2 \cdot {\un x}_{22'}} 
 \\ \notag 
 &
 \times 
 \sum_{\sigma_1, i , j} \, \sigma_1 \,  \Psi_{\lambda', \sigma_1, - \sigma_1; i, i}^{\gamma^* \to q {\bar q}} ({\un x}_{12}, z) \, \left[ \Psi_{\lambda, \sigma_1, - \sigma_1; j, j}^{\gamma^* \to q {\bar q}} ({\un x}_{1'2'}, z) \right]^* 
 \\ \notag 
 & 
 \times 
 \frac{1}{N_c}  \Bigg\{ 
 \left[ \frac{1}{z} \llangle \tord \tr \left[ V^{\textrm{pol}[1]}_{{\un 1}} \, V_{{\un 2}}^\dagger \right] \rrangle
(s)
 -
 \frac{1}{1-z} \llangle \tord \tr \left[ V_{{\un 1}} \,  V^{\textrm{pol} [1] \, \dagger}_{{\un 2}} \right] \rrangle (s)
 \right] \, 
 \Big[
  1 - S^*_{ 1^\prime 2^\prime}(s)
 \Big]
 \\ \notag 
 & \hspace{1cm} +  
 \left[ \frac{1}{z} \llangle \atord \tr \left[ V_{\un 2^\prime} V^{\textrm{pol}[1]\dagger}_{{\un 1^\prime}} \right] \rrangle (s) 
 -
 \frac{1}{1-z}
\llangle \atord \tr \left[ V^{\textrm{pol} [1]}_{{\un 2^\prime}} V_{{\un 1^\prime}}^\dagger  \right] \rrangle
 (s)
 \right] \, 
 \Big[
  1 - S_{1  2}(s)
 \Big]
 \Bigg\} .\notag
\end{align}
Here we assume that $z \sim 1-z \sim {\cal O} (1)$, such that $\as \, \ln^2 (z/(1-z)) \ll 1$ and we do not need to include quantum (DLA) evolution corrections in the rapidity interval between the produced quark and antiquark. Therefore, all the polarized dipole scattering amplitudes in the double angle brackets, along with the unpolarized dipole $S$-matrix, depend on $s$ in the argument, since, for us, $z \, s  \sim (1-z) \, s \sim s$.  

To simplify \eq{XS5} further, we will need the specific wave function overlaps that appear in each structure in \eqs{observables}. Using the wave functions from \eqs{t-psi} and (\ref{l-psi}), we write and define
\begin{subequations} \label{type1_overlaps}
\begin{align}
\label{TT1} 
& \textrm{\bf TT:} \sum_{\sigma_1, i , j} \, \sigma_1 \, \Psi_{\lambda' =T, \sigma_1, - \sigma_1; i, i}^{\gamma^* \to q {\bar q}} ({\un x}_{12}, z) \, \left[ \Psi_{\lambda =T, \sigma_1, - \sigma_1; j, j}^{\gamma^* \to q {\bar q}} ({\un x}_{1'2'}, z) \right]^* 
\\ & = \frac{2 \alpha_{EM}  Z_f^2 N_c^2}{\pi}  z^2  (1-z)^2 (1-2z) Q^2  
\, \delta_{\lambda, \lambda'} 
 \frac{i \, {\un x}_{12} \times {\un x}_{1'2'} +\lambda \, {\un x}_{12} \cdot {\un x}_{1'2'}}{x_{12} \, x_{1'2'} } \, K_1 \left( x_{12} Q \sqrt{z (1-z)} \right) \, K_1 \left( x_{1'2'} Q \sqrt{z (1-z)} \right) \notag \\ 
& \equiv \delta_{\lambda, \lambda^\prime} \Bigg[
\lambda\, 
\Phi_{\mathrm{TT}}^{[1]}(\un x_{12}, \un x_{1^\prime 2^\prime}, z) 
+
i \Phi_{\mathrm{TT}}^{\prime\,[1]}(\un x_{12}, \un x_{1^\prime 2^\prime}, z)
\Bigg],
\notag
\\ 
 & \textrm{\bf LT:}  \sum_{\sigma_1, i , j} \, \sigma_1 \, \Psi_{\lambda' =T, \sigma_1, - \sigma_1; i, i}^{\gamma^* \to q {\bar q}} ({\un x}_{12}, z) \, \left[ \Psi_{\lambda =0, \sigma_1, - \sigma_1; j, j}^{\gamma^* \to q {\bar q}} ({\un x}_{1'2'}, z) \right]^*   
 \\ \notag & 
 = -\frac{4 \, i \,  \alpha_{EM} \, Z_f^2 \, N_c^2}{\pi} \, [z (1-z)]^{5/2} \, Q^2  
 \lambda' \, 
 \frac{{\un \epsilon}_{\lambda'} \cdot {\un x}_{12}}{x_{12}}  K_1 \left( x_{12} Q \sqrt{z (1-z)} \right) \, K_0 \left( x_{1'2'} Q \sqrt{z (1-z)} \right) \notag \\
 \notag 
& \equiv 
i \lambda' \, \frac{\un \epsilon_{\lambda'} \cdot \un x_{12}}{x_{12}} \, 
\Phi_{\mathrm{LT}}^{[1]}(\un x_{12}, \un x_{1^\prime 2^\prime}, z),
 \\ 
 & \textrm{\bf TL:} \sum_{\sigma_1, i , j} \, \sigma_1 \, \Psi_{\lambda' =0, \sigma_1, - \sigma_1; i, i}^{\gamma^* \to q {\bar q}} ({\un x}_{12}, z) \, \left[ \Psi_{\lambda =T, \sigma_1, - \sigma_1; j, j}^{\gamma^* \to q {\bar q}} ({\un x}_{1'2'}, z) \right]^* = - i \lambda \, \frac{\un \epsilon_{\lambda}^* \cdot \un x_{1'2'}}{x_{1'2'}} \, 
\Phi_{\mathrm{LT}}^{[1]}(\un x_{1'2'}, \un x_{12}, z), \\
  & \textrm{\bf LL:} \sum_{\sigma_1, i , j} \, \sigma_1 \, \Psi_{\lambda' =0, \sigma_1, - \sigma_1; i, i}^{\gamma^* \to q {\bar q}} ({\un x}_{12}, z) \, \left[ \Psi_{\lambda =0 , \sigma_1, - \sigma_1; j, j}^{\gamma^* \to q {\bar q}} ({\un x}_{1'2'}, z) \right]^* = 0. \label{LL1}
\end{align}
\end{subequations}
Here the cross-product is defined by ${\un v} \times {\un w} = \epsilon^{ij} v^i w^j$. Note that the functions $\Phi^{[1]}_{\mathrm{TT}}, \Phi^{\prime \, [1]}_{\mathrm{TT}}, \Phi^{[1]}_{\mathrm{LT}}$ are real.

%%%%%%%%%%%%%%%%%%%%%%%%%%%%%%%%%%%%%%%%%%%%

\subsection{Type-2 operators}

Next, consider the terms arising from the type-2 polarized Wilson lines, $V_{{\un x}, {\un y}}^{\textrm{G} [2]} + V_{\un x}^{\textrm{q} [2]} \, \delta^2 ({\un x} - {\un y})$. The contribution of $V_{\un x}^{\textrm{q} [2]}$ to \eq{XS1} reads (again, after integrating over ${\un x}_0$, summing over $\sigma'_1, \sigma'_2$ and $\sigma_2$ while employing $\Psi \sim \delta_{\sigma_1, - \sigma_2}$)
\begin{align}\label{XS6}
z (1-z) \, \frac{1}{2} \! \sum_{S_L = \pm 1} \!\! S_L &\frac{d\sigma_{\lambda \lambda'}^{q[2] \, \gamma^* p \to q {\bar q} p'}}{d^2 p_1 \, d^2 p_2 \, d z} 
= \frac{1}{2 (2 \pi)^5} \, \int d^2 x_1 \, d^2 x_{1'} \, d^2 x_2 \, d^2 x_{2'} \,  e^{- i {\un p}_1 \cdot {\un x}_{11'} - i {\un p}_2 \cdot {\un x}_{22'}} 
\\ \notag &\times
\sum_{\sigma_1, i , j} \Psi_{\lambda', \sigma_1, -\sigma_1; i, i}^{\gamma^* \to q {\bar q}} ({\un x}_{12}, z) \, \left[ \Psi_{\lambda, \sigma_1, -\sigma_1; j, j}^{\gamma^* \to q {\bar q}} ({\un x}_{1'2'}, z) \right]^* 
\\ \notag &
\times 
 \frac{1}{N_c^2}  \Bigg\{ 
 \left[ \left\langle \tord \tr \left[ V^{\textrm{q}[2]}_{{\un 1}} \, V_{{\un 2}}^\dagger \right] \right\rangle
 \big(zs \big)
 -
 \left\langle \tord \tr \left[ V_{{\un 1}} \,  V^{\textrm{q} [2] \, \dagger}_{{\un 2}} \right] \right\rangle \big((1-z)s \big)
 \right] \, \left\langle \atord \tr \left[ V_{{\un 2'}} \, V_{{\un 1'}}^\dagger - 1 \right]  \right\rangle(s)
 \\ \notag 
 & \hspace{1cm} +  
 \left[ \left\langle \atord \tr \left[ V_{\un 2^\prime} V^{\textrm{q}[2]\dagger}_{{\un 1^\prime}} \right] \right\rangle \big(zs \big)
 -
 \left\langle \atord \tr \left[ V^{\textrm{q} [2]}_{{\un 2^\prime}} V_{{\un 1^\prime}}^\dagger    \right] \right\rangle \big((1-z)s \big)
 \right] \, \left\langle \tord \tr \left[ V_{{\un 1}} \, V_{{\un 2}}^\dagger - 1 \right]  \right\rangle(s)
 \Bigg\} .\notag
\end{align}
Again, employing \eq{unpol-dip-amp} and the double bracket notation, we arrive at 
\begin{align}\label{XS7}
 z (1-z) \, \frac{1}{2} \! \sum_{S_L = \pm 1} \!\! S_L \frac{d\sigma_{\lambda \lambda'}^{q[2] \, \gamma^* p \to q {\bar q} p'}}{d^2 p_1 \, d^2 p_2 \, d z} &= -\frac{1}{2 (2 \pi)^5 s} \, \int d^2 x_1 \, d^2 x_{1'} \, d^2 x_2 \, d^2 x_{2'} \,  e^{- i {\un p}_1 \cdot {\un x}_{11'} - i {\un p}_2 \cdot {\un x}_{22'}} 
 \\ \notag 
 &
 \times 
 \sum_{\sigma_1, i , j}  \Psi_{\lambda', \sigma_1, - \sigma_1; i, i}^{\gamma^* \to q {\bar q}} ({\un x}_{12}, z) \, \left[ \Psi_{\lambda, \sigma_1, - \sigma_1; j, j}^{\gamma^* \to q {\bar q}} ({\un x}_{1'2'}, z) \right]^* 
 \\ \notag 
 & 
 \times 
 \frac{1}{N_c}  \Bigg\{ 
 \left[ \frac{1}{z} \llangle \tord \tr \left[ V^{\textrm{q}[2]}_{{\un 1}} \, V_{{\un 2}}^\dagger \right] \rrangle
(s)
 -
 \frac{1}{1-z} \llangle \tord \tr \left[ V_{{\un 1}} \,  V^{\textrm{q} [2] \, \dagger}_{{\un 2}} \right] \rrangle (s)
 \right] \, 
 \Big[
  1 - S^*_{ 1^\prime 2^\prime}(s)
 \Big]
 \\ \notag 
 & \hspace{1cm} +  
 \left[ \frac{1}{z} \llangle \atord \tr \left[ V_{\un 2^\prime} V^{\textrm{q}[2]\dagger}_{{\un 1^\prime}} \right] \rrangle (s) 
 -
 \frac{1}{1-z}
\llangle \atord \tr \left[ V^{\textrm{q} [2]}_{{\un 2^\prime}} V_{{\un 1^\prime}}^\dagger  \right] \rrangle
 (s)
 \right] \, 
 \Big[
  1 - S_{1  2}(s)
 \Big]
 \Bigg\} .\notag
\end{align}

Continuing to the contribution of $V_{{\un x}, {\un y}}^{\textrm{G} [2]}$ to \eq{XS1} we derive (after summing over $\sigma'_1, \sigma'_2$ and $\sigma_2$ while employing $\Psi \sim \delta_{\sigma_1, - \sigma_2}$)
\begin{align}\label{XS10}
& z (1-z) \, \frac{1}{2} \! \sum_{S_L = \pm 1} \!\! S_L \, \frac{d\sigma_{\lambda \lambda'}^{\textrm{G} [2] \gamma^* p \to q {\bar q} p'}}{d^2 p_1 \, d^2 p_2 \, d z} =  \frac{1}{2 (2 \pi)^5} \, \int d^2 x_1 \, d^2 x_{1'} \, d^2 x_2 \, d^2 x_{2'} \, d^2 x_0 \, e^{- i {\un p}_1 \cdot {\un x}_{11'} - i {\un p}_2 \cdot {\un x}_{22'}} \, \sum_{\sigma_1, i , j}  
\\ 
& \times \frac{1}{N_c^2} \Bigg\{ \Psi_{\lambda', \sigma_1, -\sigma_1; i, i}^{\gamma^* \to q {\bar q}} ({\un x}_{02}, z) \, \left[ \Psi_{\lambda, \sigma_1, -\sigma_1; j, j}^{\gamma^* \to q {\bar q}} ({\un x}_{1'2'}, z) \right]^*  \left\langle \tord \tr \left[ V^{\textrm{G} [2]}_{{\un 1}, {\un 0}} \, V_{{\un 2}}^\dagger \right] \right\rangle \big( zs \big) \, \left\langle \atord \tr \left[ V_{{\un 2'}} \, V_{{\un 1'}}^\dagger - 1 \right]  \right\rangle (s) 
\notag \\
& -  \Psi_{\lambda', \sigma_1, -\sigma_1; i, i}^{\gamma^* \to q {\bar q}} ({\un x}_{10}, z) \, \left[ \Psi_{\lambda, \sigma_1, -\sigma_1; j, j}^{\gamma^* \to q {\bar q}} ({\un x}_{1'2'}, z) \right]^*  \left\langle \tord \tr \left[ V_{{\un 1}} \,  V^{\textrm{G} [2] \, \dagger}_{{\un 2}, {\un 0}} \right] \right\rangle \big((1-z)s \big)
\, \left\langle \atord \tr \left[ V_{{\un 2'}} \, V_{{\un 1'}}^\dagger - 1 \right]  \right\rangle(s)  \notag \\
& +  \Psi_{\lambda', \sigma_1, -\sigma_1; i, i}^{\gamma^* \to q {\bar q}} ({\un x}_{12}, z) \, \left[ \Psi_{\lambda, \sigma_1, -\sigma_1; j, j}^{\gamma^* \to q {\bar q}} ({\un x}_{02'}, z) \right]^*  \left\langle \tord \tr \left[   V_{{\un 1}} \, V_{{\un 2}}^\dagger - 1 \right]  \right\rangle \big( s \big)
\, \left\langle \atord \tr \left[ V_{{\un 2'}} \,  V^{\textrm{G} [2] \, \dagger}_{{\un 1'}, {\un 0}}  \right]  \right\rangle \big( zs \big) 
\notag \\
& -  \Psi_{\lambda', \sigma_1, -\sigma_1; i, i}^{\gamma^* \to q {\bar q}} ({\un x}_{12}, z) \, \left[ \Psi_{\lambda, \sigma_1, -\sigma_1; j, j}^{\gamma^* \to q {\bar q}} ({\un x}_{1'0}, z) \right]^* \left\langle \tord \tr \left[ V_{{\un 1}} \, V_{{\un 2}}^\dagger - 1 \right] \right\rangle (s) \, \left\langle \atord \tr \left[  V^{\textrm{G} [2]}_{{\un 2'}, {\un 0}} \, V_{{\un 1'}}^\dagger \right]  \right\rangle \big((1-z)s \big)   \Bigg\} .\notag
\end{align}

To simplify \eqs{XS7} and (\ref{XS10}) we will need the following wave function overlaps:
\begin{subequations} \label{type2_overlaps}
\begin{align}
\label{TT2} 
& \textrm{\bf TT \& T, -T:} \sum_{\sigma_1, i , j} \,  \Psi_{\lambda' =T, \sigma_1, - \sigma_1; i, i}^{\gamma^* \to q {\bar q}} ({\un x}_{12}, z) \, \left[ \Psi_{\lambda =T, \sigma_1, - \sigma_1; j, j}^{\gamma^* \to q {\bar q}} ({\un x}_{1'2'}, z) \right]^* 
\\ \notag & =  \frac{2 \alpha_{EM} \, Z_f^2 \, N_c^2}{\pi} \, z^2 \, (1-z)^2 \, [ (1-2z)^2 + \lambda \, \lambda' ] \, Q^2 \,
\frac{{\un \epsilon}_{\lambda'} \cdot {\un x}_{12} \, {\un \epsilon}^*_{\lambda} \cdot {\un x}_{1'2'}}{x_{12} \, x_{1'2'} } \, K_1 \left( x_{12} Q \sqrt{z (1-z)} \right) \, K_1 \left( x_{1'2'} Q \sqrt{z (1-z)} \right) \notag 
\\ \notag &
\equiv \delta_{\lambda, \lambda^\prime} \Big[ \lambda \, i \, \Phi^{[2]}_{\mathrm{TT}}(\un x_{12}, \un x_{1^\prime 2^\prime},z) + \Phi^{\prime [2]}_{\mathrm{TT}}(\un x_{12}, \un x_{1^\prime 2^\prime}, z) 
\Big] 
+ \delta_{\lambda, -\lambda^\prime} \, 
e^{i \lambda (\phi_{12} +\phi_{1'2'} )} \, 
\Phi^{[2]}_{\mathrm{T,-T}} (\un x_{12}, \un x_{1^\prime 2^\prime}, z), 
\\
& \textrm{\bf LT:}  \sum_{\sigma_1, i , j} \, \Psi_{\lambda' =T, \sigma_1, - \sigma_1; i, i}^{\gamma^* \to q {\bar q}} ({\un x}_{12}, z) \, \left[ \Psi_{\lambda =0, \sigma_1, - \sigma_1; j, j}^{\gamma^* \to q {\bar q}} ({\un x}_{1'2'}, z) \right]^* 
\\ & \notag =  - \frac{4 \, i \,  \alpha_{EM} \, Z_f^2 \, N_c^2}{\pi} \, [z (1-z)]^{5/2} \, (1-2z) \, Q^2 \, 
\frac{{\un \epsilon}_{\lambda'} \cdot {\un x}_{12}}{x_{12}} \, K_1 \left( x_{12} Q \sqrt{z (1-z)} \right) \, K_0 \left( x_{1'2'} Q \sqrt{z (1-z)} \right) \\
\notag 
& \equiv i \, \frac{\un \epsilon_{\lambda'}\cdot \un x_{12}}{x_{12}} \, \Phi^{[2]}_{\mathrm{LT}} (\un x_{12}, \un x_{1^\prime 2^\prime}, z) , \notag \\
& \textrm{\bf TL:} \sum_{\sigma_1, i , j} \, \Psi_{\lambda' =0, \sigma_1, - \sigma_1; i, i}^{\gamma^* \to q {\bar q}} ({\un x}_{12}, z) \, \left[ \Psi_{\lambda =T , \sigma_1, - \sigma_1; j, j}^{\gamma^* \to q {\bar q}} ({\un x}_{1'2'}, z) \right]^* = - i \, \frac{\un \epsilon_{\lambda}^* \cdot \un x_{1'2'}}{x_{1'2'}} \, \Phi^{[2]}_{\mathrm{LT}} (\un x_{1'2'}, \un x_{1 2}, z) , 
\\ \label{LL2}
& \textrm{\bf LL:} \sum_{\sigma_1, i , j} \, \Psi_{\lambda' =0, \sigma_1, - \sigma_1; i, i}^{\gamma^* \to q {\bar q}} ({\un x}_{12}, z) \, \left[ \Psi_{\lambda =0 , \sigma_1, - \sigma_1; j, j}^{\gamma^* \to q {\bar q}} ({\un x}_{1'2'}, z) \right]^* 
\\ \notag & = \frac{8 \,  \alpha_{EM} \, Z_f^2 \, N_c^2}{\pi} \, [z (1-z)]^{3} \, Q^2 \, 
K_0 \left( x_{12} Q \sqrt{z (1-z)} \right) \, K_0 \left( x_{1'2'} Q \sqrt{z (1-z)} \right)
\\ \notag 
& \equiv \Phi^{[2]}_{\mathrm{LL}}(\un x_{12}, \un x_{1^\prime 2^\prime}, z),
\end{align}
\end{subequations}
where $\phi_{12} (\phi_{1'2'})$ is the angle in the transverse plane associated with $\un x_{12} (\un x_{1'2'})$. The structure in \eq{TT2} is easier to understand if we notice that 
\begin{align}
{\un \epsilon}_\lambda \cdot {\un v} \ {\un \epsilon}^*_\lambda \cdot {\un w} = \frac{1}{2} \, {\un v} \cdot {\un w} + \frac{i}{2} \, \lambda \, {\un v} \times {\un w}
\end{align}
for ${\un \epsilon}_\lambda = - (1/\sqrt{2}) (-\lambda, i)$ and arbitrary transverse vectors $\un v$ and $\un w$. Similar to the above, the functions $\Phi^{[2]}_{\mathrm{TT}}, \Phi^{\prime [2]}_{\mathrm{TT}}, \Phi^{[2]}_{\mathrm{T,-T}}$, $\Phi^{[2]}_{\mathrm{LT}}$ and $\Phi^{[2]}_{\mathrm{LL}}$ are all real.

To continue, we must pick a specific structure from \eqs{observables} to focus on. Depending on the values of $\lambda$ and $\lambda^\prime$, we label the structures of \eqs{observables} accordingly. For example, when $\lambda = \lambda^\prime = \pm 1$, we call these ``TT" terms. By contrast, when $\lambda = 0, \lambda^\prime= \pm 1$, it is an ``LT" term. (We will not need to separately consider the ``TL" term with $\lambda = \pm 1, \lambda^\prime= 0$.) Finally, we denote terms with $\lambda = - \lambda' = \pm 1$ as ``T,-T" terms.

%%%%%%%%%%%%%%%%%%%%%%%%%%%%%%%%%%%%%%%%%%%%%%%%%%%%%%%%%%%%%%%%%%%%%%%%%%%%%%%
\section{Double spin asymmetry}
\label{sec:DSA}

Here we focus on the numerator of the DSA from \eq{DSA}. Specifically, we will calculate the contributions of \eqs{XS5}, (\ref{XS7}) and (\ref{XS10}) to the two (TT and LT) structures in \eq{DSA}.

%%%%%%%%%%%%%%%%%%%%%%%%%%%%%%%%

\subsection{TT terms}

The first structure we will consider is that entering the first term in \eq{DSA}, namely
\begin{align}
   z (1-z) \, \frac{1}{2} \! \sum_{S_L, \lambda = \pm 1} \!\! S_L \, \lambda   \, \frac{d\sigma_{\lambda \lambda}^{\gamma^* p \to q {\bar q} p'}}{d^2 p_1 \, d^2 p_2 \, d z}.
\end{align}
Let us start with the type-1 operators. Plugging $\lambda = \lambda^\prime$ into \eq{XS5}, and utilizing the wave function overlaps in \eqs{type1_overlaps}, we get 
\begin{align}\label{XS11}
 z (1-z)\, \frac{1}{2} \! \sum_{S_L, \lambda = \pm 1} \!\! S_L \, \lambda \ &  \frac{d\sigma_{\lambda \lambda}^{[1] \, \gamma^* p \to q {\bar q} p'}}{d^2 p_1 \, d^2 p_2 \, d z} = -\frac{1}{(2 \pi)^5 s} \, \int d^2 x_1 \, d^2 x_{1'} \, d^2 x_2 \, d^2 x_{2'} \,  e^{- i {\un p}_1 \cdot {\un x}_{11'} - i {\un p}_2 \cdot {\un x}_{22'}}  \Phi_{\mathrm{TT}}^{[1]}(\un x_{12}, \un x_{1^\prime 2^\prime}, z)
 \\ \notag 
 & \times 
 \frac{1}{N_c}  \Bigg\{ 
 \left[ \frac{1}{z} \llangle \tord \tr \left[ V^{\textrm{pol}[1]}_{{\un 1}} \, V_{{\un 2}}^\dagger \right] \rrangle
(s)
 -
 \frac{1}{1-z} \llangle \tord \tr \left[ V_{{\un 1}} \,  V^{\textrm{pol} [1] \, \dagger}_{{\un 2}} \right] \rrangle (s)
 \right] \, 
 \Big[
  1 - S^*_{ 1^\prime 2^\prime}(s)
 \Big]
 \\ \notag 
 & \hspace{1cm} +  
 \left[ \frac{1}{z} \llangle \atord \tr \left[ V_{\un 2^\prime} V^{\textrm{pol}[1]\dagger}_{{\un 1^\prime}} \right] \rrangle (s) 
 -
 \frac{1}{1-z}
\llangle \atord \tr \left[ V^{\textrm{pol} [1]}_{{\un 2^\prime}} V_{{\un 1^\prime}}^\dagger  \right] \rrangle
 (s)
 \right] \, 
 \Big[
  1 - S_{1  2}(s)
 \Big]
 \Bigg\} ,\notag
\end{align}
where we have performed the summation over $\lambda$. 

In the second term in the curly brackets of \eq{XS11}, we relabel $\un x_1 \leftrightarrow \un x_{1^\prime}$ as well as $\un x_2 \leftrightarrow \un x_{2^\prime}$. Simultaneously, we take $\un p_1 \to - \un p_1, \un p_2 \to -\un p_2$ in those terms to offset the change in sign of the exponential. The latter transformation is justified since each term in \eq{XS11} is a scalar function of two transverse vectors, $\un p_1$ and $\un p_2$. Finally, observing that $\Phi_{\mathrm{TT}}^{[1]}(\un x_{12}, \un x_{1^\prime 2^\prime}, z) = \Phi_{\mathrm{TT}}^{[1]}(\un x_{1^\prime 2^\prime}, \un x_{12}, z)$, we arrive at
\begin{align}\label{XS12}
 z (1-z) \, \frac{1}{2} \! \sum_{S_L, \lambda \pm 1} \!\! S_L \, \lambda \,  \frac{d\sigma_{\lambda \lambda}^{[1] \, \gamma^* p \to q {\bar q} p'}}{d^2 p_1 \, d^2 p_2 \, d z} &= -\frac{1}{(2 \pi)^5 s} \, \int d^2 x_1 \, d^2 x_{1'} \, d^2 x_2 \, d^2 x_{2'} \,  e^{- i {\un p}_1 \cdot {\un x}_{11'} - i {\un p}_2 \cdot {\un x}_{22'}}  \Phi_{\mathrm{TT}}^{[1]}(\un x_{12}, \un x_{1^\prime 2^\prime}, z)
 \\ \notag 
 & \hspace{-2cm}
 \times 
 \frac{1}{N_c}  \Bigg\{ 
 \left[ \frac{1}{z} \llangle \tord \tr \left[ V^{\textrm{pol}[1]}_{{\un 1}} \, V_{{\un 2}}^\dagger \right] \rrangle
(s)
 -
 \frac{1}{1-z} \llangle \tord \tr \left[ V_{{\un 1}} \,  V^{\textrm{pol} [1] \, \dagger}_{{\un 2}} \right] \rrangle (s)
 \right] \, 
 \Big[
  1 - S^*_{ 1^\prime 2^\prime}(s)
 \Big]
    + \mathrm{c.c.}
 \Bigg\}.\notag
\end{align}

Performing a similar procedure for \eq{XS7}, while now utilizing the wave function overlaps in \eqs{type2_overlaps}, we get, for the terms containing $V^{q[2]}_{\un x}$,
\begin{align}\label{XS13}
 z (1-z) \, \frac{1}{2} \!  \sum_{S_L, \lambda \pm 1} \!\! S_L \, \lambda \,  \frac{d\sigma_{\lambda \lambda}^{q[2] \, \gamma^* p \to q {\bar q} p'}}{d^2 p_1 \, d^2 p_2 \, d z} &= -\frac{i }{(2 \pi)^5 s} \, \int d^2 x_1 \, d^2 x_{1'} \, d^2 x_2 \, d^2 x_{2'} \,  e^{- i {\un p}_1 \cdot {\un x}_{11'} - i {\un p}_2 \cdot {\un x}_{22'}}  \Phi_{\mathrm{TT}}^{[2]}(\un x_{12}, \un x_{1^\prime 2^\prime}, z)
 \\ \notag 
 & \hspace{-2cm}
 \times 
 \frac{1}{N_c}  \Bigg\{ 
 \left[ \frac{1}{z} \llangle \tord \tr \left[ V^{\textrm{q}[2]}_{{\un 1}} \, V_{{\un 2}}^\dagger \right] \rrangle
(s)
 -
 \frac{1}{1-z} \llangle \tord \tr \left[ V_{{\un 1}} \,  V^{\textrm{q} [2] \, \dagger}_{{\un 2}} \right] \rrangle (s)
 \right] \, 
 \Big[
  1 - S^*_{ 1^\prime 2^\prime}(s)
 \Big]
    - \mathrm{c.c.}
 \Bigg\}.\notag
\end{align}
The difference now is that $\Phi_{\mathrm{TT}}^{[2]}(\un x_{12}, \un x_{1^\prime 2^\prime}, z) = - \Phi_{\mathrm{TT}}^{[2]}(\un x_{1^\prime 2^\prime}, \un x_{12}, z)$, i.e., the wave function overlap is antisymmetric under the interchange of the dipole sizes.

To proceed, let us consider how \eqs{XS12} and (\ref{XS13}) transform under passive parity and time-reversal (PT) transformations. Under PT, we have \cite{Cougoulic:2022gbk, Kovchegov:2023yzd} 
\begin{align} \label{PT_transform}
    \left\langle \tord \tr \left[ V_{{\un x}} \, V_{{\un y}}^\dagger \right] \right \rangle (s) \to \left \langle \atord \tr \left[ 
    V^{\dagger}_{-{\un x}} V_{-{\un y}} \right] \right \rangle (s),
\end{align}
where we can replace either of the two Wilson lines above by $V^{\textrm{pol}[1]}$, $V^{\textrm{q} [2]}$, or $V^{\textrm{G} [2]}$ without changing the PT-transformation property demonstrated in \eq{PT_transform}. Owing to the fact that we can always redefine the position variables ${\un x}_i \to - {\un x}_i$ in the integrals of \eqs{XS12} and (\ref{XS13}), the action of the PT symmetry in \eq{PT_transform} amounts to complex conjugation of the matrix element on the left to obtain the one on the right. Applying a PT transformation to \eq{XS12} and (\ref{XS13}), we see that the former is PT-even while the latter is PT-odd. Since the DSA is PT-even, we can thus discard \eq{XS13}.

Next, we turn our attention to the last remaining contribution, the terms containing $V^{\mathrm{G}[2]}_{\un x, \un y}$, summarized in \eq{XS10}. Putting $\lambda = \lambda^\prime$ in \eq{XS10}, and again utilizing the wave function overlaps from \eq{type2_overlaps}, we get 
\begin{align}\label{XS14}
z (1-z) \, \frac{1}{2} \! \sum_{S_L, \lambda =\pm1} \!\! S_L \, \lambda \ & \frac{d\sigma_{\lambda \lambda}^{\textrm{G} [2] \gamma^* p \to q {\bar q} p'}}{d^2 p_1 \, d^2 p_2 \, d z} =  -\frac{i}{(2 \pi)^5} \, \int d^2 x_1 \, d^2 x_{1'} \, d^2 x_2 \, d^2 x_{2'} \, d^2 x_0 \, e^{- i {\un p}_1 \cdot {\un x}_{11'} - i {\un p}_2 \cdot {\un x}_{22'}}  \\ 
& \times \frac{1}{N_c} \Bigg\{ 
\Phi^{[2]}_{\mathrm{TT}}(\un x_{02}, \un x_{1^\prime 2^\prime}, z) 
\left\langle \tord \tr \left[ V^{\textrm{G} [2]}_{{\un 1}, {\un 0}} \, V_{{\un 2}}^\dagger \right] \right\rangle \big( zs \big) \, 
\Big[ 1- S_{1^\prime 2^\prime}^*(s)\Big]
\notag \\
& -  
\Phi^{[2]}_{\mathrm{TT}}(\un x_{10}, \un x_{1^\prime 2^\prime}, z) 
\left\langle \tord \tr \left[ V_{{\un 1}} \,  V^{\textrm{G} [2] \, \dagger}_{{\un 2}, {\un 0}} \right] \right\rangle \big((1-z)s \big)
\, \Big[ 1- S_{1^\prime 2^\prime}^*(s) \Big]
\notag \\
& +  \Phi^{[2]}_{\mathrm{TT}}(\un x_{12}, \un x_{0 2^\prime}, z) 
\Big[ 1- S_{12}(s)\Big]
\, \left\langle \atord \tr \left[ V_{{\un 2'}} \,  V^{\textrm{G} [2] \, \dagger}_{{\un 1'}, {\un 0}}  \right]  \right\rangle \big( zs \big) \notag \\
& - \Phi^{[2]}_{\mathrm{TT}}(\un x_{12}, \un x_{1^\prime 0}, z) 
\Big[ 1- S_{12}(s)\Big]
\, \left\langle \atord \tr \left[  V^{\textrm{G} [2]}_{{\un 2'}, {\un 0}} \, V_{{\un 1'}}^\dagger \right]  \right\rangle \big((1-z)s \big)   \Bigg\},\notag
\end{align}
where we have used \eq{unpol-dip-amp} and performed the sum over $\lambda$. In the last two terms in the  curly brackets, we relabel $\un x_{1} \leftrightarrow \un x_{1^\prime}$ and $\un x_{2} \leftrightarrow \un x_{2^\prime}.$ Again we take $\un p_1 \to -\un p_1$ and $\un p_2 \to- \un p_2$ to offset the change in the exponential for those two terms. Then, employing $\Phi^{[2]}_{\mathrm{TT}}(\un x, \un y, z) = -\Phi^{[2]}_{\mathrm{TT}}(\un y, \un x, z)$ and the fact that $\Phi^{[2]}_{\mathrm{TT}}$ is real, we write \eq{XS14} as 
\begin{align}\label{XS15}
 z (1-z) \, \frac{1}{2} \! \sum_{S_L, \lambda =\pm1} \!\! S_L \, \lambda  \frac{d\sigma_{\lambda \lambda}^{\textrm{G} [2] \gamma^* p \to q {\bar q} p'}}{d^2 p_1 \, d^2 p_2 \, d z} &=  -\frac{i}{(2 \pi)^5} \, \int d^2 x_1 \, d^2 x_{1'} \, d^2 x_2 \, d^2 x_{2'} \, d^2 x_0 \, e^{- i {\un p}_1 \cdot {\un x}_{11'} - i {\un p}_2 \cdot {\un x}_{22'}}  \\ 
& \times \frac{1}{N_c} \Bigg\{ 
\Phi^{[2]}_{\mathrm{TT}}(\un x_{02}, \un x_{1^\prime 2^\prime}, z) 
\left[ \left\langle \tord \tr \left[ V^{\textrm{G} [2]}_{{\un 1}, {\un 0}} \, V_{{\un 2}}^\dagger \right] \right\rangle \big( zs \big) 
\ \Big[ 1- S_{1^\prime 2^\prime}^*(s)\Big] - \mathrm{c.c.} \right]
\notag \\
& -  
\Phi^{[2]}_{\mathrm{TT}}(\un x_{10}, \un x_{1^\prime 2^\prime}, z) 
\left[ \left\langle \tord \tr \left[ V_{{\un 1}} \,  V^{\textrm{G} [2] \, \dagger}_{{\un 2}, {\un 0}} \right] \right\rangle \big((1-z)s \big)
\ \Big[ 1- S_{1^\prime 2^\prime}^*(s) \Big]
- \mathrm{c.c.} \right] \Bigg\}. \notag
\end{align}
To further simplify this expression, we can employ the procedure from Appendix~A of \cite{Kovchegov:2023yzd}, which starts by writing
\begin{subequations}\label{VG2sub_alt}
 \begin{align}
 & \int d^2 x_1 \, d^2 x_0 \, f ({\un x}_1) \, g ({\un x}_0) \, V^{\textrm{G} [2]}_{{\un 1}, {\un 0}} = - i \, \int d^2 x_1 \, \left[ g ({\un x}_1) \, \partial^i_{\un 1} f ({\un x}_1) - f ({\un x}_1) \,  \partial^i_{\un 1} g ({\un x}_1) \right] \, V_{{\un 1}}^{i \, \textrm{G} [2]} 
 \\ \notag & \hspace{5cm}
 - \frac{i P^+}{s} \int d^2 x_1 \, L \left[g( \un x_1) \partial_{\un 1}^2 f(\un x_1)+ f(\un x_1)  \partial_{\un 1}^2 g(\un x_1) \right] V_{\un 1}
  \\ \notag & \hspace{5cm}
   - \frac{i P^+}{s} \int d^2 x_1  \int \displaylimits^\infty_{-\infty} dz^-  f(\un x_1) g(\un x_1) \left[ V_{\un 1}[\infty, z^-] \cev{D}^i(z^-, \un x_1) D^i(z^-, \un x_1) V_{\un 1}[z^-, -\infty] \right] ,
 \\
 & \int d^2 x_1 \, d^2 x_0 \, f ({\un x}_1) \, g ({\un x}_0) \, \left( V^{\textrm{G} [2]}_{{\un 1}, {\un 0}} \right)^\dagger =  i \, \int d^2 x_1 \, \left[ g ({\un x}_1) \, \partial^i_{\un 1} f ({\un x}_1) - f ({\un x}_1) \,  \partial^i_{\un 1} g ({\un x}_1) \right] \, \, \left(  V_{{\un 1}}^{i \, \textrm{G} [2]}  \right)^\dagger 
  \\ \notag & \hspace{5cm}
 + \frac{i P^+}{s} \int d^2 x_1 \, L \left[g( \un x_1) \partial_{\un 1}^2 f(\un x_1)+ f(\un x_1)  \partial_{\un 1}^2 g(\un x_1) \right] V_{\un 1}^\dagger
  \\ \notag & \hspace{5cm}
   + \frac{i P^+}{s} \int d^2 x_1  \int \displaylimits^\infty_{-\infty} dz^-  f(\un x_1) g(\un x_1) \left[ V_{\un 1}[-\infty, z^-] \cev{D}^i(z^-, \un x_1) D^i(z^-, \un x_1) V_{\un 1}[z^-, \infty] \right], 
 \end{align}
\end{subequations}
with $L = \int_{-\infty}^\infty d x^-$ and then uses PT-symmetry arguments to drop the second and third terms on the right of each expression, distilling them to 
 \begin{subequations}\label{VG2sub}
 \begin{align}
 & \int d^2 x_1 \, d^2 x_0 \, f ({\un x}_1) \, g ({\un x}_0) \, V^{\textrm{G} [2]}_{{\un 1}, {\un 0}} \to - i \, \int d^2 x_1 \, \left[ g ({\un x}_1) \, \partial^i_{\un 1} f ({\un x}_1) - f ({\un x}_1) \,  \partial^i_{\un 1} g ({\un x}_1) \right] \, V_{{\un 1}}^{i \, \textrm{G} [2]} , \\
 & \int d^2 x_1 \, d^2 x_0 \, f ({\un x}_1) \, g ({\un x}_0) \, \left( V^{\textrm{G} [2]}_{{\un 1}, {\un 0}} \right)^\dagger \to  i \, \int d^2 x_1 \, \left[ g ({\un x}_1) \, \partial^i_{\un 1} f ({\un x}_1) - f ({\un x}_1) \,  \partial^i_{\un 1} g ({\un x}_1) \right] \, \, \left(  V_{{\un 1}}^{i \, \textrm{G} [2]}  \right)^\dagger ,
 \end{align}
 \end{subequations}
with $V^{i \mathrm{G[2]}}_{\un x}$ defined above in \eq{d-d_pw}. Similar to \cite{Cougoulic:2022gbk, Kovchegov:2023yzd}, by using the passive PT-symmetry argument one can show that the terms neglected on the right-hand side of \eqs{VG2sub} are zero for the PT-even observable (DSA) in question. 
Using \eqs{VG2sub} in \eq{XS15}, we get 
\begin{align}\label{XS16}
 z (1-z) \, \frac{1}{2} \! \sum_{S_L, \lambda =\pm1} \!\! S_L \, \lambda  &\frac{d\sigma_{\lambda \lambda}^{\textrm{G} [2] \gamma^* p \to q {\bar q} p'}}{d^2 p_1 \, d^2 p_2 \, d z} =  
 \frac{1}{(2 \pi)^5 s} \, \int d^2 x_1 \, d^2 x_{1'} \, d^2 x_2 \, d^2 x_{2'} \, e^{- i {\un p}_1 \cdot {\un x}_{11'} - i {\un p}_2 \cdot {\un x}_{22'}}  \\ 
& \times \frac{1}{N_c} \Bigg\{ 
\left[
 \frac{1}{z} 
\llangle \tord \tr \left[ V^{i\textrm{G} [2]}_{{\un 1}} \, V_{{\un 2}}^\dagger \right] \rrangle (s)
\Big( 1- S_{1^\prime 2^\prime}^*(s)\Big) + \mathrm{c.c.} \right]
\Big( \partial^i_{\un 1} - i p_1^i \Big)
\notag \\
& +  
\left[
\frac{1}{1-z}
\llangle \tord \tr \left[ V_{{\un 1}} \,  V^{i\textrm{G} [2] \, \dagger}_{{\un 2}} \right] \rrangle (s)
\, \Big( 1- S_{1^\prime 2^\prime}^*(s) \Big) + \mathrm{c.c.}
\right]
\Big( \partial^i_{\un 2} - i p_2^i \Big)
 \Bigg\} \, \Phi^{[2]}_{\mathrm{TT}}(\un x_{12}, \un x_{1^\prime 2^\prime}, z).\notag
\end{align}
We have switched to the double angle bracket notation and used the fact that, in the kinematics we consider $z \, s  \sim (1-z) \, s \sim s$, to simplify the arguments of the double-angle bracket correlators to $s$ in the DLA. Adding \eq{XS16} to \eq{XS12}, we arrive at the total TT contribution to the numerator of the DSA, 
\begin{align}\label{XS17}
 z (1-z) \, \frac{1}{2} \! \sum_{S_L, \lambda =\pm1} \!\! S_L \, \lambda  \, \frac{d\sigma_{\lambda \lambda}^{\gamma^* p \to q {\bar q} p'}}{d^2 p_1 \, d^2 p_2 \, d z} &= -\frac{1}{(2 \pi)^5 s} \, \int d^2 x_1 \, d^2 x_{1'} \, d^2 x_2 \, d^2 x_{2'} \,  e^{- i {\un p}_1 \cdot {\un x}_{11'} - i {\un p}_2 \cdot {\un x}_{22'}}  
 \\ \notag 
 & \hspace{-4cm}
 \times 
 \frac{1}{N_c}  \Bigg\{ 
 \left[ \frac{1}{z} \llangle \tord \tr \left[ V^{\textrm{pol}[1]}_{{\un 1}} \, V_{{\un 2}}^\dagger \right] \rrangle
(s)
 -
 \frac{1}{1-z} \llangle \tord \tr \left[ V_{{\un 1}} \,  V^{\textrm{pol} [1] \, \dagger}_{{\un 2}} \right] \rrangle (s)
 \right] \, 
 \Big[
  1 - S^*_{ 1^\prime 2^\prime}(s)
 \Big] \Phi_{\mathrm{TT}}^{[1]}(\un x_{12}, \un x_{1^\prime 2^\prime}, z)
    + \mathrm{c.c.}
 \\ \notag & \hspace{-2cm}
 -
\left[
 \frac{1}{z} 
\llangle \tord \tr \left[ V^{i\textrm{G} [2]}_{{\un 1}} \, V_{{\un 2}}^\dagger \right] \rrangle (s)
\Big( 1- S_{1^\prime 2^\prime}^*(s)\Big) + \mathrm{c.c.} \right]
\Big( \partial^i_{\un 1} - i p_1^i \Big) \Phi^{[2]}_{\mathrm{TT}}(\un x_{12}, \un x_{1^\prime 2^\prime}, z)
\notag \\ & \hspace{-2cm}
 -
\left[
\frac{1}{1-z}
\llangle \tord \tr \left[ V_{{\un 1}} \,  V^{i\textrm{G} [2] \, \dagger}_{{\un 2}} \right] \rrangle (s)
\, \Big( 1- S_{1^\prime 2^\prime}^*(s) \Big) + \mathrm{c.c.}
\right]
\Big( \partial^i_{\un 2} - i p_2^i \Big) \Phi^{[2]}_{\mathrm{TT}}(\un x_{12}, \un x_{1^\prime 2^\prime}, z)
 \Bigg\} .\notag
\end{align}
In the dipole amplitudes entering the expressions for the helicity PDFs and OAM distributions, defined in \eqs{pdas},  the polarized Wilson lines always appear at the same transverse coordinate in both traces. Comparing those definitions with \eq{XS17}, we see that it is not the case in the latter. For example, in the second line of \eq{XS17}, the first term has the polarized Wilson line at $\un x_{1}$, while in the second term it is at $\un x_{2}$. To make contact with the definitions in \eqs{pdas}, we will ``symmetrize" the cross section between the jets, assuming that the quark and antiquark jets are indistinguishable experimentally. That is, we write 
\begin{align} \label{symmetric_XS}
 \frac{d\sigma_{\textrm{symm} \, \lambda \lambda'}^{\gamma^* p \to q {\bar q} p'}}{d^2 p_1 \, d^2 p_2 \, d z} \equiv \frac{1}{2} \left[ \frac{d\sigma_{\lambda \lambda'}^{\gamma^* p \to q {\bar q} p'}}{d^2 p_1 \, d^2 p_2 \, d z} + \frac{d\sigma_{\lambda \lambda'}^{\gamma^* p \to q {\bar q} p'}}{d^2 p_2 \, d^2 p_1 \, d (1-z)} \right] .
 \end{align}

 To compute the right hand side of \eq{symmetric_XS} for the TT terms in the DSA numerator at hand, let us define the non-singlet imaginary-part counterparts to the definitions in \eqs{pdas}, 
 \begin{subequations} \label{pdas_ns}
    \begin{align}
        Q_{10}^{\textrm{NS,Im}} (s) &= \frac{1}{2 \, N_c} \, \mbox{Im} \, \llangle - \mbox{T} \, \tr \left[ V_{\un 0} \,  V_{\un 1}^{\textrm{pol} [1] \,\dagger} \right] + \mbox{T} \,  \tr \left[ V_{\un 1}^{\textrm{pol} [1]} \, V_{\un 0}^\dagger \right]   \rrangle (s), 
        \\ 
        G_{10}^{i\, \textrm{NS,Im}} (s) &= \frac{1}{2 \, N_c} \, \mbox{Im} \, \llangle - \mbox{T} \, \tr \left[ V_{\un 0} \,  V_{\un 1}^{i\textrm{G} [2] \,\dagger} \right] + \mbox{T} \,  \tr \left[ V_{\un 1}^{i\textrm{G} [2]} \, V_{\un 0}^\dagger \right]   \rrangle (s).
    \end{align}
\end{subequations}
Furthermore, we may write \cite{Hatta:2005as,Kovchegov:2003dm}
\begin{align} \label{unpdas}
    1- S_{10}(s) &= N_{10}(s) - i O_{10}(s),
\end{align}
where 
\begin{subequations}
\begin{align}
& N_{10} (s) = 1 - \frac{1}{2 N_c} \, \left\langle \tord \tr \left[ V_{\un 1} \, V^\dagger_{\un 0} \right] + \tord \tr \left[ V_{\un 0} \, V^\dagger_{\un 1} \right] \right\rangle, \\
& O_{10} (s) = \frac{1}{2 N_c \, i} \, \left\langle \tord \tr \left[ V_{\un 1} \, V^\dagger_{\un 0} \right] - \tord \tr \left[ V_{\un 0} \, V^\dagger_{\un 1} \right] \right\rangle .
\end{align}
\end{subequations}
The dipole amplitude $N_{10}(s)$ is $\mathcal{C}$-even, $1 \leftrightarrow 0$ symmetric, Pomeron amplitude, and $O_{10}$ is $\mathcal{C}$-odd, $1 \leftrightarrow 0$ antisymmetric, odderon amplitude. These amplitudes have been extensively studied over the years, staring with \cite{Mueller:1994rr,Mueller:1994jq,Mueller:1995gb,Balitsky:1995ub,Balitsky:1998ya,Kovchegov:1999yj,Kovchegov:1999ua,Jalilian-Marian:1997dw,Jalilian-Marian:1997gr,Weigert:2000gi,Iancu:2001ad,Iancu:2000hn,Ferreiro:2001qy} for the Pomeron and  \cite{Hatta:2005as,Kovchegov:2003dm} for the odderon in the single-logarithmic approximation (SLA), i.e., resumming powers of $\as \, \ln (1/x)$, and beyond. At the leading order in $\as$ and in SLA these amplitudes are real.

Note that the imaginary part of $N_{10}$ exists and is suppressed only by one power of $\as$ compared to the real part (see e.g. \cite{Gotsman:2020mkd}), i.e., it is of the order of the odderon amplitude: we will neglect Im~$N_{10}$ in the DSA calculation we are focused on here, since it will be a higher-order in $\as$ correction to the terms we keep. In the next Section, when we consider the numerator of the SSA, we will see that Im~$N_{10}$ is part of the leading-order (in $\as$) expression, and therefore we cannot discard it. Similarly, the imaginary part of $O_{10}$ may exist; however it is suppressed by at least two powers of $\as$ compared to the real part of $N_{10}$ in the SLA power counting. Therefore, throughout this paper we will neglect $\mathrm{Im}\,O_{10}$.

To find the right hand side of \eq{symmetric_XS} for the TT terms, we use \eq{XS17} and employ the definitions in \eqs{pdas}, (\ref{pdas_ns}), and (\ref{unpdas}), while keeping only the real parts of $N_{10}$ and $O_{10}$. We also observe that the ${\un p}_1 \leftrightarrow {\un p}_2$ interchange in the second term on the right of \eq{symmetric_XS} can be implemented by switching $\un x_1 \leftrightarrow \un x_2$ and $\un x_{1'} \leftrightarrow \un x_{2'}$ in the integrand of \eq{XS17}. In addition, we employ the fact that $\Phi_{\mathrm{TT}}^{[1]}(\un x_{12}, \un x_{1^\prime 2^\prime}, z)$ is odd and $\Phi_{\mathrm{TT}}^{[2]}(\un x_{12}, \un x_{1^\prime 2^\prime}, z)$ is even under the $z \leftrightarrow 1-z$ interchange (see \eqs{PhiTT} below). Combining all these observations, we get 
\begin{align}\label{DSA_TT}
 z (1-z) \, \frac{1}{2} \! \sum_{S_L, \lambda =\pm1} \!\! S_L \, \lambda \, \frac{d\sigma_{\mathrm{symm.}\, \lambda \lambda}^{\gamma^* p \to q {\bar q} p'}}{d^2 p \, d^2 \Delta \, d z} &= -\frac{2}{(2 \pi)^5 s}  \, \int d^2 x_1 \, d^2 x_{1'} \, d^2 x_2 \, d^2 x_{2'} \,  e^{- i {\un p} \cdot ({\un x}_{12} - \un x_{1'2'}) - i \un \Delta \cdot \left[ z \, \un x_{11'} + (1-z) \un x_{22'}  \right]}  
 \\ \notag 
 & \hspace{-5cm}
 \times 
 \Bigg\{ 
 \left[ \left( \frac{1}{z} Q_{12}(s) - \frac{1}{1-z} Q_{21}(s) \right) N_{1^\prime 2^\prime}(s)
 -  \left( \frac{1}{z} Q^{\mathrm{NS,Im}}_{12}(s) + \frac{1}{1-z} Q^{\mathrm{NS,Im}}_{21}(s)
 \right)O_{1^\prime 2^\prime}(s)  \right] 
 \Phi_{\mathrm{TT}}^{[1]}(\un x_{12}, \un x_{1^\prime 2^\prime}, z)
 \\ \notag 
 & \hspace{-5cm}
- 
\left[
 \left( \frac{1}{z} G^i_{12}(s) - \frac{1}{1-z} G^i_{21}(s) \right) N_{1^\prime 2^\prime}(s)
 -  \left( \frac{1}{z} G^{i\, \mathrm{NS,Im}}_{12}(s) + \frac{1}{1-z} G^{i\,\mathrm{NS,Im}}_{21}(s)
 \right)O_{1^\prime 2^\prime}(s) 
\right]
\Big( \partial^i_{\un 1} - i p^i \Big) \Phi^{[2]}_{\mathrm{TT}}(\un x_{12}, \un x_{1^\prime 2^\prime}, z)
\notag \\
& \hspace{-5cm}
+ 
 \left[ \big(  G^i_{12}(s) +  G^i_{21}(s) \big) N_{1^\prime 2^\prime}(s)
 -  \big(  G^{i\,\mathrm{NS,Im}}_{12}(s) -  G^{i\,\mathrm{NS,Im}}_{21}(s)
 \big)O_{1^\prime 2^\prime}(s)  \right] 
\big(i \Delta^i \big) \Phi^{[2]}_{\mathrm{TT}}(\un x_{12}, \un x_{1^\prime 2^\prime}, z)
 \Bigg\} .\notag
\end{align}
In arriving at \eq{DSA_TT}, we have defined 
\begin{subequations} \label{p-d-vars}
\begin{align} \label{p-def}
    \un p &= (1-z) \, \un p_{1} - z \, \un p_{2}, \\
    \un \Delta &= \un p_1 + \un p_2 = -\un P^\prime. 
\end{align}
\end{subequations}
The evolution of the new dipole amplitude $Q_{10}^{\textrm{NS,Im}} (s)$, while not studied in the literature, should be the same as for the real part of the same operator, derived in the DLA and at large $N_c$ in \cite{Kovchegov:2016zex}. However, the initial conditions for the evolution, if nonzero, are at least order-$\as$ suppressed compared to the initial conditions for $Q_{10}$, since an additional $t$-channel gluon exchange is needed to generate the imaginary part (cf. \cite{Brodsky:2002cx, Brodsky:2002rv, Brodsky:2013oya,  Kovchegov:2020kxg, Kovchegov:2021iyc, Kovchegov:2022kyy}). Similarly, the odderon, $O_{10}$, is order-$\as$ suppressed compared to $N_{10}$, due to a minimum 3-gluon exchange in its initial conditions (see \cite{Kwiecinski:1980wb, Bartels:1999yt, Ewerz:2003xi, Hatta:2005as,Kovchegov:2003dm} and references therein). Hence, the second term in the first square brackets of \eq{DSA_TT} is at least order-$\as^2$ suppressed compared to the first term in those brackets: therefore, we will discard it. Analogously, one can argue that the dipole amplitude $G^{i\,\mathrm{NS,Im}}_{10} (s)$ is at least order-$\as$ suppressed compared to $G^i_{10}(s)$: combining this observation with the suppression of the odderon amplitude mentioned above, allows us to neglect the terms containing $G^{i\,\mathrm{NS,Im}}_{10} (s)$ in the second and third square brackets of \eq{DSA_TT}. In doing so, we arrive at our final result for the TT terms in the numerator of the DSA, 
 \begin{align}\label{DSA_TT_final}
  z (1-z) & \ \frac{1}{2} \! \sum_{S_L, \lambda =\pm1} \!\! S_L \, \lambda \, \frac{d\sigma_{\mathrm{symm.}\, \lambda \lambda}^{\gamma^* p \to q {\bar q} p'}}{d^2 p \, d^2 \Delta \, d z} = -\frac{2}{(2 \pi)^5 s} \, \int d^2 x_1 \, d^2 x_{1'} \, d^2 x_2 \, d^2 x_{2'} \, 
  e^{- i {\un p} \cdot ({\un x}_{12} - \un x_{1'2'}) - i \un \Delta \cdot \left[ z \, \un x_{11'} + (1-z) \un x_{22'}  \right] }  
 \\ \notag 
 & 
 \times 
N_{1^\prime 2^\prime}(s)
 \Bigg\{ 
 \left[ \frac{1}{z} Q_{12}(s) - \frac{1}{1-z} Q_{21}(s)  
 \right] 
 \Phi_{\mathrm{TT}}^{[1]}(\un x_{12}, \un x_{1^\prime 2^\prime}, z)
+ 
 \Big[  G^i_{12}(s) +  G^i_{21}(s) \Big] 
\big(i \Delta^i \big) \Phi^{[2]}_{\mathrm{TT}}(\un x_{12}, \un x_{1^\prime 2^\prime}, z)
\notag \\ & \hspace{3cm} 
- 
\left[
 \frac{1}{z} G^i_{12}(s) - \frac{1}{1-z} G^i_{21}(s) 
\right]
\Big( \partial^i_{\un 1} - i p^i \Big) \Phi^{[2]}_{\mathrm{TT}}(\un x_{12}, \un x_{1^\prime 2^\prime}, z)
 \Bigg\}, \notag
\end{align}
where the wave function overlaps defined in \eqs{type1_overlaps} and (\ref{type2_overlaps}) are
\begin{subequations}\label{PhiTT}
    \begin{align}
        \Phi^{[1]}_{\mathrm{TT}}(\un x_{12}, \un x_{1'2'}, z) &= \frac{2 \alpha_{EM} Z_f^2 N_c^2 Q^2}{\pi} z^2 (1-z)^2 (1-2\,z) \, \frac{\un x_{12} \cdot \un x_{1'2'}}{ x_{12} x_{1'2'}} K_1 \left(x_{12} Q \sqrt{z (1-z)} \right)
         K_1 \left(x_{1'2'} Q \sqrt{z (1-z)} \right),
         \\ 
         \Phi^{[2]}_{\mathrm{TT}}(\un x_{12}, \un x_{1'2'}, z) &= \frac{2 \alpha_{EM} Z_f^2 N_c^2 Q^2}{\pi} z^2 (1-z)^2 \Big[ z^2 + (1-z)^2 \Big]\, \frac{\un x_{12} \times \un x_{1'2'}}{ x_{12} x_{1'2'}} K_1 \left(x_{12} Q \sqrt{z (1-z)} \right)
         K_1 \left(x_{1'2'} Q \sqrt{z (1-z)} \right).
    \end{align}
\end{subequations}

%%%%%%%%%%%%%%%%%%%%%%%%%%%%%%%%%%%%%%%%%%%%%%%%%%%

\subsection{LT terms}

Now let us consider the terms in \eq{DSA} where the virtual photon has different polarizations, one transverse and one longitudinal, in the amplitude and complex conjugate amplitude. Specifically, we will compute 
\begin{align}
    z(1-z) \, \frac{1}{2} \sum_{S_L, \lambda = \pm 1} S_L \,
            \Big[ 
            e^{i \lambda \phi} \frac{d\sigma^{\gamma^*p}_{0 \lambda}}{d^2 p_1 d^2 p_2 dz} + \mathrm{c.c.} 
     \Big] .
\end{align}

Starting again with the type-1 terms, we use \eq{XS5}, and again utilize the wave function overlaps from \eqs{type1_overlaps}. The result is 
\begin{align}\label{DSA_LT11}
& z (1-z) \, \frac{1}{2}  \sum_{S_L, \lambda = \pm 1} S_L \left[ e^{i \lambda \phi} \, \frac{d\sigma_{0\lambda }^{[1] \, \gamma^* p \to q {\bar q} p'}}{d^2 p_1 \, d^2 p_2 \, d z} + \mathrm{c.c.} \right] =  -\frac{i\sqrt{2}}{2 (2 \pi)^5 \, s} \, \int d^2 x_1 \, d^2 x_{1'} \, d^2 x_2 \, d^2 x_{2'} \,  e^{- i {\un p}_1 \cdot {\un x}_{11'} - i {\un p}_2 \cdot {\un x}_{22'}} 
\\ 
& \times  \Phi_{\mathrm{LT}}^{[1]}(\un x_{12}, \un x_{1^\prime 2^\prime}, z) \frac{{\hat k} \cdot {\un x}_{12}}{x_{12}}  \frac{1}{N_c} \, \Bigg\{  \left[  \frac{1}{z} \, \llangle \tord \tr \left[ V^{\textrm{pol}[1]}_{{\un 1}} \, V_{{\un 2}}^\dagger \right] \rrangle (s) -  \frac{1}{1-z} \,  \llangle \tord \tr \left[ V_{{\un 1}} \,  V^{\textrm{pol} [1] \, \dagger}_{{\un 2}} \right]  \rrangle (s) \right] \, \left[ 1 - S_{1'2'}^* (s) \right] 
\notag \\
&  + \left[ \frac{1}{z} \, \llangle \atord \tr \left[ V_{{\un 2'}} \,  V^{\textrm{pol} [1] \, \dagger}_{{\un 1'}}  \right]  \rrangle (s) - \frac{1}{1-z} \, \llangle \atord \tr \left[  V^{\textrm{pol} [1]}_{{\un 2'}} \, V_{{\un 1'}}^\dagger \right]  \rrangle (s) \right] \, \left[ 1 - S_{12} (s) \right]  \Bigg\} + \cc, \notag
\end{align}
where again we have summed over $\lambda$. Here $\hat{k}$ is a two-dimensional unit vector in the direction of the transverse momentum of the incoming electron, $\un{k}$. 

Symmetrizing \eq{DSA_LT11} under the quark--antiquark jet interchange using \eq{symmetric_XS}, while remembering that changing ${\un p}_1 \to - {\un p}_1$ and ${\un p}_2 \to - {\un p}_2$ in some of the terms has to be accompanied by ${\hat k} \to - {\hat k}$ in the same terms to leave the cross section invariant (due to absence of other transverse vectors in the cross section at hand), we obtain
\begin{align} \label{DSA_LT1res}
    &z (1-z)  \, \frac{1}{2}  \sum_{S_L, \lambda = \pm 1} S_L \left[ e^{i \lambda \phi} \, \frac{d\sigma_{\mathrm{symm.}\,0\lambda}^{[1] \, \gamma^* p \to q {\bar q} p'}}{d^2 p \, d^2 \Delta \, d z} + \mathrm{c.c.} \right]  
    \\ \notag 
   & = -\frac{i\sqrt{2}}{(2\pi)^5s}
        \int d^2 x_1 \, d^2 x_{1'} \, d^2 x_2 \, d^2 x_{2'} \, 
          e^{- i {\un p} \cdot ({\un x}_{12} - \un x_{1'2'}) - i \un \Delta \cdot \left[ z \, \un x_{11'} + (1-z) \un x_{22'}  \right]} 
          \\ \notag & \times 
         \left[ \frac{\hat{k} \cdot \un x_{1 2}}{x_{12}}
        \Phi^{[1]}_{\mathrm{LT}}(\un x_{12}, \un x_{1^\prime 2^\prime} , z) 
         - \frac{\hat{k} \cdot \un x_{1^\prime 2^\prime}}{x_{1^\prime 2^\prime}} \Phi^{[1]}_{\mathrm{LT}}(\un x_{1^\prime 2^\prime}, \un x_{12}, z) 
        \right] \notag \\ 
    & \times \,  \Bigg\{  \left[ \frac{1}{z} \, Q_{12} (s) - \frac{1}{1-z} \, Q_{21} (s) \right] \, N_{1'2'} (s)  - \left[ \frac{1}{z} \, Q^{\textrm{NS}, Im}_{12} (s) + \frac{1}{1-z} \, Q^{\textrm{NS}, Im}_{21} (s) \right] \, O_{1'2'} (s) \Bigg\} . \notag
\end{align}

Repeating the above steps for \eq{XS7} and using \eq{PT_transform}, we see terms containing $V^{q[2]}_{\un x}$ are PT odd and therefore do not contribute to the DSA. Finally, considering the terms containing the $V_{\un x, \un y}^{\mathrm{G}[2]}$ operator, given in \eq{XS10}, we proceed analogously to the TT terms case, using \eqs{VG2sub}. Again, symmetrizing between the quark and antiquark jets, and using the definitions of various dipole amplitudes introduced above, we arrive at 
\begin{align} \label{DSA_LT2res}
   &z (1-z) \, \frac{1}{2}  \sum_{S_L, \lambda = \pm 1} S_L \left[ e^{i \lambda \phi} \, \frac{d\sigma_{\mathrm{symm.}\,0\lambda}^{\mathrm{G} [2] \, \gamma^* p \to q {\bar q} p'}}{d^2 p \, d^2 \Delta \, d z} + \mathrm{c.c.} \right]
   \\ \notag &
   = \frac{i\sqrt{2}}{(2 \pi)^5 s} \, \int d^2 x_1 \, d^2 x_{1'} \, d^2 x_2 \, d^2 x_{2'} \,  
     e^{- i {\un p} \cdot ({\un x}_{12} - \un x_{1'2'}) - i \un \Delta \cdot \left[ z \, \un x_{11'} + (1-z) \un x_{22'}  \right]} 
 \\ \notag 
 & 
 \times 
\Bigg\{
\Bigg[  \left[ \frac{1}{z} G^i_{12}(s) - \frac{1}{1-z} G^i_{21}(s) \right] N_{1^\prime 2^\prime}(s) - \left[ \frac{1}{z} G^{i\,\mathrm{NS,Im}}_{12}(s) + \frac{1}{1-z} G^{i\,\mathrm{NS,Im}}_{21}(s) \right] O_{1^\prime 2^\prime}(s) \Bigg]
 \Big( \partial^i_{\un 1} - i p^i \Big) 
\notag \\&
-  
 i \Delta^i \, \Bigg[  \left[  G^i_{12}(s) +  G^i_{21}(s)  \right] \, N_{1^\prime 2^\prime}(s) -  \left[  G^{i\,\mathrm{NS,Im}}_{12}(s) -  G^{i\,\mathrm{NS,Im}}_{21}(s)  \right]  \, O_{1^\prime 2^\prime}(s) \Bigg]
 \Bigg\} \notag \\
 & \times  \left[ \frac{\hat{k} \times \un x_{12}}{x_{12}} \Phi^{[2]}_{\mathrm{LT}}(\un x_{12}, \un x_{1^\prime 2^\prime}, z) 
+ 
\frac{\hat{k} \times \un x_{1^\prime 2^\prime}}{x_{1^\prime 2^\prime}} \Phi^{[2]}_{\mathrm{LT}}(\un x_{1^\prime 2^\prime}, \un x_{1 2}, z)
\right] . \notag
\end{align}
Adding \eqs{DSA_LT2res} and \eqref{DSA_LT1res} yields
\begin{align} \label{DSA_LT_pre_final}
  & z (1-z) \, \frac{1}{2} \sum_{S_L, \lambda = \pm 1} S_L \left[ e^{ i \lambda \phi} \, \frac{d\sigma_{\mathrm{symm.}\,0\lambda}^{ \gamma^* p \to q {\bar q} p'}}{d^2 p \, d^2 \Delta \, d z} + \mathrm{c.c.} \right]
  \\ \notag &
  = 
  -\frac{i\sqrt{2}}{(2 \pi)^5 s} \, \int d^2 x_1 \, d^2 x_{1'} \, d^2 x_2 \, d^2 x_{2'} \,    e^{- i {\un p} \cdot ({\un x}_{12} - \un x_{1'2'}) - i \un \Delta \cdot \left[ z \, \un x_{11'} + (1-z) \un x_{22'}  \right] } 
 \\ \notag 
 & \hspace{0.5cm}
 \times 
 \Bigg\{ 
 \Bigg[  \left[ \frac{1}{z} \, Q_{12} (s) - \frac{1}{1-z} \, Q_{21} (s) \right] \, N_{1'2'} (s)  - \left[ \frac{1}{z} \, Q^{\textrm{NS}, Im}_{12} (s) + \frac{1}{1-z} \, Q^{\textrm{NS}, Im}_{21} (s) \right] \, O_{1'2'} (s) \Bigg]  \notag \\
 & \hspace{2cm} \times  
\left[ \frac{\hat{k} \cdot \un x_{12}}{x_{12}} \Phi^{[1]}_{\mathrm{LT}}(\un x_{12}, \un x_{1^\prime 2^\prime}, z) 
- 
\frac{\hat{k} \cdot \un x_{1^\prime 2^\prime}}{x_{1^\prime 2^\prime}} \Phi^{[1]}_{\mathrm{LT}}(\un x_{1^\prime 2^\prime}, \un x_{1 2}, z)
\right] \notag
 \\ \notag 
 & \hspace{0.5cm}
 - \Bigg[
\left[  \left( \frac{1}{z} G^i_{12}(s) - \frac{1}{1-z} G^i_{21}(s) \right) N_{1^\prime 2^\prime}(s) - \left( \frac{1}{z} G^{i\,\mathrm{NS,Im}}_{12}(s) + \frac{1}{1-z} G^{i\,\mathrm{NS,Im}}_{21}(s) \right) \, O_{1^\prime 2^\prime}(s) \right]
 \Big( \partial^i_{\un 1} - i p^i \Big) 
\notag \\
& 
\hspace{2cm} -  
 i \Delta^i \, \Bigg[  \left[  G^i_{12}(s) +  G^i_{21}(s)  \right] \, N_{1^\prime 2^\prime}(s) -  \left[  G^{i\,\mathrm{NS,Im}}_{12}(s) -  G^{i\,\mathrm{NS,Im}}_{21}(s)  \right]  \, O_{1^\prime 2^\prime}(s) \Bigg]
 \Bigg] \notag \\
 & \hspace{2cm} \times  \left[ \frac{\hat{k} \times \un x_{12}}{x_{12}} \Phi^{[2]}_{\mathrm{LT}}(\un x_{12}, \un x_{1^\prime 2^\prime}, z) 
+ 
\frac{\hat{k} \times \un x_{1^\prime 2^\prime}}{x_{1^\prime 2^\prime}} \Phi^{[2]}_{\mathrm{LT}}(\un x_{1^\prime 2^\prime}, \un x_{1 2}, z)
\right] \Bigg\}. \notag
\end{align}

Neglecting the $\as$-suppressed amplitudes $O_{10}$, $Q^{\textrm{NS}, Im}_{10}$, and $G^{i\,\mathrm{NS,Im}}_{10}$ we arrive at our final result for the LT terms in the numerator of the DSA, 
\begin{align} \label{DSA_LT_final}
  & z (1-z) \, \frac{1}{2} \sum_{S_L, \lambda = \pm 1} S_L \left[ e^{ i \lambda \phi} \, \frac{d\sigma_{\mathrm{symm.}\,0\lambda}^{ \gamma^* p \to q {\bar q} p'}}{d^2 p \, d^2 \Delta \, d z} + \mathrm{c.c.} \right]
  \\ \notag &
  = 
  -\frac{i\sqrt{2}}{(2 \pi)^5 s} \, \int d^2 x_1 \, d^2 x_{1'} \, d^2 x_2 \, d^2 x_{2'} \,    e^{- i {\un p} \cdot ({\un x}_{12} - \un x_{1'2'}) - i \un \Delta \cdot \left[ z \, \un x_{11'} + (1-z) \un x_{22'}  \right] } 
 \\ \notag 
 & \hspace{0.5cm}
 \times 
 N_{1^\prime 2^\prime}(s)
 \Bigg\{ 
 \left[ \frac{1}{z} Q_{12}(s) - \frac{1}{1-z} Q_{21}(s)  
 \right] 
\left[ \frac{\hat{k} \cdot \un x_{12}}{x_{12}} \Phi^{[1]}_{\mathrm{LT}}(\un x_{12}, \un x_{1^\prime 2^\prime}, z) 
- 
\frac{\hat{k} \cdot \un x_{1^\prime 2^\prime}}{x_{1^\prime 2^\prime}} \Phi^{[1]}_{\mathrm{LT}}(\un x_{1^\prime 2^\prime}, \un x_{1 2}, z)
\right]
 \\ \notag 
 & \hspace{2cm}
 - 
\left[
 \frac{1}{z} G^i_{12}(s) - \frac{1}{1-z} G^i_{21}(s) 
\right]
\Big[\partial^i_{\un 1} - i p^i \Big] 
\left[ \frac{\hat{k} \times \un x_{12}}{x_{12}} \Phi^{[2]}_{\mathrm{LT}}(\un x_{12}, \un x_{1^\prime 2^\prime}, z) 
+
\frac{\hat{k} \times \un x_{1^\prime 2^\prime}}{x_{1^\prime 2^\prime}} \Phi^{[2]}_{\mathrm{LT}}(\un x_{1^\prime 2^\prime}, \un x_{1 2}, z)
\right]
\notag \\
& \hspace{2cm}
+ 
 \left[  G^i_{12}(s) +  G^i_{21}(s) \right] 
\Big(i \Delta^i \Big) 
\left[ \frac{\hat{k} \times \un x_{12}}{x_{12}} \Phi^{[2]}_{\mathrm{LT}}(\un x_{12}, \un x_{1^\prime 2^\prime}, z) 
+ 
\frac{\hat{k} \times \un x_{1^\prime 2^\prime}}{x_{1^\prime 2^\prime}} \Phi^{[2]}_{\mathrm{LT}}(\un x_{1^\prime 2^\prime}, \un x_{1 2}, z)
\right]
 \Bigg\}, \notag
\end{align}
where the wave function overlaps defined in \eqs{type1_overlaps} and (\ref{type2_overlaps}) are
\begin{subequations}
    \begin{align}
         \Phi^{[1]}_{\mathrm{LT}}(\un x_{12}, \un x_{1'2'}, z) &= -\frac{4 \alpha_{EM} Z_f^2 N_c^2 Q^2}{\pi} \big[ z (1-z) \big]^{5/2} K_1 \left(x_{12} Q \sqrt{z (1-z)} \right)
         K_0 \left(x_{1'2'} Q \sqrt{z (1-z)} \right),
         \\ 
        \Phi^{[2]}_{\mathrm{LT}}(\un x_{12}, \un x_{1'2'}, z) &= -\frac{4 \alpha_{EM} Z_f^2 N_c^2 Q^2}{\pi} \big[ z (1-z) \big]^{5/2} (1-2\,z) \, K_1 \left(x_{12} Q \sqrt{z (1-z)} \right)
         K_0 \left(x_{1'2'} Q \sqrt{z (1-z)} \right).
    \end{align}
\end{subequations}

%%%%%%%%%%%%%%%%%%%%%%%%%%%%%%%%%%%%%%%%%%%%%%%%%%%%%%%%%%%%%%%%%%%%%%%%%%%%%%%
\section{Single spin asymmetry: the LT channel and a general argument}
\label{sec:ssa}

Let us now consider the numerator of the SSA given in \eq{SSA}. It contains four different structures, dependent on the polarizations of the virtual photons: TT, T,-T, LL, and LT.  To begin, we will consider the LT terms and evaluate \eqs{XS5}, (\ref{XS7}), and (\ref{XS10}) for those terms. We will show that the polarized dipole amplitudes from \eqs{pdas} needed to calculate hPDFs and OAM distributions at small $x$ appear in the LT terms only when multiplying Im~$N_{10} (s)$. In addition, the LT contribution to the SSA numerator we will obtain below will contain other additive terms, which are the same order in $\as$ as the product of polarized dipole amplitudes and Im~$N_{10} (s)$. These terms do not contain the polarized dipole amplitudes from \eqs{pdas}. Separation of the terms containing polarized dipole amplitudes from those other additive terms does not appear to be feasible by the angular or $y$ dependence of the dijet production cross section. Since the phenomenology (or theory) of small-$x$ physics is not developed enough to give us a reliable evaluation of Im~$N_{10} (s)$ and the dipole amplitudes in those additive terms, it appears that an extraction of the polarized dipole amplitudes and their moments from the SSA is difficult at the moment. The TT, T,-T, and LL contributions to the SSA are analyzed in Appendix \ref{appendix:SSA_channels}, where the same conclusion is reached.

To evaluate the LT terms in \eq{SSA} we will compute 
\begin{align} \label{struct_LT}
    & z (1-z) \, \frac{1}{2}  \sum_{S_L, \lambda = \pm 1} S_L \, \lambda \, \left[ e^{i \lambda \phi} \, \frac{d\sigma_{0\lambda}^{ \gamma^* p \to q {\bar q} p'}}{d^2 p_1 \, d^2 p_2 \, d z} + \mathrm{c.c.} \right] . 
\end{align}

Beginning again with the type-1 terms, we use \eq{XS5} to write 
\begin{align}\label{SSA_LT11}
& z (1-z) \, \frac{1}{2}  \sum_{S_L, \lambda = \pm 1} S_L \, \lambda \, \left[ e^{i \lambda \phi} \, \frac{d\sigma_{0\lambda }^{[1] \, \gamma^* p \to q {\bar q} p'}}{d^2 p_1 \, d^2 p_2 \, d z} + \mathrm{c.c.} \right] = - \frac{\sqrt{2}}{2 (2 \pi)^5 \, s} \, \int d^2 x_1 \, d^2 x_{1'} \, d^2 x_2 \, d^2 x_{2'} \,  e^{- i {\un p}_1 \cdot {\un x}_{11'} - i {\un p}_2 \cdot {\un x}_{22'}} 
\\ 
& \times  \Phi_{\mathrm{LT}}^{[1]}(\un x_{12}, \un x_{1^\prime 2^\prime}, z) \frac{{\hat k} \times {\un x}_{12}}{x_{12}}  \frac{1}{N_c} \, \Bigg\{  \left[  \frac{1}{z} \, \llangle \tord \tr \left[ V^{\textrm{pol}[1]}_{{\un 1}} \, V_{{\un 2}}^\dagger \right] \rrangle (s) -  \frac{1}{1-z} \,  \llangle \tord \tr \left[ V_{{\un 1}} \,  V^{\textrm{pol} [1] \, \dagger}_{{\un 2}} \right]  \rrangle (s) \right] \, \left[ 1 - S_{1'2'}^* (s) \right] 
\notag \\
&  + \left[ \frac{1}{z} \, \llangle \atord \tr \left[ V_{{\un 2'}} \,  V^{\textrm{pol} [1] \, \dagger}_{{\un 1'}}  \right]  \rrangle (s) - \frac{1}{1-z} \, \llangle \atord \tr \left[  V^{\textrm{pol} [1]}_{{\un 2'}} \, V_{{\un 1'}}^\dagger \right]  \rrangle (s) \right] \, \left[ 1 - S_{12} (s) \right]  \Bigg\} + \cc .  \notag
\end{align}
Following the procedure described above for the DSA, we add the complex conjugate term in \eq{SSA_LT11} and symmetrize the resulting expression with respect to the quark--antiquark jet interchange. This yields 
\begin{align} \label{SSA_LT1res0}
    &z (1-z) \, \frac{1}{2} \sum_{S_L, \lambda = \pm 1} S_L \, \lambda \, \left[ e^{i \lambda \phi} \, \frac{d\sigma_{\mathrm{symm.}\,0\lambda }^{[1] \, \gamma^* p \to q {\bar q} p'}}{d^2 p \, d^2 \Delta \, d z} + \mathrm{c.c.} \right]   
    \\ \notag 
   & = -\frac{i \sqrt{2}}{2 (2\pi)^5s}
        \int d^2 x_1 \, d^2 x_{1'} \, d^2 x_2 \, d^2 x_{2'} \,  
        e^{- i {\un p} \cdot ({\un x}_{12} - \un x_{1'2'}) - i \un \Delta \cdot \big[ z \, \un x_{11'} + (1-z) \un x_{22'} \big] } 
    \\ \notag & \hspace{2cm}
       \times  \left[ \frac{\hat{k} \times \un x_{1 2}}{x_{12}}
        \Phi^{[1]}_{\mathrm{LT}}(\un x_{12}, \un x_{1^\prime 2^\prime} , z) 
         + \frac{\hat{k} \times \un x_{1^\prime 2^\prime}}{x_{1^\prime 2^\prime}} \Phi^{[1]}_{\mathrm{LT}}(\un x_{1^\prime 2^\prime}, \un x_{12}, z) 
        \right]
    \\ \notag & \hspace{2cm}
\times \frac{1}{N_c} \, \textrm{Im} \Bigg\{ \frac{1}{z} \, \llangle \tord \tr \left[ V^{\textrm{pol}[1]}_{{\un 1}} \, V_{{\un 2}}^\dagger \right] \rrangle (s) \, \left[ 1 - S_{1'2'}^* (s) \right] + \frac{1}{z} \, \llangle \tord \tr \left[ V_{{\un 2}} V^{\textrm{pol}[1] \dagger}_{{\un 1}} \right] \rrangle (s) \, \left[ 1 - S_{2'1'}^* (s) \right] \\ \notag & \hspace{2cm} 
- \frac{1}{1-z} \,  \llangle \tord \tr \left[ V_{{\un 1}} \,  V^{\textrm{pol} [1] \, \dagger}_{{\un 2}} \right]  \rrangle (s) \, \left[ 1 - S_{1'2'}^* (s) \right]
             - \frac{1}{1-z} \,  \llangle \tord \tr \left[ V^{\textrm{pol} [1]}_{{\un 2}} V_{{\un 1}}^\dagger \,   \right]  \rrangle (s) \, \left[ 1 - S_{2'1'}^* (s) \right]
 \Bigg\}.
\end{align}
As described above, we neglect Im~$O_{10}$ since it is suppressed by at least $\as^2$ compared to Re~$N_{10}$. However, in the present calculation, we will keep Im~$N_{10}$, which is only order-$\as$ suppressed compared to Re~$N_{10}$. \eq{SSA_LT1res0} becomes
\begin{align} \label{SSA_LT1res}
    &z (1-z) \, \frac{1}{2} \sum_{S_L, \lambda = \pm 1} S_L \, \lambda \, \left[ e^{i \lambda \phi} \, \frac{d\sigma_{\mathrm{symm.}\,0\lambda }^{[1] \, \gamma^* p \to q {\bar q} p'}}{d^2 p \, d^2 \Delta \, d z} + \mathrm{c.c.} \right]   
    \\ \notag 
   & = -\frac{i \sqrt{2}}{(2\pi)^5s}
        \int d^2 x_1 \, d^2 x_{1'} \, d^2 x_2 \, d^2 x_{2'} \,  
        e^{- i {\un p} \cdot ({\un x}_{12} - \un x_{1'2'}) - i \un \Delta \cdot \big[ z \, \un x_{11'} + (1-z) \un x_{22'} \big] } 
    \\ \notag & \hspace{2cm}
       \times  \left[ \frac{\hat{k} \times \un x_{1 2}}{x_{12}}
        \Phi^{[1]}_{\mathrm{LT}}(\un x_{12}, \un x_{1^\prime 2^\prime} , z) 
         + \frac{\hat{k} \times \un x_{1^\prime 2^\prime}}{x_{1^\prime 2^\prime}} \Phi^{[1]}_{\mathrm{LT}}(\un x_{1^\prime 2^\prime}, \un x_{12}, z) 
        \right]
    \\ \notag & \hspace{2cm}
\times \Bigg\{
            - \left[ \frac{1}{z} Q_{12}(s) - \frac{1}{1-z} Q_{21}(s) \right] \textrm{Im}\left[ N_{1^\prime 2^\prime}(s) \right] 
    \\ \notag & \hspace{2cm}
            + \left[ \frac{1}{z} Q^{\mathrm{Im}}_{12}(s) - \frac{1}{1-z} Q^{\mathrm{Im}}_{21}(s) \right] \textrm{Re}~\left[ N_{1^\prime 2^\prime}(s) \right]
 +  \left[ \frac{1}{z} Q^{\mathrm{NS}}_{12}(s) - \frac{1}{1-z} Q^{\mathrm{NS}}_{21}(s)
 \right] O_{1^\prime 2^\prime}(s)  
 \Bigg\},
\end{align}
where we have defined 
\begin{subequations} \label{ssa_amps}
\begin{align}
        Q_{10}^{\textrm{Im}} (s) &= \frac{1}{2 \, N_c} \, \mbox{Im} \, \llangle \mbox{T} \, \tr \left[ V_{\un 0} \,  V_{\un 1}^{\textrm{pol} [1] \,\dagger} \right] + \mbox{T} \,  \tr \left[ V_{\un 1}^{\textrm{pol} [1]} \, V_{\un 0}^\dagger \right]   \rrangle (s), 
        \\ 
        Q_{10}^{\textrm{NS}} (s) &= \frac{1}{2 \, N_c} \, \mbox{Re} \, \llangle - \mbox{T} \, \tr \left[ V_{\un 0} \,  V_{\un 1}^{\textrm{pol} [1] \,\dagger} \right] + \mbox{T} \,  \tr \left[ V_{\un 1}^{\textrm{pol} [1]} \, V_{\un 0}^\dagger \right]   \rrangle (s),
\end{align}
\end{subequations}
and, unlike the above notation, delineate Re and Im parts of $N_{10}$. Each term in the curly brackets of \eq{SSA_LT1res} is of the same order in SLA power counting: they are all order $\as$ suppressed compared to our expressions for the terms in the DSA numerator above.

Using PT-symmetry arguments, the terms containing $V_{\un x}^{q[2]}$ can again be shown not to contribute to the structure in \eq{struct_LT}. Finally, considering the contribution coming from the $V^{\mathrm{G}[2]}_{\un x, \un y}$ operator, we proceed analogously to the DSA above. After symmetrizing over the quark and antiquark jets, the result is 
\begin{align} \label{SSA_LTres2}
   &z (1-z) \, \frac{1}{2} \sum_{S_L, \lambda = \pm 1} S_L \, \lambda \, \left[ e^{i \lambda \phi} \, \frac{d\sigma_{\mathrm{symm.}\,0\lambda }^{\mathrm{G}[2] \, \gamma^* p \to q {\bar q} p'}}{d^2 p \, d^2 \Delta \, d z} + \mathrm{c.c.} \right] 
   \\ \notag &
    = - \frac{\sqrt{2} \, i}{(2 \pi)^5 s} \, \int d^2 x_1 \, d^2 x_{1'} \, d^2 x_2 \, d^2 x_{2'} \,  e^{- i {\un p} \cdot ({\un x}_{12} - \un x_{1'2'}) - i \un \Delta \cdot \Big[ z \, \un x_{11'} + (1-z) \un x_{22'}  \Big]}  
 \\ \notag 
 & 
 \times 
\Bigg\{\Bigg[
- \left( \frac{1}{z} G^{i}_{12}(s) - \frac{1}{1-z} G^{i}_{21}(s) \right) \textrm{Im}~\left[ N_{1^\prime 2^\prime}(s) \right] \\ \notag 
 &
 + 
 \left( \frac{1}{z} G^{i \,\mathrm{Im}}_{12}(s) - \frac{1}{1-z} G^{i \,\mathrm{Im}}_{21}(s) \right) \textrm{Re}~\left[ N_{1^\prime 2^\prime}(s) \right]
 +  \left( \frac{1}{z} G^{i\, \mathrm{NS}}_{12}(s) + \frac{1}{1-z} G^{i\,\mathrm{NS}}_{21}(s)
 \right)O_{1^\prime 2^\prime}(s) 
\Bigg] \, \Big( \partial^i_{\un 1} - i p^i \Big) 
\notag \\
& 
-  \big( i \Delta^i \big) \, 
 \Big[   - \big( G^{i}_{12}(s) +  G^{i}_{21}(s) \big) \, \textrm{Im}~\left[ N_{1^\prime 2^\prime}(s) \right] 
 +  \big( G^{i \, \mathrm{Im}}_{12}(s) +  G^{i\, \mathrm{Im}}_{21}(s) \big) \, \textrm{Re}~\left[  N_{1^\prime 2^\prime}(s) \right]
 +  \big(  G^{i\,\mathrm{NS}}_{12}(s) - G^{i\,\mathrm{NS}}_{21}(s)
 \big)O_{1^\prime 2^\prime}(s)  \Big]  \Bigg\} 
\notag \\ & \hspace{1cm}
\times  \, \left[ - \frac{\hat{k} \cdot \un x_{12}}{x_{12}} \Phi^{[2]}_{\mathrm{LT}}(\un x_{12}, \un x_{1^\prime 2^\prime}, z) 
+ 
\frac{\hat{k} \cdot  \un x_{1^\prime 2^\prime}}{x_{1^\prime 2^\prime}} \Phi^{[2]}_{\mathrm{LT}}(\un x_{1^\prime 2^\prime}, \un x_{1 2}, z)
\right] , \notag
\end{align}
where, similarly to \eqs{ssa_amps}, we have defined 
\begin{subequations} \label{ssa_amps_2}
\begin{align}
        G_{10}^{i\, \textrm{Im}} (s) &= \frac{1}{2 \, N_c} \, \mbox{Im} \, \llangle \mbox{T} \, \tr \left[ V_{\un 0} \,  V_{\un 1}^{i \textrm{G} [2] \,\dagger} \right] + \mbox{T} \,  \tr \left[ V_{\un 1}^{i\textrm{G} [2]} \, V_{\un 0}^\dagger \right]   \rrangle (s), 
        \\ 
        G_{10}^{i\, \textrm{NS}} (s) &= \frac{1}{2 \, N_c} \, \mbox{Re} \, \llangle - \mbox{T} \, \tr \left[ V_{\un 0} \,  V_{\un 1}^{i\textrm{G} [2] \,\dagger} \right] + \mbox{T} \,  \tr \left[ V_{\un 1}^{i\textrm{G} [2]} \, V_{\un 0}^\dagger \right]   \rrangle (s).
\end{align}
\end{subequations}

The full LT contribution to the numerator of the SSA for elastic dijet production is given by the sum of the terms in \eqs{SSA_LT1res} and \eqref{SSA_LTres2}. Analyzing these equations, we see that \eq{SSA_LT1res} contains the polarized dipole amplitude $Q_{10}$, while \eq{SSA_LTres2} contains $G^i_{10}$. Therefore, indeed, the LT contribution to SSA appears to depend on hPDFs and the OAM distributions. However, the dipole amplitudes $Q_{10}$ and $G^i_{10}$ enter those expressions on equal footing with other lesser-known dipole amplitudes. For instance, $Q_{10}$ is multiplied in \eq{SSA_LT1res} by Im~$N_{10}$, for which no phenomenology exists at present. Similarly, $G^i_{10}$ is multiplied in \eq{SSA_LTres2} by Im~$N_{10}$ as well. In addition, there are additive terms accompanying the amplitudes $Q_{10}$ and $G^i_{10}$ in \eqs{SSA_LT1res} and \eqref{SSA_LTres2}: those terms include the odderon amplitude $O_{10}$, which also has limited phenomenology at the moment (see \cite{Benic:2024pqe, Braun:2020vmd} for encouraging recent developments), along with the dipole amplitudes $G_{10}^{i\, \textrm{Im}}$ and $G_{10}^{i\, \textrm{NS}}$, which, to the best of our knowledge, have been introduced in the present paper for the first time ever, and are, therefore, completely unexplored. The additive structure of those corrections appears to indicate that it is impossible to separate them from the polarized dipole amplitudes we are after by, say, utilizing a difference in their dependence on the angle between $\un p$ and $\un \Delta$. We observe that while \eqs{SSA_LT1res} and \eqref{SSA_LTres2} may probe the OAM distributions, it seems hard, at the moment, to disentangle those OAM contributions from other terms, whose phenomenology is either underdeveloped or completely undeveloped. Therefore we conclude the structure in \eq{struct_LT} does not allow one, at present, to use it to extract the OAM distributions. 

In fact, we can use this specific structure as an example of a more general argument as to why none of the terms in the SSA of \eq{SSA} provide a feasible probe for the OAM distributions. All of the terms we have calculated for both the DSA and SSA thus far have the same general structure: 
\begin{align} \label{gen-struct}
    d \sigma = \Phi \otimes G \otimes S,
\end{align}
where $\Phi$ represents the virtual photon's wave function overlap between the amplitude and its complex conjugate, $G$ is a generic polarized dipole amplitude, and $S$ is a generic unpolarized dipole amplitude\footnote{Strictly speaking, the cross section in \eq{gen-struct} is in position space, and to obtain the numerator of the DSA or SSA, one needs to first Fourier transform to momentum space. However, the details of the Fourier transform do not affect the argument presented here so we omit them for clarity.}. The types of dipole amplitudes that appear on the right hand side of \eq{gen-struct} depend on the PT symmetry of the observable on the left hand side. For example, since the DSA is PT even, on the right hand side of \eq{gen-struct}, both of $G$ and $S$ must be either PT odd or PT even, since $\Phi$ does not transform under PT (after summation over all helicities, and with the transverse positions being intrinsic integration variables in the cross sections at hand). This is illustrated in \eqs{DSA_TT} and \eqref{DSA_LT_pre_final}. In all of those equations, each term has either the product of two PT-even dipole amplitudes, like $N_{1^\prime 2^\prime}  Q_{12}$, or two PT-odd dipole amplitudes, like $O_{1^\prime 2^\prime}  Q^{\mathrm{NS, Im}}_{12}$. Likewise, the SSA is PT-odd, so $G$ and $S$ in \eq{gen-struct} for the SSA must transform in opposite ways under PT: either $G$ is PT even and $S$ is PT odd, or vice versa. This is illustrated in \eqs{SSA_LT1res} and (\ref{SSA_LTres2}) above: $Q^{\mathrm{Im}}_{12}$, $G^{i\,\mathrm{Im}}_{12}$, and $O_{1^\prime 2^\prime}$ are all PT odd, while $Q^{\mathrm{NS}}_{12}$, $G^{i\,\mathrm{NS}}_{12}$, and $N_{1^\prime 2^\prime}$ are PT even. 

Without assuming that $N_{10}$ and $O_{10}$ are real, in general, there exists four kinds of unpolarized dipole amplitudes: the real and imaginary parts of $N_{10}$ and $O_{10}$. For the purposes of phenomenology, the real part of $N_{10}$ is easier to extract from unpolarized data rather than the other amplitudes, since the other amplitudes are suppressed by at least one factor of $\as$. Therefore a feasible probe for the OAM distributions should contain $\mathrm{Re}\,N_{10}$ for the ``$S$-matrix factor" in \eq{gen-struct}. This is exactly what happens in \eqs{DSA_TT_final} and (\ref{DSA_LT_final}). Since the polarized dipole amplitudes in those equations also appear in \eqs{pdfs}, the DSA is a good probe for the OAM distributions. However, for the SSA, if $\mathrm{Re}\,N_{10}$ appears in a term, it must be multiplied by a PT-odd polarized dipole amplitude, such as $Q^{\mathrm{Im}}_{12}$ or $G^{i\,\mathrm{Im}}_{12}$. Since the polarized dipole amplitudes that contribute to the OAM distributions are PT even, they have to be multiplied by the PT-odd eikonal (unpolarized) amplitude Im~$N_{10}$, as is the case in \eqs{SSA_LT1res} and \eqref{SSA_LTres2}. As mentioned above, the products $Q^{\mathrm{Im}}_{12} \, \mathrm{Re}\,N_{10}$ and $Q_{12} \, \mathrm{Im}\,N_{10}$ are the same order in $\as$ counting. We conclude that the leading (in $\as$) contribution to the SSA numerator cannot be limited to the polarized dipole amplitudes contributing to the OAM distributions, as observed above in \eqs{SSA_LT1res} and \eqref{SSA_LTres2}. Therefore, with the current status of theoretical and phenomenological developments in small-$x$ physics, the SSA for elastic dijet production in $e+p$ collisions cannot serve as a good probe for OAM. To further illustrate this argument, the remaining terms in \eq{SSA} are computed explicitly in Appendix~\ref{appendix:SSA_channels}.

We should compare our results for the SSA above and in Appendix~\ref{appendix:SSA_channels} to the results from \cite{Hatta:2016aoc}. Assuming the contribution to the SSA calculated in \cite{Hatta:2016aoc} is the TT term of the SSA, the authors of that reference find the angular structure of the SSA that couples to the gluon OAM distribution is $\un p \times \un \Delta$\footnote{Note that $\un p$, as defined in \eq{p-def} above, is slightly different than the definition employed in \cite{Hatta:2016aoc}, where $\un p = \frac{1}{2} \left( \un p_1 - \un p_2\right)$. However, the cross product $\un p \times \un \Delta$ is unchanged by this difference in definition.} (in our notation). To compare with this result, we must inspect our result for the TT term of the SSA which contains the polarized dipole amplitudes and their moments, given in \eq{SSA_TT6} below. One can show that, after expanding in $\un \Delta$ and integrating over impact parameters (as is done for the DSA in the next Section), we are left with one Levi-Civita symbol $\epsilon^{ij}$. After the integration over the dipole sizes is carried out, the only remaining transverse vectors in the expression are $\un p$ and $\un \Delta$: when combined with $\epsilon^{ij}$ they have to give $\un p \times \un \Delta$. Therefore, we conclude that, similar to \cite{Hatta:2016aoc}, our result for the TT term of the SSA contains an angular-dependent term $\sim \un p \times \un \Delta$ which couples to the OAM distributions. This is consistent with the symmetries of the problem, since the SSA is P-odd and so is the $\un p \times \un \Delta$ structure. (One can see the latter by following the arguments presented at the end of the next Section.) In addition to the $\un p \times \un \Delta$ structure in the TT term, one can see from \eq{SSA_LTres2} above that additional P-odd angular structures, $\un p \times \un \Delta \ {\un p} \cdot {\hat k}$ and $\un p \cdot \un \Delta \ {\un p} \times {\hat k}$, which couple to the OAM distributions appear in the LT term of the SSA.

%%%%%%%%%%%%%%%%%%%%%%%%%%%%%%%%%%%%%%%%%%%%%%%%%%%%%%%%%%%%%%%%%%%%%%%%%%%%%%%
\section{Expansion of double spin asymmetry in transverse momentum transfer $\Delta_\perp$}

\label{sec:DSA_exp}

To extract the OAM distributions (given in \eqs{PDF+OAM_summ}) from the data, we need the impact-parameter integrated polarized dipole amplitudes (see \eqs{integrated} and \eqref{Q_tilde_b}) and moment amplitudes (see \eqs{moments} and \eqref{object_moment}). To relate the TT and LT contributions to the DSA numerator given by \eqs{DSA_TT_final} and (\ref{DSA_LT_final}) above to both the zeroth and first impact-parameter moments of the polarized dipole amplitudes, we need to integrate over the impact parameters $\un b = z \,\un x_{1} + (1-z) \,\un x_{2}$ and $\un b' = z \,\un x_{1'} + (1-z) \,\un x_{2'}$. However, the integrals over $\un b$ and $\un b'$ are Fourier transforms, and do not immediately give us the required moments. Since we only need the first two impact-parameter moments, we will follow \cite{Hatta:2016aoc,Bhattacharya:2022vvo, Bhattacharya:2023hbq, Bhattacharya:2024sck} and expand \eqs{DSA_TT_final} and \eqref{DSA_LT_final} to the linear order in the transverse momentum $\Delta_\perp$ transferred from the outgoing proton in the elastic scattering process at hand. That is, we expand the Fourier phase as
\begin{align} \label{phase_expand}
    e^{-i \un p_{1} \cdot \un x_{1 1^\prime} - i \un p_2 \cdot x_{2 2^\prime}} = e^{-i \un p \cdot (\un x_{12} - \un x_{1^\prime 2^\prime})}
    e^{-i \un \Delta \cdot (\un b - \un b')} =
    e^{-i \un p \cdot (\un x_{12} - \un x_{1^\prime 2^\prime})} \Big[ 
    1- i \un \Delta \cdot \big( \un b - \un b' \big) + \mathcal{O}(\Delta_\perp^2) \Big],
\end{align}
where we have used \eqs{p-d-vars} and defined the impact parameters in the scattering amplitude and in the complex conjugate amplitude by
\begin{subequations}
\begin{align}
    \un b \equiv& z\, \un x_{1} + (1-z) \, \un x_{2} = {\un x}_1 - (1-z) \, {\un x}_{12} = {\un x}_2 + z \, {\un x}_{12}, 
    \\ 
    {\un b}' \equiv& z \, {\un x}_{1'} + (1-z) \, {\un x}_{2'} =
     {\un x}_{1'} - (1-z) \, {\un x}_{1'2'} = {\un x}_{2'} + z \, {\un x}_{1'2'}.
\end{align}
\end{subequations}

Note that, for the expansion in \eq{phase_expand} to be valid, we need to consider very small values of $\Delta_\perp$, namely $\Delta_\perp \ll 1/R_p$ where $R_p \sim 1/\Lambda_{QCD}$ is the proton radius and $\Lambda_{QCD}$ is the QCD confinement scale. At the same time, we need to work in the perturbative QCD regime, that is, we need $p_\perp, Q \gg \Lambda_{QCD}$. Therefore, the kinematic regime we assume here is 
\begin{align}
    p_\perp , Q \gtrsim \Lambda \gg \Lambda_{QCD} \gg \Delta_\perp,
\end{align}
where $\Lambda$ is the IR cutoff. 

To perform the integrals over $\un b$ and $\un b'$, we use \eqs{integrated}, (\ref{moments}), and 
\begin{subequations}\label{N_b}
\begin{align}
    \int d^2 x_1 \, N_{10}(s) &= N(x_{10}^2, s), \label{N_b_def} \\
    \int d^2 x_1 \,x_1^i \, N_{10}(s) &= \frac{1}{2} x_{10}^i \, N(x_{10}^2, s),
\end{align}
\end{subequations}
to write 
\begin{subequations} \label{b_ints}
\begin{align}
 \int d^2 b \, b^i \, N_{12} (s) &= \left( z - \thalf \right) \, x_{12}^i \, N (x_{12}^2, s), \\
\int d^2 b \, b^i \, Q_{12}(s) &= {x}_{12}^i \, I_3 (x_{12}^2, s) + \epsilon^{ij} x_{12}^j \, J_3 (x_{12}^2, s) - (1-z) \, x_{12}^i \, Q (x_{12}^2, s) , \\
 \int d^2 b \, b^i \, Q_{21}(s) &= - {x}_{12}^i \, I_3 (x_{12}^2, s) - \epsilon^{ij} x_{12}^j \, J_3 (x_{12}^2, s) +z \, x_{12}^i \, Q (x_{12}^2, s) , \\
\int d^2 b \, b^i \, G^j_{12}(s) &= \epsilon^{ij} \, x_{12}^2 \, I_{4}(x_{12}^2,s) + \epsilon^{ik} \, x_{12}^k \, x_{12}^j \, I_{5}(x_{12}^2 , s) + \delta^{ij} x_{12}^2 \, J_{4}(x_{12}^2, s) + x_{12}^i \, x_{12}^j \, J_{5}(x_{12}^2 , s) \\ 
& \hspace*{5cm} - (1-z) \, x_{12}^i \, \left[ x_{12}^j \, G_1(x_{12}^2, s) + \epsilon^{jl} x_{12}^l \, G_2(x_{12}^2, s) \right] , \notag \\
 \int d^2 b \, b^i \, G^j_{21}(s) &= \epsilon^{ij} \, x_{12}^2 \, I_{4}(x_{12}^2,s) + \epsilon^{ik} \, x_{12}^k \, x_{12}^j \, I_{5}(x_{12}^2 , s) + \delta^{ij} x_{12}^2 \, J_{4}(x_{12}^2, s) + x_{12}^i \, x_{12}^j \, J_{5}(x_{12}^2 , s) \\ 
& \hspace*{5cm} - z \, x_{12}^i \, \left[ x_{12}^j \, G_1(x_{12}^2, s) + \epsilon^{jl} x_{12}^l \, G_2(x_{12}^2, s) \right] . \notag 
\end{align}
\end{subequations}
Note that \eq{N_b_def} is the definition of the impact-parameter integrated unpolarized dipole amplitude $N (x_{10}^2, s)$.

To apply these equations to the DSA, we take \eqs{DSA_TT_final} and (\ref{DSA_LT_final}) and expand the Fourier phase via \eq{phase_expand}. Then, using \eqs{b_ints} along with 
\begin{align}
    d^2 x_1 \, d^2 x_{1'} \, d^2 x_2 \, d^2 x_{2'} = d^2 x_{12} \, d^2 x_{1'2'} \, d^2 b \, d^2 b' ,
\end{align}
we integrate over $\un b$ and $\un b'$, keeping $\un x_{12}$ and $\un x_{1'2'}$ constant. The result is, for the TT and LT contributions to the numerator of DSA in elastic dijet production in $e+p$ collisions, 
\begin{tcolorbox}[colback=blue!10!white]
\begin{subequations}\label{integrated_res}
\begin{align} \label{TT_result}
&z (1-z) \ \frac{1}{2} \sum_{S_L, \lambda \pm 1} S_L \, \lambda  \, \frac{d\sigma_{\mathrm{symm.}\, \lambda \lambda}^{\gamma^* p \to q {\bar q} p'}}{d^2 p \, d^2 \Delta \, d z} =
      -\frac{2}{(2\pi)^5 z (1-z) s}
        \int \displaylimits d^2 x_{12} \, d^2 x_{1^\prime 2^\prime} \,e^{-i \un p \cdot (\un x_{12} - \un x_{1^\prime 2^\prime})} \, N(x_{1^\prime 2^\prime}^2, s)
 \\ \notag & 
  \times  \Bigg\{
            \Bigg[ 
     \left(1-2z + i \un \Delta \cdot \un x_{12} \, \left( z^2 + (1-z)^2 \right) - \frac{i}{2} \, \un \Delta \cdot \un x_{1'2'} \, (1-2 z)^2  \right) Q(x_{12}^2, s) 
    - i \un \Delta \cdot \un x_{12} \, I_3(x_{12}^2, s) \\
    \notag & \hspace{1cm}
    - i \un \Delta \times \un x_{12} \, J_3(x_{12}^2, s)
    \Bigg] \, \Phi^{[1]}_{\mathrm{TT}}(\un x_{12}, \un x_{1^\prime 2^\prime}, z)
    \\ \notag & \hspace{0.5cm}
     + \Bigg[
     i (1- 2 z) \Big( \Delta^j \epsilon^{ji} x_{12}^2 I_4(x_{12}^2, s) + \un \Delta \times \un x_{12} \, x_{12}^i I_5(x_{12}^2, s)  + \Delta^i  \, x_{12}^2 J_4(x_{12}^2, s) + \un \Delta \cdot \un x_{12} \, x_{12}^i  J_5(x_{12}^2, s) 
    \Big)    
    \\ \notag & \hspace{1cm}
    - \left[ 1+ i \left(1 - 2\, z \right) \un \Delta \cdot \left( \un x_{12} - \frac{\un x_{1'2'}}{2} \right) \right] \Big( \epsilon^{ik} x_{12}^k G_2(x_{12}^2, s)  + x_{12}^i G_1(x_{12}^2, s) 
    \Big) \Bigg] \\
    \notag & \hspace{1cm} \times \, 
     \left( \partial^i_{\un 1} - ip^i \right)
     \Phi^{[2]}_{\mathrm{TT}}(\un x_{12}, \un x_{1'2'}, z)
     \Bigg\} + \mathcal{O}(\Delta_\perp^2),
     \\  \label{TL_result}
  & z (1-z) \ \frac{1}{2} \sum_{S_L, \lambda = \pm 1} S_L \, \left[ e^{i \lambda \phi} \, \frac{d\sigma_{\mathrm{symm.}\,0\lambda}^{ \gamma^* p \to q {\bar q} p'}}{d^2 p \, d^2 \Delta \, d z} + \mathrm{c.c.} \right] = 
    -\frac{2i\sqrt{2}}{2 (2\pi)^5 z (1-z) s}
        \int \displaylimits d^2 x_{12} \, d^2 x_{1^\prime 2^\prime} \, e^{-i \un p \cdot (\un x_{12} - \un x_{1^\prime 2^\prime})} 
 \\ \notag & 
  \times N(x_{1^\prime 2^\prime}^2, s)  \Bigg\{
            \Bigg[ 
 \left(1-2z + i \, \un \Delta \cdot \un x_{12} \left( z^2 + (1-z)^2 \right)  - \frac{i}{2} \un \Delta \cdot \un x_{1'2'} \, (1-2 z)^2 \right) Q(x_{12}^2, s) 
    - i \un \Delta \cdot \un x_{12} \, I_3(x_{12}^2, s) \\
    & \notag \hspace{1cm}
    - i \un \Delta \times \un x_{12} \, J_3(x_{12}^2, s)
    \Bigg] \, 
    \left[ \frac{\hat{k} \cdot \un x_{12}}{x_{12}} \Phi^{[1]}_{\mathrm{LT}}(\un x_{12}, \un x_{1^\prime 2^\prime}, z) 
- 
\frac{\hat{k} \cdot \un x_{1^\prime 2^\prime}}{x_{1^\prime 2^\prime}} \Phi^{[1]}_{\mathrm{LT}}(\un x_{1^\prime 2^\prime}, \un x_{1 2}, z)
\right]
    \\ \notag & \hspace{0.5cm}
     + \Bigg[  i (1- 2 z) \Big( \Delta^j \epsilon^{ji} \, x_{12}^2 \, I_4(x_{12}^2, s) + \un \Delta \times \un x_{12} \, x_{12}^i \, I_5(x_{12}^2, s)  + \Delta^i \, x_{12}^2 \, J_4(x_{12}^2, s) + \un \Delta \cdot \un x_{12} \, x_{12}^i \, J_5(x_{12}^2, s)
    \Big)    
    \\ \notag & \hspace{1cm}
    - \left[ 1+ i \left(1 - 2\, z \right) \un \Delta \cdot \left( \un x_{12} - \frac{\un x_{1'2'}}{2} \right) \right] \Big( \epsilon^{ik} x_{12}^k G_2(x_{12}^2, s)  + x_{12}^i G_1(x_{12}^2, s) 
    \Big)  \Bigg]
\\ \notag & \hspace{0.5cm} 
    \times  \left( \partial^i_{\un 1} - ip^i \right)
     \left[ \frac{\hat{k} \times \un x_{12}}{x_{12}} \Phi^{[2]}_{\mathrm{LT}}(\un x_{12}, \un x_{1^\prime 2^\prime}, z) 
    + 
\frac{\hat{k} \times \un x_{1^\prime 2^\prime}}{x_{1^\prime 2^\prime}} \Phi^{[2]}_{\mathrm{LT}}(\un x_{1^\prime 2^\prime}, \un x_{1 2}, z)
\right]
     \Bigg\} + \mathcal{O}(\Delta_\perp^2),
\end{align}
\end{subequations}
\end{tcolorbox}   
where the wave function overlaps, as defined above in \eqs{type1_overlaps} and (\ref{type2_overlaps}), are
\begin{subequations}
    \begin{align}
        \Phi^{[1]}_{\mathrm{TT}}(\un x_{12}, \un x_{1'2'}, z) &= \frac{2 \alpha_{EM} Z_f^2 N_c^2 Q^2}{\pi} z^2 (1-z)^2 (1-2\,z) \, \frac{\un x_{12} \cdot \un x_{1'2'}}{ x_{12} x_{1'2'}} K_1 \left(x_{12} Q \sqrt{z (1-z)} \right)
         K_1 \left(x_{1'2'} Q \sqrt{z (1-z)} \right),
         \\ 
         \Phi^{[2]}_{\mathrm{TT}}(\un x_{12}, \un x_{1'2'}, z) &= \frac{2 \, \alpha_{EM} Z_f^2 N_c^2 Q^2}{\pi} z^2 (1-z)^2 \Big[ z^2 + (1-z)^2 \Big]\, \frac{\un x_{12} \times \un x_{1'2'}}{ x_{12} x_{1'2'}} K_1 \left(x_{12} Q \sqrt{z (1-z)} \right)
         K_1 \left(x_{1'2'} Q \sqrt{z (1-z)} \right),
         \\ 
         \Phi^{[1]}_{\mathrm{LT}}(\un x_{12}, \un x_{1'2'}, z) &= -\frac{4 \alpha_{EM} Z_f^2 N_c^2 Q^2}{\pi} \big[ z (1-z) \big]^{5/2} K_1 \left(x_{12} Q \sqrt{z (1-z)} \right)
         K_0 \left(x_{1'2'} Q \sqrt{z (1-z)} \right),
         \\ 
        \Phi^{[2]}_{\mathrm{LT}}(\un x_{12}, \un x_{1'2'}, z) &= -\frac{4 \alpha_{EM} Z_f^2 N_c^2 Q^2}{\pi} \big[ z (1-z) \big]^{5/2} (1-2\,z) \, K_1 \left(x_{12} Q \sqrt{z (1-z)} \right)
         K_0 \left(x_{1'2'} Q \sqrt{z (1-z)} \right).
    \end{align}
\end{subequations}
The terms in Eqs.~\eqref{integrated_res} above containing the derivatives with respect to $\un x_1$ can be computed with the help of Eq.~(56) from \cite{Cougoulic:2022gbk}, such that
\begin{subequations}\label{derivatives}
    \begin{align}
        &\left( \partial^i_{\un 1} - i p^i \right)  \Phi_{\mathrm{TT}}^{[2]}(\un x_{12}, \un x_{1'2'}, z) = 
        \frac{2 \alpha_{EM} Z_f^2 N_c^2 Q^2}{\pi} z^2 (1-z)^2 \Big[z^2 + (1-z)^2 \Big]
        \\ \notag & \hspace{4cm}
        \times \Bigg[ \left( - \frac{\epsilon^{ik} x_{1'2'}^k}{x_{12} x_{1'2'}} + \frac{2 x_{12}^i \, \un x_{12} \times \un x_{1'2'}}{x^3_{12} x_{1'2'}} - i p^i \frac{\un x_{12} \times \un x_{1'2'}}{ x_{12} x_{1'2'}} 
        \right) K_1 \left(x_{12} Q \sqrt{z (1-z)} \right)
        \\ \notag & \hspace{5cm}
        + \frac{x_{12}^i \, \un x_{12} \times \un x_{1'2'}}{x^2_{12} x_{1'2'}} Q \sqrt{z(1-z)} \, K_0 \left(x_{12} Q \sqrt{z (1-z)} \right)
        \Bigg] K_1 \left(x_{1'2'} Q \sqrt{z (1-z)} \right),
        \\ 
&\left( \partial^i_{\un 1} - ip^i \right)
     \Bigg[ \frac{\hat{k} \times \un x_{12}}{x_{12}} \Phi^{[2]}_{\mathrm{LT}}(\un x_{12}, \un x_{1^\prime 2^\prime}, z) 
    + 
\frac{\hat{k} \times \un x_{1^\prime 2^\prime}}{x_{1^\prime 2^\prime}} \Phi^{[2]}_{\mathrm{LT}}(\un x_{1^\prime 2^\prime}, \un x_{1 2}, z)
\Bigg] = -\frac{4 \alpha_{EM} Z_f^2 N_c^2 Q^2}{\pi} \big[ z (1-z) \big]^{5/2} (1-2\,z) \notag 
\\ & \hspace{2cm}
\times \Bigg\{
\left( - \frac{\epsilon^{ji} \hat{k}^j}{x_{12}} + 
\frac{2 \, x_{12}^i \, \hat{k} \times \un x_{12} }{x_{12}^3} 
-i p^i \, \frac{\hat{k} \times \un x_{12} }{x_{12}} 
\right) K_1 \left(x_{12} Q \sqrt{z (1-z)} \right)
K_0 \left(x_{1'2'} Q \sqrt{z (1-z)} \right)
\\ \notag & \hspace{3cm}
+ \frac{x_{12}^i \, \hat{k} \times \un x_{12} }{x_{12}^2} Q \sqrt{z(1-z)}  K_0 \left(x_{12} Q \sqrt{z (1-z)} \right) K_0 \left(x_{1'2'} Q \sqrt{z (1-z)} \right) 
\\ \notag & \hspace{3cm}
+  \frac{x_{12}^i \, \hat{k} \times \un x_{1'2'} }{x_{12} x_{1'2'}} Q \sqrt{z(1-z)} K_1 \left(x_{12} Q \sqrt{z (1-z)} \right)
K_1 \left(x_{1'2'} Q \sqrt{z (1-z)} \right)
\\ \notag & \hspace{3cm}
-i p^i \, \frac{\hat{k} \times \un x_{1'2'} }{x_{1'2'}} 
K_0 \left(x_{12} Q \sqrt{z (1-z)} \right)
K_1 \left(x_{1'2'} Q \sqrt{z (1-z)} \right)
\Bigg\}.
    \end{align}
\end{subequations}
Note that in \eqs{derivatives} we have neglected $\delta^2 (\un x_{12})$, since it does not contribute to the elastic dijet production considered here (see above).

Equations~\eqref{integrated_res}, along with the numerator of the DSA in \eq{DSA} are the main results of this work. (Note, again, that \eq{DSA} was derived in the frame where the virtual photon's transverse momentum is zero, $q_\perp =0$.) We observe that the moment amplitudes $I_3, I_4$, and $I_5$ can, in principle, be extracted from the experimental data by employing the two expressions in \eqs{integrated_res} and varying the angles between $\un p$, $\un \Delta$, and $\un k$, along with $z$. Using \eqs{PDF+OAM_summ}, we can use $I_4$ and $I_5$ to construct the gluon OAM distribution $L_G (x, Q^2)$. Further, \eq{Itilde_evol} would allow us to construct the moment-amplitude $\widetilde I$ (modulo an uncertainty in the inhomogeneous term of \eqref{Itilde_evol}, which is hopefully numerically insignificant at low $x$): this, in turn, would allow us to obtain the quark OAM distribution $L_{q + \bar q} (x, Q^2)$.

Upon inspecting \eqs{integrated_res}, one notices the appearance of dipole amplitudes and moment amplitudes not related to helicity PDFs or OAM distributions: $G_1, J_3, J_4, J_5$. At first glance, it seems that these dipole amplitudes would contaminate the signal for our moment amplitudes, further complicating any phenomenological study. However, let us consider the angular structure of \eqs{integrated_res}. After the $\un x_{12}$ and $\un x_{1'2'}$ integrals are carried out, there are two remaining transverse vectors in \eq{TT_result} and three in \eq{TL_result}: these are $\{\un p, \un \Delta\}$ and $\{\un p, \un \Delta, \un k\}$, respectively. Furthermore, in \eq{TT_result}, the terms containing the polarized amplitudes $Q$ and $G_2$ and the moment amplitudes $I_3, I_4$ and $I_5$ appear multiplied by either zero or two Levi-Civita symbols $\epsilon^{ij}$. Therefore these terms, after all the integrals are done, will be proportional to $\un p \cdot \un \Delta$. Conversely, the irrelevant (for OAM and hPDF studies) terms, containing $G_1, J_3, J_4, J_5$ have exactly one $\epsilon^{ij}$ each in \eq{TT_result} and will lead to a term proportional to $\un p \times \un \Delta$ after integrating. We conclude the angular dependence of the moment amplitudes needed for OAM distributions in the TT terms of the DSA is $\un p \cdot \un \Delta$, which is, in principle, distinguishable from the $\un p \times \un \Delta$ of the terms we do not need. Similarly, in \eq{TL_result}, the relevant terms again contain zero or two $\epsilon^{ij}$'s, while the irrelevant come in with exactly one $\epsilon^{ij}$. The relevant angular structures are therefore $(\un p \cdot \un \Delta)\left( \un p \cdot \un k\right)$ and $(\un p \times \un \Delta)\left( \un p \times \un k\right)$  (there is only one factor of $\un \Delta$ and of $\un k$ in \eq{integrated_res}). Therefore, the signal of the moment amplitudes in the LT terms of the DSA has an angular dependence of either $(\un p \cdot \un \Delta)\left( \un p \cdot \un k\right)$ or $(\un p \times \un \Delta)\left( \un p \times \un k\right)$\footnote{There is an additional structure, $\un k \cdot \un \Delta$, that is possible here. However, it is related to the two structures in the main text via $(\un p \cdot \un \Delta)( \un p \cdot \un k) + (\un p \times \un \Delta)(\un p \times \un k) = p_\perp^2 (\un k \cdot \un \Delta)$.}.  This should be distinguishable from the angular structures of the irrelevant terms, $(\un p \times \un \Delta)\left( \un p \cdot \un k\right)$ and $(\un p \cdot \un \Delta)\left( \un p \times \un k\right)$\footnote{Again, there is another structure here,  $\un k \times \un \Delta$, which is related to the two structures mentioned in the text through \\ $(\un p \times \un \Delta)( \un p \cdot \un k) - (\un p \cdot \un \Delta)(\un p \times \un k) = p_\perp^2 (\un k \times \un \Delta)$.}. By performing the angular integrals over $\un x_{12}$ and $\un x_{1'2'}$, it is possible to identify exactly which linear combinations of dipole amplitudes and moment amplitudes accompany each of the three ``relevant" angular structures in \eqs{integrated_res} mentioned above ($\un p \cdot \un \Delta$, $(\un p \cdot \un \Delta)\left( \un p \cdot \un k\right)$, and $(\un p \times \un \Delta)\left( \un p \times \un k\right)$). We leave this for future work. Since there are three different relevant angular structures, one may be able to constrain the three moment amplitudes $I_3$, $I_4$, and $I_5$ needed for calculation of the OAM distributions. Isolating each of the relevant angular structures above will allow us to probe the moment amplitudes, and therefore the OAM distributions.

It appears that the moment amplitudes $J_3, J_4$, and $J_5$ could be zero for a longitudinally polarized proton target due to the following symmetry argument: suppose one can show that under a passive parity transformation, $Q_{{\un x}_1 , {\un x}_0} (s) \xrightarrow[]{\textrm{P}} Q_{-{\un x}_1 , -{\un x}_0} (s)$ and $G^i_{{\un x}_1 , {\un x}_0} (s) \xrightarrow[]{\textrm{P}} - G^i_{-{\un x}_1 , -{\un x}_0} (s)$, in addition to the $x^+ \leftrightarrow x^-$ interchange in all the operator definitions, for the terms in both amplitudes that couple to the longitudinal spin of the proton. Rewriting Eqs.~\eqref{moments} as 
\begin{subequations} \label{moments2}
    \begin{align}
        \int d^2 x_1 \, {\un y} \cdot {\un x}_1\, Q_{10}(s) &= {\un y} \cdot {\un x}_{10}\, I_3(x_{10}^2, s) + {\hat P} \cdot ({\vec y}_\perp \times {\vec x}_{10}) \, J_3(x_{10}^2, s), \label{Q_mom2}
        \\ \label{Gi_moms2}
        \int d^2 x_1 \, {\un y} \cdot {\un x}_1 \, z^i \, G^i_{10}(s) &= {\hat P} \cdot ({\vec y}_\perp \times {\vec z}_\perp) \, x_{10}^2 I_4(x_{10}^2, s) + 
        {\hat P} \cdot ({\vec y}_\perp \times {\vec x}_{10}) \, ({\un z} \cdot {\un x}_{10}) \, I_5(x_{10}^2, s) \\ & \hspace*{1cm} + 
        ({\un y} \cdot {\un z}) \, x_{10}^2 J_4(x_{10}^2, s) + ({\un y} \cdot {\un x}_{10}) \, ({\un z} \cdot {\un x}_{10}) \, J_5(x_{10}^2, s), \notag
    \end{align}
\end{subequations}
with $\un y$ and $\un z$ some arbitrary transverse position vectors, ${\vec y}_\perp = ({\un y}, 0)$ and ${\vec z}_\perp = ({\un z}, 0)$ their three-dimensional generalizations, $\hat P$ the unit vector in the direction of the proton's 3-momentum, and the vector product defined in three dimensions (with $\vec x_{10} = ({\un x}_{10}, 0)$), one can then show that the left-hand sides of Eqs.~\eqref{Q_mom2} and \eqref{Gi_moms2} are P even and P odd, respectively. Comparing this to the transformations of the right-hand sides of Eqs.~\eqref{moments2} under a passive parity transform, one concludes that $J_3 = J_4 = J_5 =0$. One can similarly show that $G_1=0$ for the longitudinally polarized proton. Alternatively, one can argue that the DSA should be P even: then, further noticing that each cross product in expressions \eqref{integrated_res} (and other expressions we have) is P odd, since it comes from a scalar triple product involving $\hat P$ (e.g., ${\un \Delta} \times {\un x}_{12} = {\hat P} \cdot ({\vec \Delta}_\perp \times {\vec x}_{12})$), one concludes that $G_1, J_3, J_4$, and $J_5$ cannot contribute to the DSA and are, therefore, zero. This conclusion is further reinforced by the above observation that the momentum structures resulting from the $G_1, J_3, J_4$, and $J_5$ terms are $\un p \times \un \Delta$, $(\un p \times \un \Delta)\left( \un p \cdot \un k\right)$, and $(\un p \cdot \un \Delta)\left( \un p \times \un k\right)$: these are all P odd and cannot contribute to DSA.

The relevant angular structures identified above for the TT and LT terms of the DSA should also be compared with the results from \cite{Bhattacharya:2022vvo, Bhattacharya:2024sck}. In \cite{Bhattacharya:2022vvo, Bhattacharya:2024sck}, the authors consider the interference between the longitudinally polarized and transversely polarized virtual photon amplitudes in the DSA. This corresponds to the LT term of the DSA we have considered above. The authors of \cite{Bhattacharya:2022vvo, Bhattacharya:2024sck} find that the angular structure which couples to the OAM distributions is proportional to $\un \Delta \cdot \un{k}$ (in our notation). As mentioned above, our result for the LT term of the DSA in \eq{TL_result} contains both angular structures 
$(\un p \cdot \un \Delta)\left( \un p \cdot \un k\right)$ and $(\un p \times \un \Delta)\left( \un p \times \un k\right)$. One can show from \eq{TL_result} that the coefficients of these two structures are not the same, so while we can extract the $\un \Delta \cdot \un{k} \propto (\un p \cdot \un \Delta)\left( \un p \cdot \un k\right) + (\un p \times \un \Delta)\left( \un p \times \un k\right)$ structure in agreement with \cite{Bhattacharya:2022vvo, Bhattacharya:2024sck}, we also obtain dependence on the moment amplitudes from a separate, independent angular structure (e.g., $(\un p \cdot \un \Delta)\left( \un p \cdot \un k\right) - (\un p \times \un \Delta)\left( \un p \times \un k\right)$). Additionally, although the TT term was not considered \cite{Bhattacharya:2022vvo, Bhattacharya:2024sck}, in that term we have found the angular structure $\sim \un p \cdot \un \Delta$ in the DSA that couples the OAM distributions as well. It is possible that this structure as well as the other structure we have identified in the LT term of the DSA which does not appear in the calculations of \cite{Bhattacharya:2022vvo, Bhattacharya:2024sck} may be suppressed at moderate or large values of $x$, or large values of $Q^2$, i.e., in the kinematic region considered in \cite{Bhattacharya:2022vvo, Bhattacharya:2024sck}. We leave such an analysis for future work. 

%%%%%%%%%%%%%%%%%%%%%%%%%%%%%%%%%%%%%%%%%%%%%%%%%%%%%%%%%%%%%%%%%%%%%%%%%%%%%%%
\section{Conclusions and outlook}

\label{sec:conclusions}

Let us summarize what we have accomplished here. We have derived the cross section for elastic dijet production for an electron scattering off of a longitudinally polarized proton target. We have studied both the DSA and SSA in terms of the virtual photon's polarization basis. Explicit expressions for the TT and LT components of the DSA numerator \eqref{DSA} are given in \eqs{DSA_TT_final} and (\ref{DSA_LT_final}), respectively. These expressions are written in terms of the polarized dipole amplitudes, $Q_{10}(s)$ and $G^i_{10}(s)$, defined above in \eqs{pdas}, and are valid to the leading (subeikonal) order and in DLA. Additionally, these expressions depend on the unpolarized dipole amplitude, $N_{10}$, defined in \eq{unpdas}, whose leading-order evolution is single-logarithmic \cite{Mueller:1994rr,Mueller:1994jq,Mueller:1995gb,Balitsky:1995ub,Balitsky:1998ya,Kovchegov:1999yj,Kovchegov:1999ua,Jalilian-Marian:1997dw,Jalilian-Marian:1997gr,Weigert:2000gi,Iancu:2001ad,Iancu:2000hn,Ferreiro:2001qy}. 

In addition, we have found the contributions to the numerator of the SSA from \eqref{SSA}. The LT component is given in \eqs{SSA_LT1res} and \eqref{SSA_LTres2}, while the LL, TT, and T,-T components are found in Appendix~\ref{appendix:SSA_channels}.

We showed that the expressions for the DSA contain leading-order dipole amplitudes, both polarized and unpolarized. By contrast, we also showed the SSA contains terms that have subleading (in $\as$) dipole amplitudes, both polarized and unpolarized. Moreover, separating the dipole amplitudes and moment amplitudes needed for hPDF and OAM calculations from the other amplitudes in the obtained SSA terms does not appear feasible at the moment. 

Following \cite{Hatta:2016aoc,Bhattacharya:2022vvo, Bhattacharya:2023hbq, Bhattacharya:2024sck}, we expanded our results for the DSA around the limit of zero transfer momentum transfer to the linear order in $\Delta_\perp = 0$. The resulting expressions are given in \eqs{integrated_res} and feature the impact-parameter integrated dipole amplitudes and their first moments, $Q(x_{10}^2, s), G_2(x_{10}^2, s), I_3(x_{10}^2,s), I_4(x_{10}^2, s), I_5(x_{10}^2, s)$, defined above in \eqs{integrated} and (\ref{moments}). These integrated dipole amplitudes determine the helicity PDFs and OAM distributions via \eqs{PDF+OAM_summ}. The integrated dipole amplitudes can be found using the DLA evolution equations constructed in \cite{Kovchegov:2015pbl, Kovchegov:2018znm,  Cougoulic:2022gbk, Borden:2024bxa}. Similarly, the moment amplitudes can be found using the DLA evolution constructed in \cite{Kovchegov:2023yzd}, whose simplified version is listed here in \eqs{Ievol} and \eqref{Gamma_evol}. 

Convoluting our expressions for the DSA-contributing cross sections \eqref{integrated_res} with the appropriate jet functions one would be able to make actual phenomenological predictions for the DSA in the elastic dijet production. Comparing such predictions with the future data, to be reported by the future EIC, may allow for experimental determination of the moment amplitudes and, via \eqs{PDF+OAM_summ}, the OAM distributions. If such an extraction of the moment amplitudes at the EIC is proven possible after a more detailed phenomenological work based on our \eqs{integrated_res}, these results will lead to the first-ever experimental signature of both the quark and the gluon OAM distributions in the small $x$ region. Once the EIC data becomes available, one may be able to use the Monte-Carlo Bayesian method pioneered in \cite{Adamiak:2021ppq} for polarized DIS at small $x$ and recently extended for polarized SIDIS in \cite{Adamiak:2023yhz} to extract the OAM distributions. Such an analysis would be the first-ever measurement of the quark and gluon OAM distributions and would allow for a complete description of all contributions to the proton spin at small $x$. 

%%%%%%%%%%%%%%%%%%%%%%%%%%%%%%%%%%%%%%%%%%%%%%%%%%%%%%%%%%%%%%%%%%%%%%%%%%%%%%%

\section*{Acknowledgments}

The authors are grateful to Elke Aschenauer, Ming Li, Alexei Prokudin, Anselm Vossen, and Feng Yuan  for informative discussions. 

This material is based upon work supported by
the U.S. Department of Energy, Office of Science, Office of Nuclear
Physics under Award Number DE-SC0004286 and within the framework of the Saturated Glue (SURGE) Topical
Theory Collaboration.
\\

%%%%%%%%%%%%%%%%%%%%%%%%%%%%%%%%%%%%%%%%%%%%%%%%%%%%%%%%%%%%%%%%%%%%%%%%%%%%%%%

\appendix

%%%%%%%%%%%%%%%%%%%%%%%%%%%%%%%%%%%%%%%%%%%%%%%%%%%%%%%%%%%%%%%%%%%%%%%%%%%%%%%

\section{Streamlined large-$N_c$ moment amplitude evolution equations}
\label{app:updated_evolution}

Here we rewrite the large-$N_c$ evolution equations derived in \cite{Kovchegov:2023yzd} in terms of the moment amplitudes defined in \eqs{moments}. To do so, we start with Eq.~(52) of \cite{Kovchegov:2023yzd}. Forming the linear combinations on the right-hand side of \eqs{translation}, we obtain 
\begin{align}\label{Ievol}
    \begin{pmatrix}
    I_3 \\ 
    I_4 \\
    I_5 
    \end{pmatrix}(x_{10}^2, zs) &= \begin{pmatrix}
    I_3^{(0)} \\ 
    I_4^{(0)} \\
    I_5^{(0)} 
    \end{pmatrix} (x_{10}^2, zs)  
    + \frac{\as N_c}{4\pi} \iUV \,  \begin{pmatrix}
    2\,\Gamma_3 - 4\,\Gamma_4 + 2 \, \Gamma_5  -2 \, \Gamma_2 \\ 
    0 \\
    0 
    \end{pmatrix}(x_{10}^2, x_{21}^2, z^\prime s) \\\notag &+ \frac{\as N_c}{4\pi} \iIR \, \begin{pmatrix}
        4 & -4 & 2 &  -4 &  - 6 \\
        0 & 4 & 2 & -2 &  -3 \\
        -2 & 2 & -1 & 4 &  7 \\
    \end{pmatrix} 
    \begin{pmatrix}
    I_3 \\ 
    I_4 \\
    I_5 \\
    G \\
    G_2
    \end{pmatrix} (x_{21}^2, z^\prime s),
\end{align}
where now the moments, $I_4, I_5$, are defined according to \eqs{moments}. Here, the $\Gamma$ amplitudes are auxiliary functions needed to enforce lifetime ordering in the evolution (see \cite{Kovchegov:2015pbl} for details).  

Performing an analogous procedure for the neighbor moment evolution equations in Eq.~(53) of \cite{Kovchegov:2023yzd}, we arrive at 
\begin{align}\label{Gamma_evol}
    &\begin{pmatrix}
    \Gamma_3 \\ 
    \Gamma_4 \\
    \Gamma_5 
    \end{pmatrix}(x_{10}^2, x_{21}^2, z^\prime s)  = \begin{pmatrix}
    I_3^{(0)} \\ 
    I_4^{(0)} \\
    I_5^{(0)} 
    \end{pmatrix} (x_{10}^2, z^\prime s)  \\ \notag 
    & 
    + \frac{\as N_c}{4\pi} \inUV \, \begin{pmatrix}
    2\,\Gamma_3 - 4\,\Gamma_4 + 2 \, \Gamma_5 -2 \, \Gamma_2 \\ 
    0 \\
    0 
    \end{pmatrix}(x_{10}^2, x_{32}^2, z^{\prime \prime} s)
    \\\notag &
    + \frac{\as N_c}{4\pi} \inIR \, \begin{pmatrix}
        4 & -4 & 2 &  -4 &  - 6 \\
        0 & 4 & 2 &  -2 &  -3 \\
        -2 & 2 & -1 &  4 &  7 \\
    \end{pmatrix} 
    \begin{pmatrix}
    I_3 \\ 
    I_4 \\
    I_5 \\
    G \\ 
    G_2
    \end{pmatrix} (x_{32}^2, z^{\prime \prime} s),
\end{align}
where the neighbor moments are defined in Eqs.~(47) of \cite{Kovchegov:2023yzd}. Note that, similar to \eq{translation},  the neighbor dipole amplitudes $\Gamma_4, \Gamma_5$ used here are related to those defined in \cite{Kovchegov:2023yzd} by
\begin{subequations} \label{translation_Gamma}
    \begin{align}
        \Gamma_4^{\mathrm{here}} (x_{10}^2, x_{21}^2, z s)  &= \Gamma_4^{\mathrm{KM}}(x_{10}^2, x_{21}^2, z s) - \Gamma_6^{\mathrm{KM}}(x_{10}^2, x_{21}^2, z s), 
        \\ 
         \Gamma_5^{\mathrm{here}}(x_{10}^2, x_{21}^2, z s) &= \Gamma_5^{\mathrm{KM}}(x_{10}^2, x_{21}^2, z s) + \Gamma_6^{\mathrm{KM}}(x_{10}^2, x_{21}^2, z s) . 
    \end{align}
\end{subequations}

%%%%%%%%%%%%%%%%%%%%%%%%%%%%%%%%%%%%%%%%%%%%%%%%%%%%%%%%%%%%%%%%%%%%%%%%%%%%%%%

\section{LL, TT, and T,-T channels in the single spin asymmetry}
\label{appendix:SSA_channels}

Here we discuss the remaining terms in the SSA from \eq{SSA} not calculated in the main text. Namely, we will consider here the three structures 
\begin{align} \label{other_structs}
    \sigma^{\gamma^*p}_{00}, \hspace{0.25cm}
    \sum_{\lambda = \pm 1} \sigma^{\gamma^*p}_{\lambda \lambda},
    \hspace{0.25cm} 
e^{-2i\phi} \sigma^{\gamma^*p}_{1,-1} +\mathrm{c.c.}
\end{align}
which enter \eq{SSA}. We will show that these terms are also suppressed by factors of $\as$ compared to the DSA as was shown above in Section \ref{sec:ssa}. 

%%%%%%%%%%%%%%%%%%%%%%%%%%%
\subsection{LL terms}

We begin with the LL term, which is the first term in \eq{other_structs}. Using \eq{LL1}, we see that the type-1 terms vanish here. Therefore, we consider the type-2 terms only. Again, using PT-symmetry arguments one can show that $V^{\mathrm{q}[2]}_{\un x}$ operator does not contribute to the LL term in question. Moving on then to the contribution from $V^{\mathrm{G}[2]}_{\un x, y}$, we use \eqs{XS10} and (\ref{LL2}), setting $\lambda = \lambda' = 0$, to write 
\begin{align}\label{SSA_LL1}
& z (1-z)  \, \frac{1}{2} \, \sum_{S_L} S_L  \frac{d\sigma_{00}^{\textrm{G} [2] \gamma^* p \to q {\bar q} p'}}{d^2 p_1 \, d^2 p_2 \, d z} =  -\frac{1}{2 (2 \pi)^5} \, \int d^2 x_1 \, d^2 x_{1'} \, d^2 x_2 \, d^2 x_{2'} \, d^2 x_0 \, e^{- i {\un p}_1 \cdot {\un x}_{11'} - i {\un p}_2 \cdot {\un x}_{22'}}  \\ 
& \times \frac{1}{N_c} \Bigg\{ 
\Phi^{[2]}_{\mathrm{LL}}(\un x_{02}, \un x_{1^\prime 2^\prime}, z) 
\left\langle \tord \tr \left[ V^{\textrm{G} [2]}_{{\un 1}, {\un 0}} \, V_{{\un 2}}^\dagger \right] \right\rangle \big( zs \big) \, 
\Big[ 1- S_{1^\prime 2^\prime}^*(s)\Big]
\notag \\
& -  
\Phi^{[2]}_{\mathrm{LL}}(\un x_{10}, \un x_{1^\prime 2^\prime}, z) 
\left\langle \tord \tr \left[ V_{{\un 1}} \,  V^{\textrm{G} [2] \, \dagger}_{{\un 2}, {\un 0}} \right] \right\rangle \big((1-z)s \big)
\, \Big[ 1- S_{1^\prime 2^\prime}^*(s) \Big]
\notag \\
& +  \Phi^{[2]}_{\mathrm{LL}}(\un x_{12}, \un x_{0 2^\prime}, z) 
\Big[ 1- S_{12}(s)\Big]
\, \left\langle \atord \tr \left[ V_{{\un 2'}} \,  V^{\textrm{G} [2] \, \dagger}_{{\un 1'}, {\un 0}}  \right]  \right\rangle \big( zs \big) \notag \\
& - \Phi^{[2]}_{\mathrm{LL}}(\un x_{12}, \un x_{1^\prime 0}, z) 
\Big[ 1- S_{12}(s)\Big]
\, \left\langle \atord \tr \left[  V^{\textrm{G} [2]}_{{\un 2'}, {\un 0}} \, V_{{\un 1'}}^\dagger \right]  \right\rangle \big((1-z)s \big)   \Bigg\} . \notag
\end{align}
Using $\Phi^{[2]}_{\mathrm{LL}}(\un x, \un x^\prime, z) = \Phi^{[2]}_{\mathrm{LL}}(\un x', \un x, z)$, we relabel $\un x_1 \leftrightarrow \un x_{1'}, \un x_2 \leftrightarrow \un x_{2'}$ in the last two terms of \eq{SSA_LL1}, again taking $\un p_1, \un p_2 \to - \un p_1,- \un p_2$, to obtain 
\begin{align}\label{SSA_LL2}
& z (1-z)  \, \frac{1}{2} \, \sum_{S_L} S_L  \frac{d\sigma_{00}^{\textrm{G} [2] \gamma^* p \to q {\bar q} p'}}{d^2 p_1 \, d^2 p_2 \, d z} =  -\frac{1}{2 (2 \pi)^5} \, \int d^2 x_1 \, d^2 x_{1'} \, d^2 x_2 \, d^2 x_{2'} \, d^2 x_0 \, e^{- i {\un p}_1 \cdot {\un x}_{11'} - i {\un p}_2 \cdot {\un x}_{22'}}  \\ 
& \times \frac{1}{N_c} \Bigg\{ 
\Phi^{[2]}_{\mathrm{LL}}(\un x_{02}, \un x_{1^\prime 2^\prime}, z) 
\Bigg[ 
\left\langle \tord \tr \left[ V^{\textrm{G} [2]}_{{\un 1}, {\un 0}} \, V_{{\un 2}}^\dagger \right] \right\rangle \big( zs \big) \, 
\Big[ 1- S_{1^\prime 2^\prime}^*(s)\Big]
 + \mathrm{c.c.}
\Bigg]
\notag \\
& -  
\Phi^{[2]}_{\mathrm{LL}}(\un x_{10}, \un x_{1^\prime 2^\prime}, z) 
\Bigg[
\left\langle \tord \tr \left[ V_{{\un 1}} \,  V^{\textrm{G} [2] \, \dagger}_{{\un 2}, {\un 0}} \right] \right\rangle \big((1-z)s \big)
\, \Big[ 1- S_{1^\prime 2^\prime}^*(s) \Big]
+ \mathrm{c.c.}
\Bigg]
\Bigg\}. \notag
\end{align}
To proceed, we use \eqs{VG2sub} while noting that the remaining terms on the right-hand side of \eqs{VG2sub} vanish for the PT-odd observable at hand (SSA). The result is 
\begin{align}\label{SSA_LL3}
 z (1-z) \, \frac{1}{2} \, \sum_{S_L} \,  S_L  \, \frac{d\sigma_{00}^{\mathrm{G}[2] \gamma^* p \to q {\bar q} p'}}{d^2 p_1 \, d^2 p_2 \, d z} &=  
 - 
 \frac{1}{2  (2 \pi)^5 s} \, \int d^2 x_1 \, d^2 x_{1'} \, d^2 x_2 \, d^2 x_{2'} \,  e^{- i {\un p}_1 \cdot {\un x}_{11'} - i {\un p}_2 \cdot {\un x}_{22'}}  
 \\ \notag 
 & \hspace{-4cm}
 \times 
 \frac{1}{N_c}  \Bigg\{
\left[
 \frac{i}{z} 
\llangle \tord \tr \left[ V^{i\textrm{G} [2]}_{{\un 1}} \, V_{{\un 2}}^\dagger \right] \rrangle (s)
\Big( 1- S_{1^\prime 2^\prime}^*(s)\Big) + \mathrm{c.c.} \right]
\Big( \partial^i_{\un 1} - i p_1^i \Big) \Phi^{[2]}_{\mathrm{LL}}(\un x_{12}, \un x_{1^\prime 2^\prime}, z)
\notag \\ & \hspace{-3cm}
 + 
\left[
\frac{i}{1-z}
\llangle \tord \tr \left[ V_{{\un 1}} \,  V^{i\textrm{G} [2] \, \dagger}_{{\un 2}} \right] \rrangle (s)
\, \Big( 1- S_{1^\prime 2^\prime}^*(s) \Big) + \mathrm{c.c.}
\right]
\Big( \partial^i_{\un 2} - i p_2^i \Big) \Phi^{[2]}_{\mathrm{LL}}(\un x_{12}, \un x_{1^\prime 2^\prime}, z)
 \Bigg\} .\notag
\end{align}

Symmetrizing with respect to the quark-antiquark jet interchange and utilizing the amplitudes defined in \eqs{pdas}, (\ref{unpdas}) and (\ref{ssa_amps_2}), we find
\begin{align} \label{SSA_LL4}
    z (1-z) \, \frac{1}{2} \, \sum_{S_L} \, S_L  \, \frac{d\sigma_{\mathrm{symm.}\,00}^{\mathrm{G}[2] \, \gamma^* p \to q {\bar q} p'}}{d^2 p \, d^2 \Delta \, d z}
 &= 
  -
 \frac{1}{(2\pi)^5 s}
        \int d^2 x_1 \, d^2 x_{1'} \, d^2 x_2 \, d^2 x_{2'} \,  
        e^{- i {\un p} \cdot ({\un x}_{12} - \un x_{1'2'}) - i \un \Delta \cdot \big[ z \, \un x_{11'} + (1-z) \un x_{22'} \big] } 
    \\ \notag & \hspace{-4cm}
\times \Bigg\{
\Bigg[ 
    \left( \frac{1}{z} G^{i}_{12}(s) - \frac{1}{1-z} G^{i}_{21}(s) \right) \textrm{Im} \left[ N_{1^\prime 2^\prime}(s) \right] - \left( \frac{1}{z} G^{i, \mathrm{Im}}_{12}(s) - \frac{1}{1-z} G^{i, \mathrm{Im}}_{21}(s) \right) \textrm{Re}\left[ N_{1^\prime 2^\prime}(s) \right] 
    \\ \notag & \hspace{-2cm}
- \left( \frac{1}{z} G^{i,\mathrm{NS}}_{12}(s) + \frac{1}{1-z} G^{i,\mathrm{NS}}_{21}(s)
 \right) O_{1^\prime 2^\prime}(s)  \Bigg]
 \left(\partial_{\un 1}^i - i p^i  \right)
        \Phi^{[2]}_{\mathrm{LL}}(\un x_{12}, \un x_{1^\prime 2^\prime}, z)
        \\ \notag & \hspace{-3cm}
-\Bigg[ 
    \left( G^{i}_{12}(s) +  G^{i}_{21}(s) \right) \textrm{Im} \left[ N_{1^\prime 2^\prime}(s) \right] - \left( G^{i, \mathrm{Im}}_{12}(s) +  G^{i, \mathrm{Im}}_{21}(s) \right) \textrm{Re}\left[ N_{1^\prime 2^\prime}(s) \right] 
    \\ \notag & \hspace{-1cm}
 -  \left(  G^{i,\mathrm{NS}}_{12}(s) - G^{i,\mathrm{NS}}_{21}(s)
 \right) O_{1^\prime 2^\prime}(s)  \Bigg]
 \left( i \Delta^i  \right)
        \Phi^{[2]}_{\mathrm{LL}}(\un x_{12}, \un x_{1^\prime 2^\prime}, z)
         \Bigg\}.
\end{align}
Just as the LT term of the SSA considered in the main text, \eq{SSA_LL4} contains only terms suppressed by a factor of $\as$ compared to the DSA. While the polarized dipole amplitudes $G^i$ do enter \eq{SSA_LL4}, it appears very hard at this point to separate them from other terms, which are not needed for the hPDF and OAM distributions calculations. Therefore, at the moment, the LL term of the SSA does not appear to be a good probe for the moment amplitudes and the polarized dipole amplitudes. 

%%%%%%%%%%%%%%%%%%%%%%%%%%%
\subsection{TT terms}
 
Next, we consider the TT term in the SSA from \eq{SSA}, 
\begin{align}
    \sum_{\lambda = \pm 1} \sigma^{\gamma^*p}_{\lambda \lambda}.
\end{align}
Starting with the type-1 terms, we again use \eqs{XS5} and (\ref{type1_overlaps}) to write 
\begin{align}\label{SSA_TT1}
 z (1-z) \, \frac{1}{2}  \sum_{S_L, \lambda \pm 1} S_L \,  \frac{d\sigma_{\lambda \lambda}^{[1] \, \gamma^* p \to q {\bar q} p'}}{d^2 p_1 \, d^2 p_2 \, d z} &= -\frac{ i}{ (2 \pi)^5 s} \, \int d^2 x_1 \, d^2 x_{1'} \, d^2 x_2 \, d^2 x_{2'} \,  e^{- i {\un p}_1 \cdot {\un x}_{11'} - i {\un p}_2 \cdot {\un x}_{22'}}  \Phi_{\mathrm{TT}}^{'[1]}(\un x_{12}, \un x_{1^\prime 2^\prime}, z)
 \\ \notag 
 & \hspace{-2cm}
 \times 
 \frac{1}{N_c}  \Bigg\{ 
 \left[ \frac{1}{z} \llangle \tord \tr \left[ V^{\textrm{pol}[1]}_{{\un 1}} \, V_{{\un 2}}^\dagger \right] \rrangle
(s)
 -
 \frac{1}{1-z} \llangle \tord \tr \left[ V_{{\un 1}} \,  V^{\textrm{pol} [1] \, \dagger}_{{\un 2}} \right] \rrangle (s)
 \right] \, 
 \Big[
  1 - S^*_{ 1^\prime 2^\prime}(s)
 \Big]
 \\ \notag 
 & \hspace{-2cm} +  
 \left[ \frac{1}{z} \llangle \atord \tr \left[ V_{\un 2^\prime} V^{\textrm{pol}[1]\dagger}_{{\un 1^\prime}} \right] \rrangle (s) 
 -
 \frac{1}{1-z}
\llangle \atord \tr \left[ V^{\textrm{pol} [1]}_{{\un 2^\prime}} V_{{\un 1^\prime}}^\dagger  \right] \rrangle
 (s)
 \right] \, 
 \Big[
  1 - S_{1  2}(s)
 \Big]
 \Bigg\} . \notag
\end{align}
In the second term in the curly brackets of \eq{SSA_TT1} we replace $\un x_1 \leftrightarrow \un x_{1'}, \un x_2 \leftrightarrow \un x_{2'}$, along with $\un p_1, \un p_2 \to - \un p_1,- \un p_2$. We end up with 
\begin{align}\label{SSA_TT2}
 z (1-z) \, \frac{1}{2}  \sum_{S_L, \lambda \pm 1} S_L \,  \frac{d\sigma_{\lambda \lambda}^{[1] \, \gamma^* p \to q {\bar q} p'}}{d^2 p_1 \, d^2 p_2 \, d z} &= -\frac{ i}{ (2 \pi)^5 s} \, \int d^2 x_1 \, d^2 x_{1'} \, d^2 x_2 \, d^2 x_{2'} \,  e^{- i {\un p}_1 \cdot {\un x}_{11'} - i {\un p}_2 \cdot {\un x}_{22'}}  \Phi_{\mathrm{TT}}^{'[1]}(\un x_{12}, \un x_{1^\prime 2^\prime}, z)
 \\ \notag 
 & \hspace{-2cm}
 \times 
 \frac{1}{N_c}  \Bigg\{ 
 \left[ \frac{1}{z} \llangle \tord \tr \left[ V^{\textrm{pol}[1]}_{{\un 1}} \, V_{{\un 2}}^\dagger \right] \rrangle
(s)
 -
 \frac{1}{1-z} \llangle \tord \tr \left[ V_{{\un 1}} \,  V^{\textrm{pol} [1] \, \dagger}_{{\un 2}} \right] \rrangle (s)
 \right] \, 
 \Big[
  1 - S^*_{ 1^\prime 2^\prime}(s)
 \Big]
    - \mathrm{c.c.}
 \Bigg\},\notag
\end{align}
where we have used $\Phi^{'[1]}(\un x, \un x^\prime, z) = - \Phi^{'[1]}(\un {x}^\prime, \un x, z)$. Symmetrizing with respect to the quark-antiquark jet exchange and using \eqs{pdas}, (\ref{unpdas}) and (\ref{ssa_amps}), we end up with 
\begin{align} \label{SSA_TT1res}
    &z (1-z) \, \frac{1}{2} \sum_{S_L, \lambda = \pm 1} S_L  \,  \frac{d\sigma_{\mathrm{symm.}\,\lambda \lambda }^{[1] \, \gamma^* p \to q {\bar q} p'}}{d^2 p \, d^2 \Delta \, d z}   
 = \frac{2}{ (2\pi)^5s}
        \int d^2 x_1 \, d^2 x_{1'} \, d^2 x_2 \, d^2 x_{2'} \,  
        e^{- i {\un p} \cdot ({\un x}_{12} - \un x_{1'2'}) - i \un \Delta \cdot \big[ z \, \un x_{11'} + (1-z) \un x_{22'} \big] }
    \\ \notag & \hspace{2cm}
\times \Bigg\{
        - \left[ \frac{1}{z} Q_{12}(s) - \frac{1}{1-z} Q_{21}(s) \right] \textrm{Im}\left[ N_{1^\prime 2^\prime}(s) \right]
             + \left[ \frac{1}{z} Q^{\mathrm{Im}}_{12}(s) - \frac{1}{1-z} Q^{\mathrm{Im}}_{21}(s) \right] \textrm{Re}~\left[ N_{1^\prime 2^\prime}(s) \right]
    \\ \notag & \hspace{3cm}
 +  \left[ \frac{1}{z} Q^{\mathrm{NS}}_{12}(s) + \frac{1}{1-z} Q^{\mathrm{NS}}_{21}(s)
 \right] O_{1^\prime 2^\prime}(s)  
 \Bigg\}\Phi^{\prime[1]}_{\mathrm{TT}}(\un x_{12}, \un x_{1^\prime 2^\prime} , z).
\end{align}
We see that all the terms appearing in \eq{SSA_TT1res} are $\as$-suppressed compared to the DSA. More importantly, we again see a mix of OAM- and hPDF-related dipole amplitudes with the amplitudes unrelated to those quantities while unexplored in theory and phenomenology. Once again, separation of those terms appears to be impractical at this time. 

Proceeding to the terms containing $V^{\mathrm{G}[2]}_{\un x, \un y}$, we use \eqs{XS10} and (\ref{type2_overlaps}) to write 
\begin{align}\label{SSA_TT3}
& z (1-z) \, \half \sum_{S_L, \lambda = \pm 1} S_L  \, \frac{d\sigma_{\lambda \lambda }^{\textrm{G} [2] \gamma^* p \to q {\bar q} p'}}{d^2 p_1 \, d^2 p_2 \, d z} =  -\frac{1}{(2 \pi)^5} \, \int d^2 x_1 \, d^2 x_{1'} \, d^2 x_2 \, d^2 x_{2'} \, d^2 x_0 \, e^{- i {\un p}_1 \cdot {\un x}_{11'} - i {\un p}_2 \cdot {\un x}_{22'}}  \\ 
& \times \frac{1}{N_c} \Bigg\{ 
\Phi^{'[2]}_{\mathrm{TT}}(\un x_{02}, \un x_{1^\prime 2^\prime}, z) 
\left\langle \tord \tr \left[ V^{\textrm{G} [2]}_{{\un 1}, {\un 0}} \, V_{{\un 2}}^\dagger \right] \right\rangle \big( zs \big) \, 
\Big[ 1- S_{1^\prime 2^\prime}^*(s)\Big]
\notag \\
& -  
\Phi^{'[2]}_{\mathrm{TT}}(\un x_{10}, \un x_{1^\prime 2^\prime}, z) 
\left\langle \tord \tr \left[ V_{{\un 1}} \,  V^{\textrm{G} [2] \, \dagger}_{{\un 2}, {\un 0}} \right] \right\rangle \big((1-z)s \big)
\, \Big[ 1- S_{1^\prime 2^\prime}^*(s) \Big]
\notag \\
& +  \Phi^{'[2]}_{\mathrm{TT}}(\un x_{12}, \un x_{0 2^\prime}, z) 
\Big[ 1- S_{12}(s)\Big]
\, \left\langle \atord \tr \left[ V_{{\un 2'}} \,  V^{\textrm{G} [2] \, \dagger}_{{\un 1'}, {\un 0}}  \right]  \right\rangle \big( zs \big) \notag \\
& - \Phi^{'[2]}_{\mathrm{TT}}(\un x_{12}, \un x_{1^\prime 0}, z) 
\Big[ 1- S_{12}(s)\Big]
\, \left\langle \atord \tr \left[  V^{\textrm{G} [2]}_{{\un 2'}, {\un 0}} \, V_{{\un 1'}}^\dagger \right]  \right\rangle \big((1-z)s \big)   \Bigg\}.\notag
\end{align}
We use $\Phi^{'[2]}_{\mathrm{TT}}(\un x, \un x^\prime, z) = \Phi^{'[2]}_{\mathrm{TT}}(\un x', \un x, z)$, and relabel $\un x_1 \leftrightarrow \un x_{1'}, \un x_2 \leftrightarrow \un x_{2'}$, taking $\un p_1, \un p_2 \to - \un p_1,- \un p_2$ in the last two terms in the curly brackets. We get 
\begin{align}\label{SSA_TT4}
& z (1-z) \, \half \sum_{S_L, \lambda = \pm 1} S_L  \,\frac{d\sigma_{\lambda \lambda}^{\textrm{G} [2] \gamma^* p \to q {\bar q} p'}}{d^2 p_1 \, d^2 p_2 \, d z} =  -\frac{1}{(2 \pi)^5} \, \int d^2 x_1 \, d^2 x_{1'} \, d^2 x_2 \, d^2 x_{2'} \, d^2 x_0 \, e^{- i {\un p}_1 \cdot {\un x}_{11'} - i {\un p}_2 \cdot {\un x}_{22'}}  \\ 
& \times \frac{1}{N_c} \Bigg\{ 
\Phi^{'[2]}_{\mathrm{TT}}(\un x_{02}, \un x_{1^\prime 2^\prime}, z) 
\Bigg[ 
\left\langle \tord \tr \left[ V^{\textrm{G} [2]}_{{\un 1}, {\un 0}} \, V_{{\un 2}}^\dagger \right] \right\rangle \big( zs \big) \, 
\Big[ 1- S_{1^\prime 2^\prime}^*(s)\Big]
 + \mathrm{c.c.}
\Bigg]
\notag \\
& -  
\Phi^{'[2]}_{\mathrm{TT}}(\un x_{10}, \un x_{1^\prime 2^\prime}, z) 
\Bigg[
\left\langle \tord \tr \left[ V_{{\un 1}} \,  V^{\textrm{G} [2] \, \dagger}_{{\un 2}, {\un 0}} \right] \right\rangle \big((1-z)s \big)
\, \Big[ 1- S_{1^\prime 2^\prime}^*(s) \Big]
+ \mathrm{c.c.}
\Bigg]
\Bigg\}. \notag 
\end{align}
Using the simplification in \eqs{VG2sub}, we end up with 
\begin{align} \label{SSA_TT5}
    z (1-z) \, \frac{1}{2} \, \sum_{S_L, \lambda = \pm1} \!\! S_L  \, \frac{d\sigma_{\lambda \lambda}^{\mathrm{G}[2] \gamma^* p \to q {\bar q} p'}}{d^2 p_1 \, d^2 p_2 \, d z} &=  -\frac{1}{(2 \pi)^5 s} \, \int d^2 x_1 \, d^2 x_{1'} \, d^2 x_2 \, d^2 x_{2'} \,  e^{- i {\un p}_1 \cdot {\un x}_{11'} - i {\un p}_2 \cdot {\un x}_{22'}}  
 \\ \notag 
 & \hspace{-4cm}
 \times 
 \frac{1}{N_c}  \Bigg\{
\left[
 \frac{i}{z} 
\llangle \tord \tr \left[ V^{i\textrm{G} [2]}_{{\un 1}} \, V_{{\un 2}}^\dagger \right] \rrangle (s)
\Big( 1- S_{1^\prime 2^\prime}^*(s)\Big) + \mathrm{c.c.} \right]
\Big( \partial^i_{\un 1} - i p_1^i \Big) \Phi^{'[2]}_{\mathrm{TT}}(\un x_{12}, \un x_{1^\prime 2^\prime}, z)
\notag \\ & \hspace{-3cm}
 + 
\left[
\frac{i}{1-z}
\llangle \tord \tr \left[ V_{{\un 1}} \,  V^{i\textrm{G} [2] \, \dagger}_{{\un 2}} \right] \rrangle (s)
\, \Big( 1- S_{1^\prime 2^\prime}^*(s) \Big) + \mathrm{c.c.}
\right]
\Big( \partial^i_{\un 2} - i p_2^i \Big) \Phi^{'[2]}_{\mathrm{TT}}(\un x_{12}, \un x_{1^\prime 2^\prime}, z)
 \Bigg\} .\notag
\end{align}
Symmetrizing with respect to the quark-antiquark jet interchange, and employing \eqs{pdas}, (\ref{unpdas}), and (\ref{ssa_amps_2}), we get
\begin{align} \label{SSA_TT6}
    z (1-z) \, \frac{1}{2} \, \sum_{S_L, \lambda = \pm 1} S_L  \, \frac{d\sigma_{\mathrm{symm.}\,\lambda \lambda }^{\mathrm{G}[2] \, \gamma^* p \to q {\bar q} p'}}{d^2 p \, d^2 \Delta \, d z}
 &= -\frac{2}{(2\pi)^5s}
        \int d^2 x_1 \, d^2 x_{1'} \, d^2 x_2 \, d^2 x_{2'} \,  
        e^{- i {\un p} \cdot ({\un x}_{12} - \un x_{1'2'}) - i \un \Delta \cdot \big[ z \, \un x_{11'} + (1-z) \un x_{22'} \big] } 
    \\ \notag & \hspace{-4cm}
\times \Bigg\{
\Bigg[ 
    \left( \frac{1}{z} G^{i}_{12}(s) - \frac{1}{1-z} G^{i}_{21}(s) \right) \textrm{Im} \left[ N_{1^\prime 2^\prime}(s) \right] - \left( \frac{1}{z} G^{i, \mathrm{Im}}_{12}(s) - \frac{1}{1-z} G^{i, \mathrm{Im}}_{21}(s) \right) \textrm{Re}\left[ N_{1^\prime 2^\prime}(s) \right] 
    \\ \notag & \hspace{-2cm}
 -  \left( \frac{1}{z} G^{i,\mathrm{NS}}_{12}(s) + \frac{1}{1-z} G^{i,\mathrm{NS}}_{21}(s)
 \right) O_{1^\prime 2^\prime}(s)  \Bigg]
 \left(\partial_{\un 1}^i - i p^i  \right)
        \Phi^{'[2]}_{\mathrm{TT}}(\un x_{12}, \un x_{1^\prime 2^\prime}, z)
        \\ \notag & \hspace{-3cm}
-\Bigg[ 
    \left( G^{i}_{12}(s) +  G^{i}_{21}(s) \right) \textrm{Im} \left[ N_{1^\prime 2^\prime}(s) \right] - \left( G^{i, \mathrm{Im}}_{12}(s) +  G^{i, \mathrm{Im}}_{21}(s) \right) \textrm{Re}\left[ N_{1^\prime 2^\prime}(s) \right] 
    \\ \notag & \hspace{-1cm}
 -  \left(  G^{i,\mathrm{NS}}_{12}(s) - G^{i,\mathrm{NS}}_{21}(s)
 \right) O_{1^\prime 2^\prime}(s)  \Bigg]
 \left( i \Delta^i  \right)
        \Phi^{'[2]}_{\mathrm{TT}}(\un x_{12}, \un x_{1^\prime 2^\prime}, z)
         \Bigg\}.
\end{align}
Again we find a mix of amplitudes that are related to the hPDFs and OAM distributions as well as some that are not. Therefore, separation of the relevant dipole amplitudes appears infeasible at the moment.

%%%%%%%%%%%%%%%%%%%%%%%%%%%
\subsection{T,-T terms}

Finally, we consider the T,-T term from the SSA in \eq{SSA}, 
\begin{align}
  e^{-2i\phi} \sigma^{\gamma^*p}_{1,-1} +\mathrm{c.c.}.
\end{align}
Using \eq{TT1}, we see the type-1 terms do not contribute to the T,-T terms. Thus, we are left to consider the type-2 terms, specifically those coming from the operator associated with $V^{\mathrm{G}[2]}_{\un x, \un y}$. Using \eqs{XS10} and (\ref{type2_overlaps}), we write 
\begin{align}\label{SSA_T-T1}
& z (1-z)\,   \frac{1}{2}\,  \sum_{S_L} \, S_L \, e^{-2i\phi} \, \frac{d\sigma_{1, -1}^{\textrm{G} [2] \gamma^* p \to q {\bar q} p'}}{d^2 p_1 \, d^2 p_2 \, d z} + \mathrm{c.c.} =  -\frac{1}{2 (2 \pi)^5} \, \int d^2 x_1 \, d^2 x_{1'} \, d^2 x_2 \, d^2 x_{2'} \, d^2 x_0 \, e^{- i {\un p}_1 \cdot {\un x}_{11'} - i {\un p}_2 \cdot {\un x}_{22'}}  \\ 
& \times \frac{1}{N_c} \Bigg\{ 
\Phi^{[2]}_{\mathrm{T, -T}}(\un x_{02}, \un x_{1^\prime 2^\prime}, z) \,e^{i(\phi_{02} + \phi_{1'2'} - 2 \phi)} 
\left\langle \tord \tr \left[ V^{\textrm{G} [2]}_{{\un 1}, {\un 0}} \, V_{{\un 2}}^\dagger \right] \right\rangle \big( zs \big) \, 
\Big[ 1- S_{1^\prime 2^\prime}^*(s)\Big]
\notag \\ & \hspace{1cm}
-  
\Phi^{[2]}_{\mathrm{T,-T}}(\un x_{10}, \un x_{1^\prime 2^\prime}, z) \,
e^{i(\phi_{10} + \phi_{1'2'} - 2 \phi)}
\left\langle \tord \tr \left[ V_{{\un 1}} \,  V^{\textrm{G} [2] \, \dagger}_{{\un 2}, {\un 0}} \right] \right\rangle \big((1-z)s \big)
\, \Big[ 1- S_{1^\prime 2^\prime}^*(s) \Big]
\notag \\ & \hspace{1cm}
+  \Phi^{[2]}_{\mathrm{T,-T}}(\un x_{12}, \un x_{0 2^\prime}, z) \,
e^{i(\phi_{02'} + \phi_{12} - 2 \phi)}
\Big[ 1- S_{12}(s)\Big]
\, \left\langle \atord \tr \left[ V_{{\un 2'}} \,  V^{\textrm{G} [2] \, \dagger}_{{\un 1'}, {\un 0}}  \right]  \right\rangle \big( zs \big) \notag \\ & \hspace{1cm}
- \Phi^{[2]}_{\mathrm{T,-T}}(\un x_{12}, \un x_{1^\prime 0}, z) \,
e^{i(\phi_{1'0} + \phi_{12} - 2 \phi)}
\Big[ 1- S_{12}(s)\Big]
\, \left\langle \atord \tr \left[  V^{\textrm{G} [2]}_{{\un 2'}, {\un 0}} \, V_{{\un 1'}}^\dagger \right]  \right\rangle \big((1-z)s \big)   \Bigg\}\notag + \mathrm{c.c.},
\end{align}
where, as in the main text, $\phi_{ij}$ corresponds to the angle in the transverse plane associated with the vector $\un x_{ij}$ and $\phi$ is the angle associated with the outgoing (and incoming) electron's transverse momentum $\un k$. In the last two terms in the curly brackets we relabel $\un x_{1} \leftrightarrow \un x_{1'}, \un x_{2} \leftrightarrow \un x_{2'}$ and take $\un p_1, \un p_2, \hat{k} \to - \un p_1, - \un p_2, - \hat{k}$. The resulting expression is 
\begin{align}\label{SSA_T-T2}
& z (1-z)\, \frac{1}{2}\,  \sum_{S_L} \, S_L \, e^{-2i\phi} \frac{d\sigma_{1, -1}^{\textrm{G} [2] \gamma^* p \to q {\bar q} p'}}{d^2 p_1 \, d^2 p_2 \, d z} + \mathrm{c.c.} =  -\frac{1}{(2 \pi)^5} \, \int d^2 x_1 \, d^2 x_{1'} \, d^2 x_2 \, d^2 x_{2'} \, d^2 x_0 \, e^{- i {\un p}_1 \cdot {\un x}_{11'} - i {\un p}_2 \cdot {\un x}_{22'}}  \\ 
& \times \frac{1}{N_c} \Bigg\{ 
\Phi^{[2]}_{\mathrm{T, -T}}(\un x_{02}, \un x_{1^\prime 2^\prime}, z) \,\cos ( \phi_{02} + \phi_{1'2'} - 2 \phi ) 
\Bigg[
\left\langle \tord \tr \left[ V^{\textrm{G} [2]}_{{\un 1}, {\un 0}} \, V_{{\un 2}}^\dagger \right] \right\rangle \big( zs \big) \, 
\Big[ 1- S_{1^\prime 2^\prime}^*(s)\Big]
+ \mathrm{c.c.}
\Bigg]
\notag \\ & \hspace{1cm}
-  
\Phi^{[2]}_{\mathrm{T,-T}}(\un x_{10}, \un x_{1^\prime 2^\prime}, z) \,
\cos(\phi_{10} + \phi_{1'2'} - 2 \phi)
\Bigg[
\left\langle \tord \tr \left[ V_{{\un 1}} \,  V^{\textrm{G} [2] \, \dagger}_{{\un 2}, {\un 0}} \right] \right\rangle \big((1-z)s \big)
\, \Big[ 1- S_{1^\prime 2^\prime}^*(s) \Big]
+ \mathrm{c.c.}
\Bigg] \Bigg\}, \notag 
\end{align}
where we have used $\Phi^{[2]}_{\mathrm{T,-T}}(\un x, \un x', z) = \Phi^{[2]}_{\mathrm{T,-T}}(\un x', \un x, z)$ and added the complex conjugate explicitly. To proceed we again need to use the substitutions in \eqs{VG2sub}. Similar to the above calculations, we symmetrize with respect to the quark-antiquark jet interchange. We obtain 
\begin{align} \label{SSA_T-T3}
  & z (1-z)\, \frac{1}{2}\,  \sum_{S_L} \, S_L \, e^{-2i\phi} \frac{d\sigma_{1, -1}^{\textrm{G} [2] \gamma^* p \to q {\bar q} p'}}{d^2 p_1 \, d^2 p_2 \, d z} + \mathrm{c.c.} = -\frac{2}{(2\pi)^5s}
        \int d^2 x_1 \, d^2 x_{1'} \, d^2 x_2 \, d^2 x_{2'} \,  
        e^{- i {\un p} \cdot ({\un x}_{12} - \un x_{1'2'}) - i \un \Delta \cdot \big[ z \, \un x_{11'} + (1-z) \un x_{22'} \big] } 
    \\ \notag & \hspace{0cm}
\times \Bigg\{
\Bigg[ 
    \left( \frac{1}{z} G^{i}_{12}(s) - \frac{1}{1-z} G^{i}_{21}(s) \right) \textrm{Im} \left[ N_{1^\prime 2^\prime}(s) \right] - \left( \frac{1}{z} G^{i, \mathrm{Im}}_{12}(s) - \frac{1}{1-z} G^{i, \mathrm{Im}}_{21}(s) \right) \textrm{Re}\left[ N_{1^\prime 2^\prime}(s) \right] 
    \\ \notag & \hspace{0.5cm}
 -  \left( \frac{1}{z} G^{i,\mathrm{NS}}_{12}(s) + \frac{1}{1-z} G^{i,\mathrm{NS}}_{21}(s)
 \right) O_{1^\prime 2^\prime}(s)  \Bigg]
 \left(\partial_{\un 1}^i - i p^i  \right)
        \\ \notag & \hspace{0.5cm}
-\Bigg[ 
    \left( G^{i}_{12}(s) +  G^{i}_{21}(s) \right) \textrm{Im} \left[ N_{1^\prime 2^\prime}(s) \right] - \left( G^{i, \mathrm{Im}}_{12}(s) +  G^{i, \mathrm{Im}}_{21}(s) \right) \textrm{Re}\left[ N_{1^\prime 2^\prime}(s) \right] 
    \\ \notag & \hspace{0.5cm}
 -  \left(  G^{i,\mathrm{NS}}_{12}(s) - G^{i,\mathrm{NS}}_{21}(s)
 \right) O_{1^\prime 2^\prime}(s)  \Bigg]
 \left( i \Delta^i  \right)
         \Bigg\} \cos \left(\phi_{12} + \phi_{1'2'} - 2\phi \right) \Phi^{[2]}_{\mathrm{T,-T}}(\un x_{12}, \un x_{1'2'}, z).
\end{align}

Once more, we see the dipole amplitudes relevant to the OAM distributions and hPDFs are seemingly inseparable from other amplitudes with underdeveloped phenomenology. Therefore, we see explicitly, that the general argument presented in Section~\ref{sec:ssa} holds for the other channels in the SSA from \eq{SSA} not considered explicitly in that Section. We conclude that at the present time, none of the terms constituting the SSA in elastic dijet production from polarized $e+p$ collisions make a feasible probe for the OAM distributions.

%%%%%%%%%%%%%%%%%%%%%%%%%%%%%%%%%%%%%%%%%%%%%%%%%%%%%%%%%%%%%%%%%%%%%%%%%%%%%%%

\providecommand{\href}[2]{#2}\begingroup\raggedright\endgroup

\end{document}